\documentclass[journal,romanappendices]{IEEEtran}
\IEEEoverridecommandlockouts

\usepackage{amsmath,amssymb,amsthm}
\usepackage{stmaryrd}
\usepackage{enumerate}
\usepackage{stfloats}
\usepackage{flushend}
\usepackage{hyperref}

\usepackage{tikz}
\usetikzlibrary{arrows,shapes,chains,matrix,positioning,scopes}
\usepackage{pgfplots}
\usepgflibrary{shapes}
\pgfplotsset{compat=1.4}

\newtheoremstyle{custom}
{} 
{} 
{} 
{} 
{\bfseries} 
{:} 
{.25em} 
{} 
\theoremstyle{custom}

\newtheorem{theorem}{Theorem}
\newtheorem{lemma}[theorem]{Lemma}
\newtheorem{proposition}[theorem]{Proposition}
\newtheorem{definition}[theorem]{Definition}

\newtheorem{remark}[theorem]{Remark}

\newtheorem*{theorem*}{Theorem}
\newtheorem*{lemma*}{Lemma}
\newtheorem*{proposition*}{Proposition}
\newtheorem*{definition*}{Definition}
\newtheorem*{example*}{Example}
\newtheorem*{remark*}{Remark}
\newtheorem*{corollary*}{Corollary}

\renewcommand{\epsilon}{\varepsilon}

\newcommand{\h}{\texttt{h}}
\newcommand{\hbp}{\h^{\mathrm{BP}}}
\newcommand{\hmap}{\h^{\mathrm{MAP}}}
\newcommand{\hstab}{\h^{\mathrm{stab}}}
\newcommand{\harea}{\h^{A}}

\newcommand{\expt}{\mathbb{E}}

\newcommand{\abs}[1]{\left\lvert#1\right\rvert}

\newcommand{\mb}[1]{\mathbf{#1}}
\newcommand{\mbb}[1]{\mathbb{#1}}
\newcommand{\mr}[1]{\mathrm{#1}}
\newcommand{\mc}[1]{\mathcal{#1}}
\newcommand{\ms}[1]{\mathsf{#1}}
\newcommand{\mf}[1]{\mathfrak{#1}}

\newcommand{\mse}{\mathsf{e}}
\newcommand{\msx}{\mathsf{x}}
\newcommand{\msxvn}{\tilde{\mathsf{x}}}
\newcommand{\msy}{\mathsf{y}}
\newcommand{\msz}{\mathsf{z}}
\newcommand{\msa}{\mathsf{a}}
\newcommand{\msb}{\mathsf{b}}
\newcommand{\msc}{\mathsf{c}}
\newcommand{\msd}{\mathsf{d}}
\newcommand{\msbx}{\underline{\mathsf{x}}}
\newcommand{\msby}{\underline{\mathsf{y}}}

\newcommand{\msbz}{\underline{\mathsf{z}}}
\newcommand{\msba}{\underline{\mathsf{a}}}

\newcommand{\bop}{\ast}
\newcommand{\vnop}{\varoast}
\newcommand{\disth}{d_{\mathrm{H}}}
\newcommand{\cnop}{\boxast}
\newcommand{\diff}[1]{d#1}
\newcommand{\deri}[1]{\mathrm{d}_{ #1 }\hspace{0.05cm}}
\newcommand{\dderi}[1]{\mathrm{d}_{ #1 }^2\hspace{0.05cm}}
\newcommand{\bvert}[1]{\,\Big{\vert}_{ #1  }}
\newcommand{\degr}{\succ}
\newcommand{\degreq}{\succeq}
\newcommand{\upgr}{\prec}
\newcommand{\upgreq}{\preceq}
\newcommand{\extR}{\overline{\mathbb{R}}}

\newcommand{\des}{\mathsf{T}_\mathrm{s}}
\newcommand{\dec}{\mathsf{T}_\mathrm{c}}
\newcommand{\pots}{U_\mathrm{s}}
\newcommand{\potc}{U_\mathrm{c}}
\newcommand{\shft}{\mathsf{S}}

\newcommand{\vnunit}{\Delta_0}
\newcommand{\cnunit}{\Delta_\infty}

\newcommand{\ent}[1]{ \mathrm{H} \left( #1 \right) }

\newcommand{\meass}{\mathcal{M}}
\newcommand{\probs}{\mathcal{X}}
\newcommand{\dpros}{\mathcal{X}_{\mathrm{d}}}

\newcommand{\chend}{N_{w}}

\newcommand{\minf}{\mathsf{f}_{0}}
\newcommand{\minfb}{\underline{\minf}}

\newlength\tikzwidth
\newlength\tikzheight

\begin{document}

\title{Threshold Saturation for Spatially-Coupled LDPC and LDGM Codes on BMS Channels}

\author{Santhosh Kumar, Andrew J. Young, Nicolas Macris, and Henry D. Pfister
\thanks{This material is based upon work supported in part by the National Science Foundation (NSF) under Grants No. 0747470 and No. 1320924.
The work of N.~Macris was supported by Swiss National Foundation Grant No. 200020-140388.
Any opinions, findings, conclusions, and recommendations expressed in this material are those of the authors and do not necessarily reflect the views of these sponsors.
This work was presented in part at the Allerton Conference on Communication, Control, and Computing, 2012.
}
\thanks{S.~Kumar is with the Department of Electrical and Computer Engineering, Texas A\&M University, College Station (email: santhosh.kumar@tamu.edu).
}
\thanks{A.~J.~Young is with the Department of Electrical Engineering and Computer Science, Massachusetts Institute of Technology (email: ajy@mit.edu).
}
\thanks{N.~Macris is with the School of Computer and Communication Sciences, \'{E}cole Polytechnique F\'{e}d\'{e}rale de Lausanne, Switzerland (email: nicolas.macris@epfl.ch).
}
\thanks{H.~D.~Pfister is with the Department of Electrical and Computer Engineering, Duke University (email: henry.pfister@duke.edu).
}
}

\maketitle

\begin{abstract}
  Spatially-coupled low-density parity-check (LDPC) codes, which were first introduced as LDPC convolutional codes, have been shown to exhibit excellent performance under low-complexity belief-propagation decoding.
  This phenomenon is now termed threshold saturation via spatial coupling.
  Spatially-coupled codes have been successfully applied in numerous areas.
  In particular, it was proven that spatially-coupled regular LDPC codes universally achieve capacity over the class of binary memoryless symmetric (BMS) channels under belief-propagation decoding.

  Recently, potential functions have been used to simplify threshold saturation proofs for scalar and vector recursions.
  In this paper, potential functions are used to prove threshold saturation for \emph{irregular} LDPC and low-density generator-matrix (LDGM) codes on BMS channels, extending the simplified proof technique to BMS channels.
  The corresponding potential functions are closely related to the average Bethe free entropy of the ensembles in the large-system limit.
  These functions also appear in statistical physics when the replica method is used to analyze optimal decoding.
 \end{abstract}


\section{Introduction}
\label{section:introduction}

Low-density parity-check (LDPC) convolutional codes were introduced in \cite{Felstrom-it99} and shown to have outstanding performance under belief-propagation (BP) decoding in \cite{Sridharan-aller04,Lentmaier-isit05,Lentmaier-it10}.
The fundamental principle behind this phenomenon is described by Kudekar, Richardson, and Urbanke in \cite{Kudekar-it11} and coined \emph{threshold saturation via spatial coupling}.
Roughly speaking, multiple LDPC ensembles are placed next to each other, locally coupled together, and then terminated at the boundaries.
The number of LDPC ensembles is called the \emph{chain length} and the range of local coupling is determined by the \emph{coupling width}.
This termination at the boundary can be regarded as perfect side information for decoding.
Under iterative decoding, this ``perfect'' information propagates inward and dramatically improves performance.
See \cite{Kudekar-scc-tutorial-13} for a tutorial introduction, \cite{Kudekar-it11} for a rigorous construction of spatially-coupled codes, and \cite{Kudekar-it13} for a comprehensive discussion of these codes.

For the binary erasure channel (BEC),  spatially coupling a collection of $(d_v,d_c)$-regular LDPC ensembles produces a new ensemble that is nearly regular.
Moreover, the BP threshold of the coupled ensemble approaches the maximum a posteriori (MAP) threshold of the original ensemble~\cite{Kudekar-it11}. 
Recently, a proof of saturation to the \emph{area threshold} has been given for $(d_v,d_c)$-regular LDPC ensembles on binary memoryless symmetric (BMS) channels under mild conditions~\cite{Kudekar-it13}.
This result implies that spatially-coupled LDPC codes achieve capacity \emph{universally} over the class of BMS channels because the area threshold of regular LDPC codes can approach the Shannon limit uniformly over this class.

The idea of threshold saturation via spatial coupling has started a small revolution in coding theory, and spatially-coupled codes have now been observed to universally approach the capacity regions of many systems~\cite{Lentmaier-it10,Kudekar-istc10,Rathi-isit11,Yedla-isit11,Kudekar-isit11-DEC,Kudekar-isit11-MAC,Nguyen-arxiv11,Nguyen-icc12}.
For spatially-coupled systems with suboptimal component decoders, such as message-passing decoding of code-division multiple access (CDMA)~\cite{Takeuchi-isit11,Schlegel-isit11} or iterative hard-decision decoding of spatially-coupled generalized LDPC codes~\cite{Jian-isit12}, the threshold saturates instead to an intrinsic threshold defined by the suboptimal component decoders.

Spatial-coupling has also led to new results for $K$-SAT, graph coloring, and the Curie-Weiss model in statistical physics~\cite{Hassani-itw10,Hassani-jsm12,Hassani-jsp13}.
For compressive sensing, spatially-coupled measurement matrices were introduced in~\cite{Kudekar-aller10}, shown to give large improvements with Gaussian approximated BP reconstruction in~\cite{Krzakala-physrevx}, and finally proven to achieve the theoretical limit in~\cite{Donoho-arxiv11}.
Recent results based on spatial-coupling are now too numerous to cite thoroughly.

Recently, a simple approach, based on potential functions, is used in \cite{Yedla-istc12,Yedla-itw12} to prove that the BP threshold of spatially-coupled irregular LDPC ensembles over a BEC saturates to the conjectured MAP threshold (known as the Maxwell threshold) of the underlying irregular ensembles.
This technique was motivated by~\cite{Takeuchi-ieice12} and is also related to the continuum approach to density evolution (DE) in which potential functions are used to prove threshold saturation for compressed sensing \cite{Donoho-arxiv11}.

In this paper, the threshold saturation proof based on potential functions in~\cite{Yedla-istc12,Yedla-itw12} is extended to spatially-coupled irregular LDPC and LDGM codes on BMS channels.
The main results are summarized, rather informally, in the following theorems whose proofs comprise the majority of this paper.
See the main text for precise statements and conditions under which the results hold.
Moreover, for LDPC codes, we actually show threshold saturation to a quantity called the \emph{potential threshold}.
For many LDPC ensembles, it is known that the MAP threshold $\hmap$ is upper bounded by the potential threshold.
In some cases, they are actually equal (e.g., see Remark~\ref{remark:threshold_discussion}).

\begin{theorem*}
Consider a spatially-coupled LDPC ensemble and a family of BMS channels that is ordered by degradation, and parameterized by entropy, $\h$.
If $\h<\hmap$, then, for any sufficiently large coupling width, the spatially-coupled DE converges to the perfect decoding solution.
Conversely, if $\h>\hmap$, then for a fixed coupling width and sufficiently large chain length, the spatially-coupled DE \emph{does not} converge to the perfect decoding solution.
\end{theorem*}
Thus, the spatially-coupled BP threshold saturates to $\hmap$ for LDPC codes.

For LDGM codes, message-passing decoding always results in non-negligible error floors.
Even when DE is initialized with perfect information, it converges to a nontrivial \emph{minimal fixed point}.
When a certain quantity, which we call the \emph{energy gap}, is positive, the spatially-coupled DE converges to a fixed point which is elementwise better than the minimal fixed point.
Also, it is conjectured that the MAP decoding performance is governed by the region where the energy gap is positive (e.g., see~Section \ref{subsection:single_system_ldgm}).

\begin{theorem*}
Consider a spatially-coupled LDGM ensemble and a BMS channel.
If the \emph{energy gap} for the channel is positive, then, for sufficiently large coupling width, the spatially-coupled DE converges to a fixed point which is elementwise better than the minimal fixed point of the underlying LDGM ensemble.
\end{theorem*}

A variety of observations, formal proofs, and applications now bear evidence to the generality of threshold saturation.
The technique in \cite{Yedla-istc12,Yedla-itw12} is based on defining a potential function.
The average Bethe free entropy in the large-system limit~\cite{Macris-it07,Mori-isit11} serves as our potential function.
The crucial properties of the free entropy that we leverage are 1) stationary points of the free entropy are related to the fixed points of DE, 2) there exists a spatially-coupled potential, defined by a spatial average of the free entropy, where the fixed points of spatially-coupled DE are stationary points of the spatially-coupled potential.
It is tempting to conjecture that this approach can be applied to more general graphical models by computing their average Bethe free entropy.

\section{Preliminaries}
\label{section:preliminaries}
\subsection{Measures and Algebraic Structure}
\label{subsection:Operators}
Any output $Y$ of a binary-input communication channel, with input $X$, can be represented by the log-likelihood ratio (LLR)
\begin{align*}
  Q = \log \frac{ P_{Y | X}( \alpha | 1 ) }{ P_{Y | X}( \alpha | -1)},
\end{align*}
which is a sufficient statistic for $X$ given $Y$.
Therefore, a communication channel can be associated with a LLR distribution.
If the channel is output symmetric, then it suffices to compute the LLR distribution conditional on $X=1$.
For mathematical convenience, we represent these distributions by measures on the extended real numbers $\extR$.
Thus, $Q$ is represented by a measure $\msx$ where
\begin{align*}
  \Pr (Q \leq t) = \msx([-\infty,t]).
\end{align*}

We call a finite signed Borel measure $\msx$ on $\extR$ \emph{symmetric} if
\begin{align*}
  \msx(-E) = \int_{-E} \msx(\diff{\alpha}) = \int_{E} e^{-\alpha} \msx(\diff{\alpha}),
\end{align*}
for all Borel sets $E \subseteq \extR$, where $\extR$ is a compact metric space under $\tanh(\cdot)$.
This necessarily implies that for any finite symmetric measure $\msx$, $\msx(\{-\infty\}) = e^{-\infty}\msx(\{\infty\})=0$. 
Equivalently, a more operational definition, a finite signed Borel measure $\msx$ is symmetric if
\begin{align*}
  \int_{-E} f( \alpha) \msx(\diff{\alpha}) = \int_{E} f(-\alpha) e^{- \alpha} \msx(\diff{\alpha}),
\end{align*}
for all bounded measurable real-valued functions $f$ and Borel sets $E \subseteq \extR$.
An immediate consequence is the following Proposition.
\begin{proposition}
  \label{proposition:tanh_symmetry}
  Let $\msx$ be a symmetric measure and $f\colon \extR \to \mbb{R}$ be an odd function that is bounded and measurable, then
  \begin{align*}
    \int f(\alpha) \msx(\diff{\alpha}) = \int f(\alpha)\tanh\left( \tfrac{\alpha}{2} \right) \msx(\diff{\alpha}).
  \end{align*}
\end{proposition}
\begin{IEEEproof}
  See Appendix \ref{appendix:proof_tanh_symmetry}.
\end{IEEEproof}
In particular, for a symmetric measure $\msx$ and any natural number $k$,
\begin{align*}
  \int \tanh \left( \tfrac{\alpha}{2} \right)^{2k-1} \msx(\diff{\alpha})
  = \int \tanh \left( \tfrac{\alpha}{2} \right)^{2k} \msx(\diff{\alpha}).
\end{align*}
This last relation is a well-known result and its utility will become apparent in the section on entropy.

Let $\meass$ denote the set of finite signed symmetric Borel measures on the extended real numbers $\extR$.
In this work, the primary focus is on convex combinations and differences of symmetric probability measures, which inherit many of their properties from $\meass$.
Let $\probs \subset \meass$ be the convex subset of symmetric probability measures.
Also, let $\dpros \subset \meass$ be the subset of differences of symmetric probability measures:
\begin{align*}
  \dpros \triangleq \left\{ \msx_1 - \msx_2 \mid \msx_1, \msx_2 \in \probs \right\} .
\end{align*}
In the interest of notational consistency, $\msx$ is reserved for both finite signed symmetric Borel measures and symmetric probability measures, and $\msy$, $\msz$ denote differences of symmetric probability measures.
Also, all logarithms that appear in this article are \emph{natural}, unless the base is explicitly mentioned.

In this space, there are two important binary operators, $\vnop$ and $\cnop$, that denote the variable-node operation and the check-node operation for LLR message distributions, respectively.
Below, we give an explicit integral characterization of the operators $\vnop$ and $\cnop$.
For $\msx_{1}, \msx_2 \in \meass$, and any Borel set $E \subset \extR$, define
\begin{align*}
  ( \msx_{1} \vnop \msx_{2} )(E) &\triangleq \int \msx_{1}( E-\alpha ) \, \msx_{2} ( \diff{\alpha} ), \\
  ( \msx_{1} \cnop \msx_{2} )(E) &\triangleq \int \msx_{1} \left( 2 \tanh^{-1} \left( \frac{\tanh(\frac{E}{2})}{\tanh(\frac{\alpha}{2})} \right) \right) \msx_{2} ( \diff{\alpha} ).
\end{align*}
Equivalently, for any bounded measurable real-valued function $f$,
\begin{align*}
  \int f \diff{(\msx_{1}\vnop\msx_{2})} &= \iint f(\alpha_1 + \alpha_2) \, \msx_{1}(\diff{\alpha_1}) \, \msx_{2}(\diff{\alpha_2}), \\
  \int f \diff{(\msx_{1}\cnop\msx_{2})} &= \iint f(\tau^{-1}( \tau(\alpha_1) \tau(\alpha_2))) \, \msx_{1}(\diff{ \alpha_1}) \, \msx_{2}(\diff{\alpha_2}),
\end{align*}
where $\tau\colon\extR \rightarrow [-1,1]$, $\tau(\alpha) = \tanh \left( \frac{ \alpha}{2} \right)$.
Associativity, commutativity, and linearity of the operators $\vnop$, $\cnop$ are inherited from the underlying algebraic structure of $(\extR,+)$, $([-1,1],\cdotp)$, respectively.
Moreover, the space of symmetric probability measures is closed under these binary operations~\cite[Theorem 4.29]{RU-2008}.

In a more abstract sense, the measure space $\meass$ along with either multiplication operator ($\vnop$, $\cnop$) forms a commutative monoid, and this algebraic structure is induced on the space of symmetric probability measures $\probs$.
There is also an intrinsic connection between the algebras defined by each operator and one consequence is the duality (or conservation) result in Proposition \ref{proposition:duality}.
The identities in these algebras, $\mse_{\vnop} = \vnunit$ and $\mse_{\cnop} = \cnunit$, also exhibit an annihilator property under the dual operation
\begin{align*}
  \vnunit \cnop \ms{x} &= \vnunit, & \cnunit \vnop \ms{x} &= \cnunit.
\end{align*}

The wildcard $\bop$ is used to represent either operator in statements that apply to both operations.
For example, the shorthand $\msx^{\bop n}$ is used to denote $n$ fold operations
\begin{align*}
  \msx^{\bop n} = \underbrace{\msx \bop \cdots \bop \msx}_{n}\, ,
\end{align*}
and this notation is extended to polynomials.
In particular, for a polynomial $p(t) = \sum_{n=0}^{\mathrm{deg}(p)} p_{n} t^{n}$ with real coefficients, we define
\begin{align*}
  p^{\bop}(\msx) \triangleq \sum_{n=0}^{\mathrm{deg}(p)} p_{n} \msx^{\bop n} ,
\end{align*}
where we define $\msx^{\bop 0} \triangleq \mse_{\bop}$.
For the formal derivative $p'(t) = \frac{\diff p}{\diff t}$, we have
\begin{align*}
  p'^{\bop}(\msx) = \sum_{n=0}^{\mathrm{deg}(p)} n p_{n} \msx^{\bop n-1} .
\end{align*}

In general, the operators $\vnop$, $\cnop$ do not associate
\begin{align*}
  \msx_{1} \vnop (\msx_{2} \cnop \msx_{3}) &\neq (\msx_{1} \vnop \msx_{2}) \cnop \msx_{3} \\
  \msx_{1} \cnop (\msx_{2} \vnop \msx_{3}) &\neq (\msx_{1} \cnop \msx_{2}) \vnop \msx_{3},
\end{align*}
nor distribute
\begin{align*}
  \msx_{1} \vnop (\msx_{2} \cnop \msx_{3}) &\neq (\msx_{1} \vnop \msx_{2}) \cnop ( \msx_{1} \vnop \msx_{3} )\\
  \msx_{1} \cnop (\msx_{2} \vnop \msx_{3}) &\neq (\msx_{1} \cnop \msx_{2}) \vnop ( \msx_{1} \cnop \msx_{3} ).
\end{align*}

\subsection{Partial Ordering by Degradation}
\label{subsection:Degradation}
Degradation is an important concept that allows one to compare some LLR message distributions.
The order imposed by degradation is indicative of relating probability measures through a communication channel \cite[Definition 4.69]{RU-2008}.
The following is one of several equivalent definitions and is the most suitable for our purposes.
\begin{definition}
  \label{definition:degradation}
  For $\msx \in \probs$ and $f\colon [0,1]\to\mbb{R}$, define
  \begin{align*}
    I_f(\msx)  \triangleq \int f \left( \abs{ \tanh\left(\tfrac{\alpha}{2}\right) } \right) \msx( \diff{\alpha} ) .
  \end{align*}
  For $\msx_1, \msx_2 \in \probs$, $\msx_1$ is said to be \emph{degraded} with respect to $\msx_2$ (denoted $\msx_1 \degreq \msx_2$), if $I_f(\msx_1) \geq I_f(\msx_2)$ for all concave non-increasing $f$.
  Furthermore, $\msx_1$ is said to be \emph{strictly degraded} with respect to $\msx_2$ (denoted $\msx_1 \degr \msx_2$) if $\msx_1 \degreq \msx_2$ and $\msx_1 \neq \msx_2$.
  We also write $\msx_2 \upgreq \msx_1$ (respectively, $\msx_2 \upgr \msx_1$) to mean $\msx_1 \degreq \msx_2$ (respectively, $\msx_1 \degr \msx_2$).
\end{definition}
Recall that two measures $\msx_1$, $\msx_2$ are equal if $\msx_1(E)=\msx_2(E)$ for all Borel sets $E \subseteq \extR$.
The class of concave non-increasing functions is rich enough to capture the notion of non-equality.
That is, if $\msx_1 \neq \msx_2$, then there exists a concave non-increasing $f \colon [0,1] \to \mbb{R}$ such that $I_f(\msx_1) \neq I_f(\msx_2)$.

Degradation defines a partial order on the space of symmetric probability measures, with the greatest element $\vnunit$ and the least element $\cnunit$.
Thus  
\begin{align*}
  \text{$\msx \degr \cnunit$ if $\msx \neq \cnunit$, and $\msx \upgr \vnunit$ if $\msx \neq \vnunit$.}
\end{align*}

This partial ordering is also preserved under the binary operations as follows.
\begin{proposition}
  \label{proposition:degradation_preservation}
  Suppose $\msx_1, \msx_2, \msx_3 \in \probs$.
  \begin{enumerate}[i)]
    \item 
    If $\msx_{1} \degreq \msx_{2}$, then
    \begin{align*}
      \msx_{1} \bop \msx_{3} \degreq \msx_{2} \bop \msx_{3} , \quad \text{for all $\msx_3 \in \probs$}.
    \end{align*}
    \item The operators $\vnop$ and $\cnop$ also preserve a strict ordering for non-extremal measures.
      That is, if $\msx_1 \degr \msx_2$, then
      \begin{align*}
        \msx_1 \vnop \msx_3 \degr \msx_2 \vnop \msx_3 &\quad \text{for $\msx_3 \neq \cnunit$}, \\
        \msx_1 \cnop \msx_3 \degr \msx_2 \cnop \msx_3 &\quad \text{for $\msx_3 \neq \vnunit$}.
      \end{align*}
  \end{enumerate}
\end{proposition}
\begin{IEEEproof}
\begin{enumerate}[i)]
\item
Direct application of \cite[Lemma 4.80]{RU-2008}.
\item
It suffices to show that $\msx_1 \bop \msx_3 \neq \msx_2 \bop \msx_3$ under the stated conditions.
  For this, it is sufficient to construct a functional which gives different values under $\msx_1 \bop \msx_3$ and $\msx_2 \bop \msx_3$.
  The entropy functional (e.g., see Proposition \ref{proposition:entropy_dpros_properties}(iv)) provides such a property.
\end{enumerate}
\end{IEEEproof}
Order by degradation is also preserved, much like the standard order of real numbers, under nonnegative multiplications and additions, i.e. for $0 \leq \alpha \leq 1$ and $\msx_{1} \degreq \msx_{2}$, $\msx_{3} \degreq \msx_{4}$,
\begin{equation*}
  \alpha \msx_{1} + (1-\alpha) \msx_{3} \degreq \alpha \msx_{2} + (1-\alpha) \msx_{4}.
\end{equation*}
This ordering is our primary tool in describing relative channel quality.
For further information see  \cite[pp.~204-208]{RU-2008}.

\subsection{Entropy Functional for Symmetric Measures}
To explicitly quantify the difference between two symmetric measures, one can employ the entropy functional.
The entropy functional is the linear functional $\mathrm{H}\colon \meass \rightarrow \mbb{R}$ defined by
\begin{align*}
  \ent{\msx} \triangleq \int \log_2 \left(1 + e^{-\alpha} \right) \msx(\diff{\alpha}).
\end{align*}
This is the primary functional used in our analysis.
It preserves the partial order under degradation and for $\msx_{1}, \msx_{2} \in \probs$, we have
\begin{align*}
  \ent{ \msx_{1} } & > \ent{ \msx_{2} } \quad \text{for $\msx_{1} \degr \msx_{2}$.}
\end{align*}
The restriction to symmetric probability measures also implies the bound
\begin{equation*}
  0 \leq \ent{\msx} \leq 1, \qquad \text{if $\msx \in \probs$}.
\end{equation*}

The operators $\vnop$ and $\cnop$ admit a number of relationships under the entropy functional. 
The following results will prove invaluable in the ensuing analysis.
Proposition \ref{proposition:duality} provides an important conservation result (also known as the duality rule for entropy) and Proposition \ref{proposition:duality_difference} extends this relation to encompass differences of symmetric probability measures.
\begin{proposition}[{\cite[Lemma 4.41]{RU-2008}}]
  \label{proposition:duality}
  For $\msx_{1},\msx_{2} \in \probs$,
  \begin{align*}
    \ent{ \msx_{1} \vnop \msx_{2} } + \ent{ \msx_{1} \cnop \msx_{2} } = \ent{ \msx_{1} } + \ent{ \msx_{2} }.
  \end{align*}
\end{proposition}

\begin{proposition}
  \label{proposition:duality_difference}
  For $\msx_{1}$, $\msx_{2}$, $\msx_{3}$, $\msx_{4} \in \probs$,
  \begin{align*}
    \quad \ent{ \msx_{1} \vnop (\msx_{3} - \msx_{4}) }  + \ent{ \msx_{1} \cnop (\msx_{3} - \msx_{4}) } 
    = \ent{ \msx_{3} - \msx_{4} },
  \end{align*}
  \begin{align*}
    \ent{  (\msx_{1} - \msx_{2}) \vnop (\msx_{3} - \msx_{4}) }  + \ent{ (\msx_{1} - \msx_{2}) \cnop (\msx_{3} - \msx_{4}) } = 0.
  \end{align*}
\end{proposition}

\begin{IEEEproof}
  Consider the LHS of the first equality,
  \begin{align*}
    & \ent{ \msx_{1} \vnop (\msx_{3} - \msx_{4}) } + \ent{ \msx_{1} \cnop (\msx_{3} - \msx_{4}) } \\
    &=  \ent{ \msx_{1} \vnop \msx_{3} } + \ent{ \msx_{1} \cnop \msx_{3} } - \ent{ \msx_{1} \vnop \msx_{4} } - \ent{ \msx_{1} \cnop \msx_{4} } \\
    &= \ent{ \msx_{1} } + \ent{ \msx_{3} } - \ent{ \msx_{1} } - \ent{ \msx_{4} }  \quad \text{(Proposition \ref{proposition:duality})}  \\
    &= \ent{ \msx_{3}- \msx_{4} }.
  \end{align*}
  The second equality follows by expanding the LHS and applying the first equality twice.
\end{IEEEproof}

For $k \in \mbb{N}$, let $M_k\colon \meass \to \mbb{R}$ denote the linear functional that maps $\msx \in \meass$ to its $2k$-th moment under $\tanh$,
\begin{align*}
  M_{k}(\msx) \triangleq \int  \tanh^{2k} \left( \tfrac{\alpha}{2} \right) \msx(\diff{\alpha}) .
\end{align*}
\begin{proposition}
  \label{proposition:Mk_properties}
  The following results hold.
  \begin{enumerate}[i)]
  \item For $\msx \in \probs$, $0 \leq M_k(\msx) \leq 1$.
  \item For $\msx_1,\msx_2 \in \probs$ with $\msx_1 \degreq \msx_2$, $M_k(\msx_1) \leq M_k(\msx_2)$.
  \item $M_k$ satisfies the following product form identity for the operator $\cnop$,
    \begin{align*}
      M_{k}(\msx_1 \cnop \msx_2) = M_{k}(\msx_1) M_{k}(\msx_2) .
    \end{align*}
  \item If $\msx=\cnunit$ (respectively, $\msx=\vnunit$), $M_k(\msx)=1$ (respectively, $M_k(\msx)=0$) for all $k$.
    Conversely, for some $\msx \in \probs$, if $M_k(\msx)=1$ (respectively, $M_k(\msx)=0$) for some $k$, then $\msx=\cnunit$ (respectively, $\msx=\vnunit$).
 \end{enumerate}
\end{proposition}
\begin{IEEEproof}
  See Appendix \ref{appendix:proof_Mk_properties}.
\end{IEEEproof}

Due to the symmetry of the measures, the entropy functional has an equivalent series representation in terms of the moments $M_k$.
\begin{proposition}[{\cite[Lemma 3]{Montanari-it05}}]
  \label{proposition:entropy_symmetric_measures}
  If $\msx \in \meass$, then
  \begin{align*}
    \ent{\msx} = \msx \left( \extR \right) - \sum_{k=1}^{\infty} \gamma_k M_{k}(\msx) , \quad \text{where $\gamma_k=\frac{\left( \log 2 \right)^{-1}}{2k (2k-1)}$}.
  \end{align*}
\end{proposition}
\begin{IEEEproof}
  The main idea is to observe that
  \begin{align*}
    \log_2(1+e^{-\alpha}) = 1 - \log_2(1+\tanh(\tfrac{\alpha}{2})) .
  \end{align*}
  From there, use the series expansion of $\log_2(1+t)$ and Proposition \ref{proposition:tanh_symmetry} to combine the odd and even $\tanh$ moments.
  For a detailed proof, see \cite[Lemma 3]{Montanari-it05} and \cite[pp. 267-268]{RU-2008}.
\end{IEEEproof}

\begin{proposition}
  \label{proposition:entropy_dpros_properties}
  From the series expansion for symmetric measures, the entropy functional satisfies the following properties.
  \begin{enumerate}[i)]
  \item For $\msy_1, \msy_2 \in \dpros$,
    \begin{align*}
      \ent{\msy_{1}} &= - \sum_{k=1}^{\infty} \gamma_k M_{k}(\msy_{1}), \\
      \ent{\msy_1 \cnop \msy_2} &= - \sum_{k=1}^{\infty} \gamma_k M_{k}(\msy_1) M_{k}(\msy_2) .
    \end{align*}
  \item For $\msy \in \dpros$,
    \begin{align*}
      \ent{\msy \cnop \msy} = - \sum_{k=1}^{\infty} & \gamma_k M_{k}(\msy)^{2} \leq 0 , &  \ent{\msy \vnop \msy} &\geq 0 .
    \end{align*}
    with equality iff $\msy=0$.
    Additionally if $\msx \in \probs$,
    \begin{align*}
      \ent{\msy \cnop \msy \cnop \msx} &\leq 0 ,
    \end{align*}
    with equality iff $\msy=0$ or $\msx=\vnunit$.
  \item If $\msy_1=\msx'_1-\msx_1$, $\msy_2=\msx'_2-\msx_2$ with $\msx'_1 \degreq \msx_1$, $\msx'_2 \degreq \msx_2$,
    \begin{align*}
      \ent{\msy_1 \cnop \msy_2} &\leq 0,  & \ent{\msy_1 \vnop \msy_2} &\geq 0.
    \end{align*}
  \item If $\msx_1 \degr \msx_2$, then
    \begin{align*}
      \ent{\msx_1 \vnop \msx_3} > \ent{\msx_2 \vnop \msx_3} \quad \text{if $\msx_3 \neq \cnunit$}  \\
      \ent{\msx_1 \cnop \msx_3} > \ent{\msx_2 \cnop \msx_3} \quad \text{if $\msx_3 \neq \vnunit$} .
    \end{align*}
  \end{enumerate}
\end{proposition}
\begin{IEEEproof}
  See Appendix \ref{appendix:proof_entropy_dpros_properties}.
\end{IEEEproof}

Proposition \ref{proposition:entropy_dpros_properties} also implies the following upper bound on the entropy functional for differences of symmetric probability measures under the operators $\vnop$ and $\cnop$. 
\begin{proposition}
  \label{proposition:entropy_bound}
  For $\msx_1, \msx_1', \msx_2, \msx_3, \msx_4 \in \probs$ with $\msx_1' \degreq \msx_1$,
  \begin{align*}
    \abs{\ent{ \left( \msx_1' - \msx_1 \right) \bop \left( \msx_2 - \msx_3 \right) }} \leq \ent{ \msx_1' - \msx_1 }, \\
    \abs{\ent{\left( \msx_1' - \msx_1 \right) \bop \left( \msx_2 - \msx_3 \right) \bop \msx_4}} \leq \ent{\msx_1' - \msx_1} .
  \end{align*}
\end{proposition}
\begin{IEEEproof}
  Consider the first inequality with the operator $\cnop$.
  From Proposition \ref{proposition:entropy_dpros_properties}(i),
  \allowdisplaybreaks{
    \begin{align*}
      & \abs{\ent{\left( \msx_1' - \msx_1 \right)  \cnop \left( \msx_2 - \msx_3 \right) }} \\
      & \qquad \le \sum_{k=1}^{\infty} \gamma_k \abs{ M_{k}(\msx_1' - \msx_1)}\abs{ M_{k}(\msx_2 - \msx_3)} \\
      & \qquad \overset{(a)}{=} - \sum_{k=1}^{\infty} \gamma_k M_{k}(\msx_1' - \msx_1) \abs{ M_{k}(\msx_2 - \msx_3) } \\
      & \qquad \overset{(b)}{\leq} - \sum_{k=1}^{\infty} \gamma_k M_{k}(\msx_1' - \msx_1) \\
      & \qquad = \ent{\msx_1' - \msx_1} ,
    \end{align*}
    where $(a)$ follows from $M_{k}(\msx_1') \leq M_{k}(\msx_1)$ and $(b)$ follows since $0 \leq M_{k}(\msx_2), M_{k}(\msx_3) \leq 1$.
    The result for the operator $\vnop$ then follows from Proposition \ref{proposition:duality_difference}.
    The second inequality follows from the first by replacing $\msx_2$, $\msx_3$ with $\msx_2 \bop \msx_4$, $\msx_3 \bop \msx_4$.
  }
\end{IEEEproof}

The series expansion in Proposition \ref{proposition:entropy_symmetric_measures} leads us to define the following metric on the set of symmetric probability measures.
\begin{definition}
  \label{definition:entropy_distance}
  For $\msx_1,\msx_2 \in \probs$, the \emph{entropy distance} is defined as
  \begin{align*}
    \disth(\msx_1,\msx_2) = \sum_{k=1}^{\infty} \gamma_k \abs{M_k(\msx_1)-M_k(\msx_2)} .
  \end{align*}
\end{definition}
When $\msx_2 \degreq \msx_1$, observe that $\disth(\msx_1,\msx_2)=\ent{\msx_2-\msx_1}$; hence the name entropy distance.
Thus, $\disth(\cnunit,\msx)=\ent{\msx}$ and $\disth(\msx,\vnunit)=1-\ent{\msx}$.
Moreover, for any $\msx_1,\msx_2 \in \probs$, $\disth(\msx_1,\msx_2) \geq \abs{\ent{\msx_1-\msx_2}}$, and for $\msx_{3}\degreq \msx_{2} \degreq \msx_{1}$, $\disth(\msx_{1},\msx_{3}) \ge \disth(\msx_{1},\msx_{2})$.

\begin{proposition}
  \label{proposition:entropy_distance_properties}
  We have the following topological results related to the entropy distance.
  \begin{enumerate}[i)]
  \item The entropy distance $\disth$ is a metric on the set of symmetric probability measures, $\probs$.
  \item The metric topology $(\probs,\disth)$ is compact and hence complete.
  \item The entropy functional $\mathrm{H}\colon \probs \rightarrow [0,1]$ is continuous.
  \item With the product topology on $\probs \times \probs$, the operators $\vnop\colon \probs \times \probs \rightarrow \probs$ and $\cnop\colon \probs \times \probs \rightarrow \probs$ are continuous.
  \item If a sequence of measures $\{\msx_n\}_{n=1}^{\infty}$ in $\probs$ satisfies $\msx_n \degreq \msx_{n-1}$ (respectively, $\msx_n \upgreq \msx_{n-1}$), then $\msx_n \xrightarrow{\disth} \msx$, for some $\msx \in \probs$, and $\msx \degreq \msx_n$ (respectively, $\msx \upgreq \msx_n$) for all $n$.
  \item If $\msx'_n \degreq \msx_n$ and $\msx'_n \xrightarrow{\disth} \msx'$, $\msx_n \xrightarrow{\disth} \msx$, then $\msx' \degreq \msx$.   
  \end{enumerate}
\end{proposition}
\begin{IEEEproof}
  See Appendix \ref{appendix:topology}.
\end{IEEEproof}
We use these topological results minimally.
The compactness of $\probs$ and the continuity of $\ent{\cdot}$, $\vnop$ and $\cnop$ are used to establish the existence of minimizing measures for some functionals.
These minima are used to show the threshold saturation converse for LDPC ensembles.
For the achievability result (Theorems \ref{theorem:threshold_saturation_ldpc} and \ref{theorem:threshold_saturation_ldgm}), we require properties (v) and (vi) in the above proposition, which appear in \cite[Section 4.1]{RU-2008}.
We note that our previous article, \cite{Kumar-aller12}, shows the achievability of threshold saturation for LDPC ensembles using only existing convergence results from \cite[Section 4.1]{RU-2008}.

\subsection{Bhattacharyya Functional for Symmetric Measures}
The quantity that characterizes the stability of LDPC ensembles is the Bhattacharyya functional, $\mf{B}\colon \meass \rightarrow \mbb{R}$,
\begin{align*}
  \mf{B}(\msx) \triangleq \int e^{-\alpha/2} \msx(\diff{\alpha}) .
\end{align*}
Since this is a Laplace transform of the measure evaluated at $1/2$, Bhattacharyya functional is multiplicative under the convolution operator $\vnop$,
\begin{align*}
  \mf{B}(\msx^{\vnop n}) = \mf{B}(\msx)^{n} .
\end{align*}
Like the entropy functional, the Bhattacharyya functional also preserves the degradation order,
\begin{align*}
  \mf{B}(\msx_1) &> \mf{B}(\msx_2), \quad \text{if $\msx_1 \degr \msx_2$} .
\end{align*}
It also satisfies the bound
\begin{align*}
  0 \leq \mf{B}(\msx) \leq 1, \quad \text{if $\msx \in \probs$} .
\end{align*}
Importantly, the Bhattacharyya functional characterizes the logarithmic decay rate of the entropy functional under the operator $\vnop$.
\begin{proposition}
  \label{proposition:entropy_vnop_decay_rate}
  For $\msx \in \probs$, 
  \begin{align*}
    \lim_{n \rightarrow \infty} \frac{1}{n} \log \ent{\msx^{\vnop n}} = \log \mf{B}(\msx) .
  \end{align*}
\end{proposition}
\begin{IEEEproof}
  See Appendix \ref{appendix:proof_entropy_vnop_decay_rate}.
\end{IEEEproof}

\subsection{Directional Derivatives}
\label{subsection:DirectionalDerivative}
The main result in this paper is derived using potential theory and differential relations.
One can avoid some technical challenges of differentiation in the abstract space of measures by focusing on directional derivatives of functionals that map measures to real numbers.
\begin{definition}
  \label{definition:directional_derivative}
  Let $F\colon\meass \rightarrow \mbb{R}$ be a functional on $\mc{M}$.
  The \emph{directional derivative} of $F$ at $\msx$ in the direction $\msy$ is
  \begin{align*}
    \deri{\msx} F(\msx)[\msy] \triangleq \lim_{\delta \rightarrow 0} \frac{F(\msx + \delta \msy ) - F(\msx)}{\delta} ,
  \end{align*}
  whenever the limit exists.
  For $G\colon \meass \to \meass$, define
  \begin{align*}
    \deri{\msx} F(G(\msx))[\msy] &\triangleq \deri{\msx} (F \circ G)(\msx)[\msy] \\
    &= \lim_{\delta \to 0} \frac{F(G(\msx+\delta \msy)) - F(G(\msx))}{\delta} ,
  \end{align*}
  whenever the limit exists.
  For convenience, we sometimes write
  \begin{align*}
    \deri{\msx} F(\msx)[\msy] \bvert{\msx=\msx_1} \triangleq \deri{\msx_1} F(\msx_1) [\msy] .
  \end{align*}
\end{definition}
This definition is naturally extended to higher-order directional derivatives using
\begin{align*}
  \mathrm{d}_{\msx}^{n}\hspace{0.05cm} F(\msx) [ \msy_{1}, \ldots, \msy_{n} ] \triangleq \deri{\msx} \left( \cdots \deri{\msx} \left( \deri{\msx}  F \left( \msx \right) [ \msy_{1} ] \right) [\msy_{2}]  \cdots \right) [\msy_{n}],
\end{align*}
and vectors of measures using, for $\msbx = [\msx_1,\cdots,\msx_m]$, 
\begin{align*}
  \deri{\msbx} F(\msbx)[\msby] &\triangleq \lim_{\delta \rightarrow 0} \frac{F(\msbx + \delta \msby ) - F(\msbx)}{\delta},
\end{align*}
whenever the limit exists.
Similarly, we can define higher-order directional derivatives for the composition of functions and functionals on vectors of measures.

The utility of directional derivatives for linear functionals is evident from the following result.
\begin{proposition}
  \label{proposition:directional_derivative_single_operator}
  Let $F\colon \meass \rightarrow \mbb{R}$ be a linear functional, and $\bop$ be either $\vnop$ or $\cnop$.
  Then, for $\msx,\msy, \msz \in \meass$, we have
  \begin{align*}
    \deri{\msx} F(\msx^{\bop n})[\msy] &= n F( \msx^{\bop (n-1)} \bop \msy ) , \\
    \dderi{\msx} F(\msx^{\bop n})[\msy,\msz] &= n \left( n-1 \right) F \left( \msx^{\bop (n-2)} \bop \msy \bop \msz  \right) .
  \end{align*}
\end{proposition}
\begin{IEEEproof}
  Associativity, commutativity, and linearity of the binary operator $\bop$ allow a binomial expansion of $\left( \msx + \delta \msy \right)^{\bop n}$:
  \begin{align*}
    \left( \msx + \delta \msy \right)^{\bop n} = \sum_{i=0}^{n} \delta^{i} \binom{n}{i} \msx^{\bop (n-i)} \bop \msy^{\bop i}.
  \end{align*}
  Then, the linearity of $F$ implies that
  \begin{align*}
    & F \left( (\msx + \delta \msy)^{\bop n} \right) - F \left( \msx^{\bop n} \right)  \\
    & \quad = \delta n F( \msx^{\bop (n-1)} \bop \msy ) + \sum_{i=2}^{n} \delta^{i} \binom{n}{i} F( \msx^{\bop (n-i)} \bop \msy^{\bop i} ).
  \end{align*}
  Dividing by $\delta$ and taking a limit gives
  \begin{align*}
    \deri{\msx} F( \msx^{\bop n} )  [\msy] = n F( \msx^{\bop (n-1)} \bop \msy ).
  \end{align*}
  An analogous argument shows that
  \begin{align*}
    \dderi{\msx} F( \msx^{\bop n} ) [\msy,\msz]  = n (n-1) F(\msx^{\bop (n-2)} \bop \msy \bop \msz ).
  \end{align*}
\end{IEEEproof}

In the following proposition, we evaluate the directional derivative of a linear functional which contains both the operators $\vnop$ and $\cnop$.
\begin{proposition}
  \label{proposition:directional_derivative_multiple_operators}
  Suppose $F\colon \meass \to \mbb{R}$ is a linear functional and $p$, $q$ are polynomials.
  Then
  \begin{align*}
    \deri{\msx} F(p^{\vnop}(q^{\cnop}(\msx)))[\msy] = F\left(p'^{\vnop}\left( q^{\cnop}(\msx) \right) \vnop \left(q'^{\cnop}(\msx) \cnop \msy \right) \right) .
  \end{align*}
\end{proposition}
\begin{IEEEproof}
  Since $F$ is a linear functional, it suffices to show the result when $p(\alpha)=\alpha^n$.
  In view of the proof of previous proposition, the coefficient of $\delta$ in
  \begin{align*}
    (q^{\cnop}(\msx+\delta \msy))^{\vnop n} - (q^{\cnop}(\msx))^{\vnop n} 
  \end{align*}
  determines the first-order directional derivative.
  Again, from the binomial expansion,
  \begin{align*}
    & (q^{\cnop}(\msx+\delta \msy))^{\vnop n} - (q^{\cnop}(\msx))^{\vnop n} \\
    & = \Big{(}\sum_{k=0}^{\mathrm{deg(q)}} q_k (\msx+\delta \msy)^{\cnop k} \Big{)}^{\vnop n} - (q^{\cnop}(\msx))^{\vnop n} \\
    & = \Big{(}q^{\cnop}(\msx) \!+\! \Big{(} \sum_{k=1}^{\mathrm{deg}(q)} k q_k \msx^{\cnop k-1}\!\cnop \msy \Big{)} \delta + o(\delta) \Big{)}^{\vnop n} \!-\! (q^{\cnop}(\msx))^{\vnop n} \\
    &= \left(q^{\cnop}(\msx) + \left( q'^{\cnop}(\msx) \cnop \msy \right) \delta  + o(\delta)  \right)^{\vnop n} - (q^{\cnop}(\msx))^{\vnop n} 
  \end{align*}
  A direct inspection from the multinomial expansion of the first term gives the coefficient of $\delta$,
  \begin{align*}
    n \left(\left(q^{\cnop}(\msx)\right)^{\vnop n-1}\right) \vnop (q'^{\cnop}(\msx) \cnop \msy) .
  \end{align*}
  Thus, when $p(\alpha)=\alpha^n$,
  \begin{align*}
    \deri{\msx} F(p^{\vnop}(q^{\cnop}(\msx)))[\msy] = F\left(p'^{\vnop}\left( q^{\cnop}(\msx) \right) \vnop \left(q'^{\cnop}(\msx) \cnop \msy \right) \right) .
  \end{align*}
  The general result follows.
\end{IEEEproof}

One recurring theme in this article when relating two quantities $F(\msx_1)$, $F(\msx_2)$ is to consider a parameterized path from $\msx_1$ to $\msx_2$, of the form $\msx_1 + t(\msx_2-\msx_1)=(1-t)\msx_1+t \msx_2$, in the set of symmetric probability measures, and analyze the directional derivative of $F(\cdot)$ at $\msx_1 + t(\msx_2-\msx_1)$, in the direction $\msx_2-\msx_1$.
The following proposition formalizes this idea.
\begin{proposition}
  \label{proposition:polynomial_structure_example}
  Let $F \colon \probs \rightarrow \mbb{R}$ be a linear functional, $\bop$ either $\vnop$ or $\cnop$, $p$ a polynomial, and $G\colon \probs \rightarrow \mbb{R}$, $G(\msx)=F(p^*(\msx))$. 
For $\msx_1,\msx_2 \in \probs$, let $\phi\colon [0,1] \rightarrow \mbb{R}$,
  \begin{align*}
    \phi(t) = G(\msx_1 + t(\msx_2-\msx_1)).
  \end{align*}
  Then, $\phi(t)$ is a polynomial in $t$, 
  \begin{align*}
    \phi'(t)&=\deri{\msx} G(\msx) [\msx_2-\msx_1] \bvert{\msx=\msx_1+t(\msx_2-\msx_1)} , \, \text{and}\\
    \phi''(t)&=\dderi{\msx} G(\msx) [\msx_2-\msx_1,\msx_2-\msx_1] \bvert{\msx=\msx_1+t(\msx_2-\msx_1)}.
  \end{align*}
\end{proposition}
\begin{IEEEproof}
  Since $\msx_1+t(\msx_2-\msx_1)=(1-t)\msx_1+t \msx_2$, from the binomial expansion,
  \begin{align*}
    (\msx_1+t(\msx_2-\msx_1))^{\bop n} = \sum_{k=0}^{n} \binom{n}{k} \left( \msx_1^{\bop n-k} \bop \msx_2^{\bop k} \right) (1-t)^{n-k}t^{k} . 
  \end{align*}
  Since $F$ is a linear functional,
  \begin{align*}
    \phi(t) &= G(\msx_1+t(\msx_2-\msx_1)) \\
    &= F(p^*(\msx_1+t(\msx_2-\msx_1)))\\
    &= \sum_{n=0}^{\mathrm{deg}(p)} p_n F\left( (\msx_1+t(\msx_2-\msx_1)^{\bop n} \right) \\
    &= \sum_{n=0}^{\mathrm{deg}(p)} p_n \sum_{k=0}^{n}  \binom{n}{k} F\left( \msx_1^{\bop n-k} \bop \msx_2^{\bop k} \right) (1-t)^{n-k}t^{k} ,
  \end{align*}
  is a polynomial of degree at most $\mathrm{deg}(p)$.
  Moreover,
  \begin{align*}
    \phi'(t) &= \lim_{\delta \to 0} \frac{G(\msx_1+(t+\delta)(\msx_2-\msx_1))-G(\msx_1+t(\msx_2-\msx_1))}{\delta} \\
    &= \deri{\msx}G(\msx)[\msx_2-\msx_1] \bvert{\msx=\msx_1+t(\msx_2-\msx_1)},
  \end{align*}
  by Definition \ref{definition:directional_derivative}.
  The expression for second derivative $\phi''(t)$ follows similarly.
\end{IEEEproof}
As such, if $\phi'(t) \leq 0$ in the above proposition for all $t \in (0,1)$, we find that $G(\msx_1) \leq G(\msx_2)$ because $\phi(0)=G(\msx_1)$, $\phi(1)=G(\msx_2)$.

\begin{remark}
  In general, applying Taylor's theorem to some mapping $F\colon \probs \to \probs$ requires Fr\'{e}chet derivatives.
  However, the linearity of the entropy functional and its interplay with the operators $\vnop$ and $\cnop$ impose a polynomial structure on the functions of interest, obviating the need for advanced mathematical machinery.
  Therefore, Taylor's theorem becomes quite simple for parameterized linear functionals $\phi\colon [0,1] \rightarrow \mathbb{R}$ of the form
  \begin{align*}
    \phi(t) = F\left(\msx_1 + t (\msx_2 -\msx_1)\right).
  \end{align*}
\end{remark}

\section{Low-Density Parity-Check Ensembles}

\label{section:ldpc}

\subsection{Single System}
\label{subsection:single_system_ldpc}

Let LDPC($\lambda,\rho$) denote the LDPC ensemble with variable-node degree distribution $\lambda$ and check-node degree distribution $\rho$.
The edge perspective degree distributions $\lambda,\rho$ have an equivalent representation in terms of the node perspective degree distributions $L$, $R$ given by
\begin{align*}
  \lambda(t) &= \frac{L'(t)}{L'(1)}, & \rho(t) &= \frac{R'(t)}{R'(1)}.
\end{align*}
It is important to note that the distributions $\lambda$, $\rho$, $L$ and $R$ are all \emph{polynomials}.
We assume that the LDPC($\lambda,\rho$) ensemble does not have any degree-one variable-nodes, as these ensembles exhibit non-negligible error floors.
We also refer to this ensemble as a single system to differentiate from its coupled variant introduced later.

\emph{Density evolution} (DE) characterizes the asymptotic performance of the LDPC$(\lambda,\rho)$ ensemble under message-passing decoding by describing the evolution of message distributions with iteration.
Under locally optimal processing, the message-passing decoder is equivalent to the belief-propagation (BP) decoder.
For the LDPC($\lambda,\rho$) ensemble, the DE under BP decoding is described by
\begin{align}
  \label{equation:de_singlesystem_ldpc}
  \msxvn^{(\ell+1)} = \msc \vnop \lambda^{\vnop}(\rho^{\cnop}(\msxvn^{(\ell)})) , 
\end{align}
where $\msxvn^{(\ell)}$ is the variable-node output distribution after $\ell$ iterations of message passing~\cite{Richardson-it01,RU-2008}.
If the iterative system in (\ref{equation:de_singlesystem_ldpc}) is initialized with $\msx^{(0)}=\msa$, the variable-node output-distribution after $\ell$ iterations of message-passing is denoted by $\des^{(\ell)}(\msa;\msc)$.
The variable-node output after one iteration is also denoted by
\begin{align*}
  \des(\msa;\msc) \triangleq \des^{(1)}(\msa;\msc) = \msc \vnop \lambda^{\vnop}(\rho^{\cnop}(\msa)) .
\end{align*}
If the sequence of measures $\{\des^{(\ell)}(\msa;\msc)\}$ converges in $(\probs,\disth)$, then its limit is denoted by $\des^{(\infty)}(\msa;\msc)$.

The DE update operator $\des$ satisfies certain monotonicity properties.
These properties play a crucial role in the analysis of LDPC ensembles.
\begin{lemma}[{\cite[Section 4.6]{RU-2008}}]
  \label{lemma:de_monotonicity_ldpc}
  The operator $\des^{(\ell)}\colon \probs\times\probs \to \probs$ satisfies the following monotonicity properties for all $1 \leq \ell < \infty$. 
  \begin{enumerate}[i)]
  \item If $\msa_1 \degreq \msa_2$, then $\des^{(\ell)}(\msa_1;\msc) \degreq \des^{(\ell)}(\msa_2;\msc)$ for all $\msc \in \probs$.
  \item If $\msc_1 \degreq \msc_2$, $\des^{(\ell)}(\msa;\msc_1) \degreq \des^{(\ell)}(\msa;\msc_2)$ for all $\msa \in \probs$.
  \item If $\des(\msa;\msc) \upgreq \msa$, then $\des^{(\ell+1)}(\msa;\msc) \upgreq \des^{(\ell)}(\msa;\msc)$.
    Moreover, $\des^{(\infty)}(\msa;\msc)$ exists and satisfies $\des^{(\infty)}(\msa;\msc) \upgreq \des^{(\ell)}(\msa;\msc)$,
    \begin{align*}
      \des(\des^{(\infty)}(\msa;\msc);\msc)=\des^{(\infty)}(\msa;\msc) .
    \end{align*}
  \item If $\des(\msa;\msc) \degreq \msa$, then $\des^{(\ell+1)}(\msa;\msc) \degreq \des^{(\ell)}(\msa;\msc)$.
    Moreover, $\des^{(\infty)}(\msa;\msc)$ exists and satisfies $\des^{(\infty)}(\msa;\msc) \degreq \des^{(\ell)}(\msa;\msc)$,
    \begin{align*}
      \des(\des^{(\infty)}(\msa;\msc);\msc)=\des^{(\infty)}(\msa;\msc) .
    \end{align*}
  \end{enumerate}
\end{lemma}
\begin{IEEEproof}
  The monotonicity properties can be derived from Proposition \ref{proposition:degradation_preservation}, while the existence of the limit in $(\probs,\disth)$ and its properties follow from Proposition \ref{proposition:entropy_distance_properties}.
  That the limit satisfies
  \begin{align*}
    \des(\des^{(\infty)}(\msa;\msc);\msc)=\des^{(\infty)}(\msa;\msc)
  \end{align*}
  follows from the continuity of $\vnop$, $\cnop$, and the fact that $\lambda,\rho$ are polynomials. 
\end{IEEEproof}
Thus, when (\ref{equation:de_singlesystem_ldpc}) is initialized with $\vnunit$, the sequence of measures $\{\des^{(\ell)}(\vnunit;\msc)\}$, satisfies $\des(\vnunit;\msc) \upgreq \vnunit$, and converges to a limit $\msx$, which satisfies
\begin{align*}
  \msx = \msc \vnop \lambda^{\vnop}(\rho^{\cnop}(\msx)) .
\end{align*}

\begin{definition}
  A measure $\msx \in \probs$ is a DE \emph{fixed point} for the LDPC($\lambda,\rho$) ensemble if 
  \begin{align*}
    \msx = \msc \vnop \lambda^{\vnop} \left( \rho^{\cnop}(\msx)  \right) .
  \end{align*}
\end{definition}

We now state some necessary definitions for the single system potential framework.
Included are the potential functional, stationary points, the directional derivative of the potential functional, and thresholds.

\begin{definition}
  \label{definition:uncoupled_potential_ldpc}
  The \emph{potential functional}, $\pots\colon \mc{X} \times \mc{X} \to \mbb{R}$, of the LDPC$(\lambda,\rho)$ ensemble and a channel $\msc \in \probs$ is
  \begin{align*}
    \pots(\msx; \msc) &\triangleq  \tfrac{L'(1)}{R'(1)} \ent{R^{\cnop}(\msx)} + L'(1) \ent{ \rho^{\cnop}(\msx) } \\
    & \quad - L'(1) \ent{\msx \cnop \rho^{\cnop}(\msx)} - \ent{ \msc \vnop L^{\vnop} \left( \rho^{\cnop}(\msx) \right) } .
  \end{align*}
\end{definition}

\begin{figure}[!tb]
  \centering
  \setlength\tikzheight{5cm}
  \setlength\tikzwidth{6cm} 
  \input{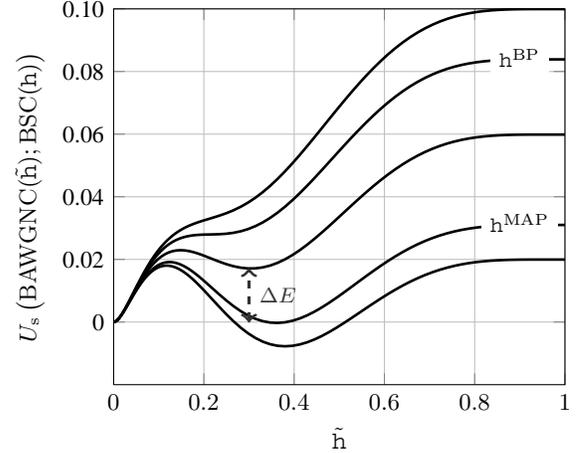}
  \caption{Potential functional for the LDPC$(\lambda,\rho)$ ensemble with $\lambda(t)=t^2$ and $\rho(t)=t^5$ over a binary symmetric channel (BSC), with entropy $\h$.
    The values of $\h$ for these curves are, from the top to bottom, $0.40, 0.416, 0.44, 0.469, 0.48$.
    The other input to the potential functional is the LLR distribution for the binary AWGN channel (BAWGNC) with entropy $\tilde{\h}$.
    The choice of BAWGNC distribution for the first argument in $\pots(\cdotp;\cdot)$ is arbitrary.}
  \label{figure:potential_functional_ldpc}
\end{figure}

\begin{remark}
  The potential functional is essentially the negative of the trial entropy, formally known as the replica-symmetric free entropy, calculated in~\cite{Montanari-it05,Kudekar-it09,Macris-it07}.\footnote{While it is possible to use the term replica-symmetric free entropy instead of `potential', our terminology is consistent with \cite{Yedla-istc12,Yedla-itw12,Takeuchi-ieice12}. Moreover, we later define \emph{coupled potential}; this brings both definitions together. In addition, for general systems, potential function need not be defined from the free entropy (e.g., see \cite{Jian-isit12}).}
  In Appendix \ref{appendix:potential_free_entropy}, we describe the Bethe formalism to obtain the free entropy and detail the calculations involved to derive the potential in Definition \ref{definition:uncoupled_potential_ldpc}.
  When applied to the binary erasure channel, $\pots$ is a constant multiple of the potential function defined in \cite{Yedla-istc12}.
  An example of $\pots(\msx;\msc)$ is shown in Fig.~\ref{figure:potential_functional_ldpc}.
\end{remark}

It is hard to define precisely what conditions are required for a potential functional, that operates on measures, to prove threshold saturation.
But, the crucial properties of the single system potential that we leverage are 1) the fixed points of the single system DE are the stationary points of the single system potential (Lemma \ref{lemma:firstderivative_singlesystem_ldpc}), 2) there exists a spatially-coupled potential, defined by a spatial average of the single system potential (Definition \ref{definition:coupled_potential_ldpc}), where the fixed points of spatially-coupled DE are stationary points of the spatially-coupled potential (Lemma \ref{lemma:first_derivative_coupled_system_ldpc}).

The entropy functional and the operators ($\vnop$, $\cnop$) are continuous.
Hence, the potential functional $\pots(\cdotp;\msc)$ for a fixed $\msc$ is continuous.
Since the metric topology $(\probs,\disth)$ is compact, $\pots(\cdotp;\msc)$ achieves its minimum and maximum on $\probs$.
Though we also have the joint continuity of $\pots(\cdotp;\cdotp)$, it is not used in this work.

\begin{definition}
  A measure $\msx \in \probs$ is a \emph{stationary point} of the potential if, for all $\msy \in \dpros$,
  \begin{align*}
    \deri{\msx} \pots(\msx ; \msc) [\msy] = 0.
  \end{align*}
\end{definition}

\begin{lemma}
  \label{lemma:firstderivative_singlesystem_ldpc}
  For $\msx$, $\msc \in \probs$ and $\msy \in \dpros$, the directional derivative of the potential functional with respect to $\msx$ in the direction $\msy$ is
  \begin{align*}
    \deri{\msx} \pots( \msx ; \msc ) [\msy] &= L'(1) \ent{ \left[ \des(\msx;\msc) - \msx \right] \cnop \left[ \rho'^{\cnop}(\msx) \cnop \msy \right] } .
  \end{align*}
\end{lemma}
\begin{IEEEproof}
  Since the distributions $\lambda,\rho,L,R$ are polynomials, the directional derivative for each of the four terms can be calculated following the procedure outlined in the proof of Proposition \ref{proposition:directional_derivative_single_operator}.
  The directional derivatives of the first three terms are
  \begin{align*}
    \deri{\msx} \ent{ R^{\cnop}(\msx)} [\msy] &= R'(1) \ent{ \rho^{\cnop}(\msx) \cnop \msy }, \\
    \deri{\msx} \ent{ \rho^{\cnop}(\msx) }[\msy] &= \ent{ \rho'^{\cnop}(\msx) \cnop \msy }, \\    
    \deri{\msx}\ent{ \msx \cnop \rho^{\cnop}(\msx) } [\msy] &=  \ent{ \rho^{\cnop}(\msx) \cnop \msy }  + \ent{ \msx \cnop \rho'^{\cnop}(\msx) \cnop \msy } \\
    & \overset{(a)}{=}   \ent{ \rho^{\cnop}(\msx) \cnop \msy } + \ent{  \rho'^{\cnop}(\msx) \cnop \msy } \\
    & \qquad - \ent{ \msx \vnop \left[ \rho'^{\cnop}(\msx) \cnop \msy \right]},
  \end{align*}
  where $(a)$ follows from Proposition \ref{proposition:duality_difference} with the observation that $\rho'^{\cnop}(\msx) \cnop \msy$ is a difference of probability measures multiplied by the scalar $\rho'(1)$.
  Since the operators $\vnop$ and $\cnop$ do not associate, one must exercise care in analyzing the last term.
  From Proposition \ref{proposition:directional_derivative_multiple_operators},
  \begin{align*}
    & \deri{\msx} \ent{ \msc \vnop L^{\vnop} \left( \rho^{\cnop}(\msx) \right) } [\msy] \\
    & \quad = L'(1) \ent{  \left[ \msc \vnop \lambda^{\vnop}(\rho^{\cnop}(\msx)) \right] \vnop \left[ \rho'^{\cnop}(\msx) \cnop \msy \right] }.
  \end{align*}
  Consolidating the four terms,
  \begin{align*}
    \deri{\msx} \pots( \msx ; \msc ) [\msy] &= L'(1) \ent{ \left[ \msx - \des(\msx;\msc)  \right] \vnop \left[ \rho'^{\cnop}(\msx) \cnop \msy \right] } .
  \end{align*}
  Using Proposition \ref{proposition:duality_difference}, we have the desired result.
\end{IEEEproof}

\begin{lemma}
  \label{lemma:fixed_stationary_ldpc}
  If $\msx \in \probs$ is a fixed point of single system DE, then it is also a stationary point of the potential functional.
  Moreover, for a fixed channel $\msc$, the minimum of the potential functional,
  \begin{align*}
    \min_{\msx \in \probs} \pots(\msx ; \msc),
  \end{align*}
  occurs only at a fixed point of single system DE.
\end{lemma}
\begin{IEEEproof}
  See Appendix \ref{appendix:proof_fixed_stationary_ldpc}.
\end{IEEEproof}

\begin{definition}
  \label{definition:basinofattraction_energygap_ldpc}
  For the LDPC($\lambda,\rho$) ensemble and a channel $\msc \in \probs$, define 
  \begin{enumerate}[i)]
  \item The \emph{basin of attraction to $\cnunit$} as
    \begin{align*}
      \mc{V}(\msc) \triangleq \left\{ \msa \in \probs \mid \des^{(\infty)}(\msa ; \msc) = \cnunit \right\}.
    \end{align*}
  \item The \emph{energy gap} as
    \begin{align*}
      \Delta E (\msc) \triangleq \inf_{\msx \in \probs \setminus \mc{V}(\msc) } \pots(\msx;\msc) ,
    \end{align*}
    with the convention that the infimum over the empty set is $\infty$.
  \end{enumerate}
\end{definition}
The only fixed point contained in $\mc{V}(\msc)$ is the trivial $\cnunit$ fixed point.
Therefore, all other fixed points are in the complement, $\probs \setminus \mc{V}(\msc)$.

\begin{lemma}
  \label{lemma:monotonicity_ldpc}
  Suppose $\msc_1 \degr \msc_2$.
  Then
  \begin{enumerate}[i)]
    \item $\pots(\msx ; \msc_1) < \pots(\msx ; \msc_2)$ if $\msx \neq \cnunit$
    \item $\mc{V}(\msc_1) \subseteq \mc{V}(\msc_2)$ and $\probs\setminus\mc{V}(\msc_1) \supseteq \probs\setminus\mc{V}(\msc_2) $
    \item $\Delta E(\msc_1) \leq \Delta E(\msc_2)$
  \end{enumerate}
\end{lemma}
\begin{IEEEproof}
  See Appendix \ref{appendix:proof_monotonicity_ldpc}.
\end{IEEEproof}

\begin{definition}
  A \emph{family of BMS channels} is a function $\msc(\cdot)\colon [0,1] \to \probs$ that is 
  \begin{enumerate}[i)]
  \item ordered by degradation, $\msc(\h_{1}) \degreq \msc(\h_{2})$ for $\h_{1} \geq \h_{2}$,
  \item parameterized by entropy $\ent{\msc(\h)}=\h$.
  \end{enumerate}
\end{definition}

\begin{definition}
  \label{definition:thresholds_ldpc}
  Consider a family of BMS channels and the LDPC($\lambda,\rho$) ensemble.
  Define
  \begin{enumerate}[i)]
  \item The \emph{BP threshold} as
    \begin{align*} 
      \hbp \triangleq \sup \left\{ \h \in [0,1] \mid \des^{(\infty)}(\vnunit ; \msc(\h))=\cnunit \right\} . 
    \end{align*}
  \item The \emph{MAP threshold} as $\hmap \triangleq$
    \begin{align*} 
      \inf \left\{ \h \in [0,1] \mid \liminf_{n\to\infty} \tfrac{1}{n} \expt \left[ \ent{ X^n \mid Y^n (\msc(\h)) } \right] > 0 \right\} ,
    \end{align*}
    where the expectation $\expt [\cdotp]$ is over the LDPC ensemble.
  \item The \emph{potential threshold} as
    \begin{align*}
      \h^*  \triangleq \sup \{ \h \in [0,1] \mid \Delta E (\msc(\h)) > 0 \} .
    \end{align*}
  \item The \emph{stability threshold} as
    \begin{align*}
      \hstab \triangleq \sup \{ \h \in [0,1] \mid \mf{B}(\msc(\h)) \lambda'(0)\rho'(1) < 1 \} .
    \end{align*}
  \end{enumerate}
\end{definition}

In the sequel, the potential threshold and its role in connecting the BP and MAP thresholds are paramount.
In particular, the region where $\Delta E(\msc(\h)) >0$ characterizes the BP performance of the spatially-coupled ensemble, and, by definition of the potential threshold and Lemma \ref{lemma:monotonicity_ldpc}(iii), if $\h < \h^*$, then $\Delta E(\msc(\h)) > 0$.

The stability threshold establishes an important technical property of the potential functional.
When $\hstab=1$, any constraints involving $\hstab$ are \emph{superfluous}. 
For LDPC ensembles with no degree-two variable-nodes, $\hstab=1$.
For ensembles with degree-two variable-nodes\footnote{We exclude ensembles with degree-one variable-nodes.}, $0< \hstab \leq 1$.

\begin{lemma}
  \label{lemma:stability_threshold_properties}
  The following properties regarding the stability threshold hold.
  \begin{enumerate}[i)]
  \item $\h^* \leq \hstab$
  \item If $\h < \hstab$, $\cnunit \in (\mc{V}(\msc(\h)))^{o}$, the interior of the set $\mc{V}(\msc(\h))$ in $(\probs,\disth)$.
  \end{enumerate}
\end{lemma}
\begin{IEEEproof}
  See Appendix \ref{appendix:proof_stability_threshold_properties}.
\end{IEEEproof}

\begin{lemma}
  \label{lemma:beyond_potential_threshold_negativity}
  If $\h^* < \hstab$, then for $\h>\h^*$ there exists an $\msx \in \probs$ such that $\pots(\msx;\msc(\h))<0$.
\end{lemma}
\begin{IEEEproof}
  See Appendix \ref{appendix:proof_beyond_potential_threshold_negativity}.
\end{IEEEproof}

\begin{remark}
  Negativity of the potential functional beyond the potential threshold is important.
  This allows us to relate the potential and MAP threshold (Lemma \ref{lemma:mapandpotential_connection}).
  Negativity is also used in the converse of the threshold saturation result (Theorem \ref{theorem:converse_threshold_saturation_ldpc}).
  For a family of BEC or binary AWGN channels, Lemma \ref{lemma:beyond_potential_threshold_negativity} can be extended to include the case $\h^*=\hstab$.
  We conjecture that this holds for \emph{any} family of BMS channels.
  See Appendix \ref{appendix:negativity_potential_functional} for a further discussion.
\end{remark}

\begin{lemma}
  \label{lemma:mapandpotential_connection}
  For an LDPC ensemble without odd-degree check-nodes over any BMS channel, or any LDPC ensemble over the BEC or the binary AWGN channel,
  \begin{enumerate}[i)]
  \item $\displaystyle{\liminf_{n \to \infty}} \tfrac{1}{n} \expt \left[ \ent{ X^n | Y^n (\msc(\h)) } \right] \geq -\inf_{\msx \in \probs} \pots(\msx;\msc(\h)) ,$
  \item If $\h^*<\hstab$, then $\hmap \leq \h^*$.
  \end{enumerate}
\end{lemma}
\begin{IEEEproof}
  \begin{enumerate}[i)]
  \item Since the potential functional is the negative of the replica-symmetric free entropy calculated in~\cite{Montanari-it05,Kudekar-it09,Macris-it07}, the main result of these papers translates directly into the desired result.
  \item Let $\h>\h^*$.
    Since $\h^*<\hstab$ by assumption, from Lemma \ref{lemma:beyond_potential_threshold_negativity} and part i,
    \begin{align*}
      \liminf_{n \to \infty} \tfrac{1}{n} \expt \left[ \ent{ X^n | Y^n (\msc(\h)) } \right] \geq -\inf_{\msx \in \probs} \pots(\msx;\msc(\h))  > 0.
    \end{align*}
    Thus, by Definition \ref{definition:thresholds_ldpc}(ii), $\h \geq \hmap$.
    Hence $\h^* \geq \hmap$.
  \end{enumerate}
\end{IEEEproof}

The following remark discusses, rather informally, further connections between single and spatially-coupled system thresholds, based on results from \cite{Giurgiu-arxiv13}, \cite{Kudekar-it13}.
\begin{remark}
  \label{remark:threshold_discussion}
  Let $\hbp_{\mathrm{c}}$ and $\hmap_{\mathrm{c}}$ denote the BP and MAP thresholds, respectively, of the spatially-coupled system by first letting the chain length and then the coupling width go to infinity. 
  This article establishes (Theorems \ref{theorem:threshold_saturation_ldpc} and \ref{theorem:converse_threshold_saturation_ldpc})  that 
  \begin{align}
    \label{equation:article_results}
    \hbp_{\mathrm{c}} = \h^*.
  \end{align}
  
  In \cite{Giurgiu-arxiv13} it is shown that, under some restrictions on the degree distributions\footnote{Requires regular check-nodes with even degree; this can be relaxed to $R(t)$ convex on $[-1,1]$.}, $\hmap_{\mathrm{c}} = \hmap$.
  By Lemma \ref{lemma:mapandpotential_connection}, for any ensemble with $\hstab = 1$, e.g. an ensemble with no degree-two variable nodes, $\hmap \le \h^*$.
  Combining these results with optimality of the MAP decoder and (\ref{equation:article_results})
  \begin{align*}
    \hmap \le \h^* = \hbp_{\mathrm{c}} \leq \hmap_{\mathrm{c}} = \hmap .
  \end{align*}
  This shows that $\h^* = \hmap$, for an ensemble satisfying the aforementioned conditions.
  
  The threshold saturation result shown in \cite{Kudekar-it13} can be summarized as follows.
  For regular codes with left-degree $d_v$, right-degree $d_c$, and a \emph{smooth} family of channels, the BP threshold is equal to the \emph{area} threshold $\hbp_{\mathrm{c}}=\harea$, where the area threshold is
  \begin{align*}
    \harea \triangleq \sup \left\{ \h \in [0,1] \mid A(\des^{(\infty)}(\vnunit;\msc(\h)),d_v,d_c) \leq 0 \right\},
  \end{align*}
  and
  \begin{align*}
    A(\msx,d_v,d_c) &\triangleq \ent{\msx} + \Big{(}d_v-1-\frac{d_v}{d_c}\Big{)} \ent{\msx^{\cnop d_c}} \\
    &\qquad - (d_v-1)\ent{\msx^{\cnop d_c-1}} .
  \end{align*}
  At the DE fixed point $\des^{(\infty)}(\vnunit;\msc(\h))$, using the duality rule for entropy (Proposition \ref{proposition:duality}), it is also easy to show that
  \begin{align*}
    A(\des^{(\infty)}(\vnunit;\msc(\h)),d_v,d_c) = - \pots(\des^{(\infty)}(\vnunit;\msc(\h)) ; \msc(\h)) .
  \end{align*}
  This immediately implies that $\h^* \leq \harea$.
  Therefore, by \cite[Theorem 41]{Kudekar-it13}, $\harea = \hbp_{\mathrm{c}}$, and the results of this article, (\ref{equation:article_results}), $\h^* = \harea$.
  Hence, the thresholds $\hmap$, $\h^*$ and $\harea$ are all equal under suitable conditions.
  
  In particular, for regular codes with even-degree checks, it has been shown rigorously that $\hmap=\harea$.
  However, it is instructive to note that the Maxwell conjecture \cite[Conjecture 1]{Measson-it09}, which states that the MAP GEXIT function is obtained by applying the Maxwell construction to the EBP GEXIT curve, is yet to be established for BMS channels.
\end{remark}

\subsection{Coupled System}
\label{subsection:coupled_system_ldpc}

\begin{figure*}[!tb]
  \centering
  \begin{tikzpicture}[
  xscale=0.33,yscale=1,
  bitnode/.style={circle,minimum size=10pt,thick,draw=black,fill=white},
  checknode/.style={rectangle,minimum size=8pt,thick,draw=black,fill=white},
  permnode/.style={rectangle,very thin,minimum width=30pt,minimum height=18pt,fill=white,draw=black},
  permedge/.style={black!65},
  socketedge/.style={black},
  ]

  \foreach \i/\text in {0/1,3/\cdots,6/i-2,10/i-1,14/i,18/i+1,22/i+2,25/\cdots,28/2N} {
    \node at (\i,2.5) {\footnotesize{$\text$}};
  }

  \def \vndist {0.75}
  \def \cndist {0.75}
  \def \vny {2}
  \def \cny {-2}

  \def \midx   {0.5}

  \foreach \x/\i/\text in {0/1/1} {
    \draw[socketedge] (\x-\vndist, \vny) +(0,0) -- +(240:1);
    \draw[socketedge] (\x-\vndist, \vny) +(0,0) -- +(255:1);
    \draw[socketedge] (\x-\vndist, \vny) +(0,0) -- +(270:1);
    \draw[socketedge] (\x-\vndist, \vny) +(0,0) -- +(285:1);
    \draw[socketedge] (\x-\vndist, \vny) +(0,0) -- +(300:1);

    \draw[socketedge] (\x-\vndist, \cny) +(0,0) -- +(52.5:1);
    \draw[socketedge] (\x-\vndist, \cny) +(0,0) -- +(67.5:1);
    \draw[socketedge] (\x-\vndist, \cny) +(0,0) -- +(82.5:1);
    \draw[socketedge] (\x-\vndist, \cny) +(0,0) -- +(97.5:1);
    \draw[socketedge] (\x-\vndist, \cny) +(0,0) -- +(112.5:1);
    \draw[socketedge] (\x-\vndist, \cny) +(0,0) -- +(127.5:1);

    \draw[socketedge] (\x+\vndist, \vny) +(0,0) -- +(240:1);
    \draw[socketedge] (\x+\vndist, \vny) +(0,0) -- +(255:1);
    \draw[socketedge] (\x+\vndist, \vny) +(0,0) -- +(270:1);
    \draw[socketedge] (\x+\vndist, \vny) +(0,0) -- +(285:1);
    \draw[socketedge] (\x+\vndist, \vny) +(0,0) -- +(300:1);

    \draw[socketedge] (\x+\vndist, \cny) +(0,0) -- +(52.5:1);
    \draw[socketedge] (\x+\vndist, \cny) +(0,0) -- +(67.5:1);
    \draw[socketedge] (\x+\vndist, \cny) +(0,0) -- +(82.5:1);
    \draw[socketedge] (\x+\vndist, \cny) +(0,0) -- +(97.5:1);
    \draw[socketedge] (\x+\vndist, \cny) +(0,0) -- +(112.5:1);
    \draw[socketedge] (\x+\vndist, \cny) +(0,0) -- +(127.5:1);

    \node[bitnode] (v1\i) at (\x-\vndist, \vny) {};
    \node[bitnode] (v2\i) at (\x+\vndist, \vny) {};
    \node[checknode] (c1\i) at (\x-\cndist, \cny) {};
    \node[checknode] (c2\i) at (\x+\cndist, \cny) {};

    \foreach \j in {-1.7,1.7} {
      \node at (\x,\j) {\tiny{$...$}};
    }

    \node[permnode] (perm1_node\i) at (\x,\vny-1) {\footnotesize{$\pi_{\text}$}};
    \node[permnode] (perm2_node\i) at (\x,\cny+1) {\footnotesize{$\pi_{\text}'$}};
  }

  \foreach \x in {3,25} {
    \foreach \y in {1,-1} {
      \node at (\x,\y) {\footnotesize{$\cdots$}};
    }
  }

  \foreach \x/\i/\text in {6/2/i-2,10/3/i-1,14/4/i,18/5/i+1,22/6/i+2,28/7/2N} {
    \draw[socketedge] (\x-\vndist, \vny) +(0,0) -- +(240:1);
    \draw[socketedge] (\x-\vndist, \vny) +(0,0) -- +(255:1);
    \draw[socketedge] (\x-\vndist, \vny) +(0,0) -- +(270:1);
    \draw[socketedge] (\x-\vndist, \vny) +(0,0) -- +(285:1);
    \draw[socketedge] (\x-\vndist, \vny) +(0,0) -- +(300:1);

    \draw[socketedge] (\x-\vndist, \cny) +(0,0) -- +(52.5:1);
    \draw[socketedge] (\x-\vndist, \cny) +(0,0) -- +(67.5:1);
    \draw[socketedge] (\x-\vndist, \cny) +(0,0) -- +(82.5:1);
    \draw[socketedge] (\x-\vndist, \cny) +(0,0) -- +(97.5:1);
    \draw[socketedge] (\x-\vndist, \cny) +(0,0) -- +(112.5:1);
    \draw[socketedge] (\x-\vndist, \cny) +(0,0) -- +(127.5:1);

    \draw[socketedge] (\x+\vndist, \vny) +(0,0) -- +(240:1);
    \draw[socketedge] (\x+\vndist, \vny) +(0,0) -- +(255:1);
    \draw[socketedge] (\x+\vndist, \vny) +(0,0) -- +(270:1);
    \draw[socketedge] (\x+\vndist, \vny) +(0,0) -- +(285:1);
    \draw[socketedge] (\x+\vndist, \vny) +(0,0) -- +(300:1);

    \draw[socketedge] (\x+\vndist, \cny) +(0,0) -- +(52.5:1);
    \draw[socketedge] (\x+\vndist, \cny) +(0,0) -- +(67.5:1);
    \draw[socketedge] (\x+\vndist, \cny) +(0,0) -- +(82.5:1);
    \draw[socketedge] (\x+\vndist, \cny) +(0,0) -- +(97.5:1);
    \draw[socketedge] (\x+\vndist, \cny) +(0,0) -- +(112.5:1);
    \draw[socketedge] (\x+\vndist, \cny) +(0,0) -- +(127.5:1);

    \node[bitnode] (v1\i) at (\x-\vndist, \vny) {};
    \node[bitnode] (v2\i) at (\x+\vndist, \vny) {};
    \node[checknode] (c1\i) at (\x-\cndist, \cny) {};
    \node[checknode] (c2\i) at (\x+\cndist, \cny) {};

    \foreach \j in {-1.7,1.7} {
      \node at (\x,\j) {\tiny{$...$}};
    }

    \node[permnode] (perm1_node\i) at (\x,\vny-1) {\footnotesize{$\pi_{\text}$}};
    \node[permnode] (perm2_node\i) at (\x,\cny+1) {\footnotesize{$\pi_{\text}'$}};
  }

  \foreach \x/\i/\text in {32/8/2N\!+\!1,36/9/\chend} {

    \draw[socketedge] (\x-\vndist, \cny) +(0,0) -- +(52.5:1);
    \draw[socketedge] (\x-\vndist, \cny) +(0,0) -- +(67.5:1);
    \draw[socketedge] (\x-\vndist, \cny) +(0,0) -- +(82.5:1);
    \draw[socketedge] (\x-\vndist, \cny) +(0,0) -- +(97.5:1);
    \draw[socketedge] (\x-\vndist, \cny) +(0,0) -- +(112.5:1);
    \draw[socketedge] (\x-\vndist, \cny) +(0,0) -- +(127.5:1);

    \draw[socketedge] (\x+\vndist, \cny) +(0,0) -- +(52.5:1);
    \draw[socketedge] (\x+\vndist, \cny) +(0,0) -- +(67.5:1);
    \draw[socketedge] (\x+\vndist, \cny) +(0,0) -- +(82.5:1);
    \draw[socketedge] (\x+\vndist, \cny) +(0,0) -- +(97.5:1);
    \draw[socketedge] (\x+\vndist, \cny) +(0,0) -- +(112.5:1);
    \draw[socketedge] (\x+\vndist, \cny) +(0,0) -- +(127.5:1);

    \node[checknode] (c1\i) at (\x-\cndist, \cny) {};
    \node[checknode] (c2\i) at (\x+\cndist, \cny) {};

    \foreach \j in {-1.7} {
      \node at (\x,\j) {\tiny{$...$}};
    }

    \node[permnode] (perm2_node\i) at (\x,\cny+1) {\footnotesize{$\pi_{\text}'$}};
  }

  \foreach \x/\i in {0/1,6/2,10/3,14/4,18/5,22/6,28/7} {
    \draw[permedge] (perm1_node\i) .. controls (\x-0.2,0) .. (perm2_node\i);
    \draw[permedge] (perm1_node\i) .. controls (\x+0.2,0) .. (perm2_node\i);
  }
  \foreach \x/\xn/\i/\j in {0/6/1/2,6/10/2/3,10/14/3/4,14/18/4/5,18/22/5/6,22/28/6/7,28/32/7/8} {
    \draw[permedge] (perm1_node\i) .. controls (0.5*\x+0.5*\xn-0.2,0) .. (perm2_node\j);
    \draw[permedge] (perm1_node\i) .. controls (0.5*\x+0.5*\xn+0.2,0) .. (perm2_node\j);
  }
  \node[fill=white,rotate=-38] at (3,0) {\footnotesize{$\cdots$}};
  \node[fill=white,rotate=-38] at (25,0) {\footnotesize{$\cdots$}};

  \foreach \x/\xn/\i/\j in {0/10/1/3,6/14/2/4,10/18/3/5,14/22/4/6,18/28/5/7,22/32/6/8,28/36/7/9} {
    \draw[permedge] (perm1_node\i) .. controls (0.5*\x+0.5*\xn-0.2,0) .. (perm2_node\j);
    \draw[permedge] (perm1_node\i) .. controls (0.5*\x+0.5*\xn+0.2,0) .. (perm2_node\j);
  }
  \node[fill=white,rotate=-30] at (4,0.2) {\footnotesize{$\cdots$}};
  \node[fill=white,rotate=-30] at (24,-0.2) {\footnotesize{$\cdots$}};
  \node[fill=white,rotate=-30] at (26,0.2) {\footnotesize{$\cdots$}};

  \draw[permedge] (perm2_node1) -- (-2,0.04);
  \draw[permedge] (perm2_node1) -- (-2,-0.04);

  \draw[permedge] (perm2_node1) -- (-3,0.04);
  \draw[permedge] (perm2_node1) -- (-3,-0.04);

  \draw[permedge] (perm2_node2) -- (2,-0.44);
  \draw[permedge] (perm2_node2) -- (2,-0.36);
  \node[rotate=-26] at (1.4,-0.3) {\footnotesize{$\cdots$}};

  \draw[permedge] (perm2_node8) -- (32.1,-0.2);
  \draw[permedge] (perm2_node8) -- (31.9,-0.2);

  \draw[permedge] (perm2_node9) -- (36.1,0);
  \draw[permedge] (perm2_node9) -- (35.9,0);

  \draw[permedge] (perm2_node9) -- (37.1,0);
  \draw[permedge] (perm2_node9) -- (36.9,0);

\end{tikzpicture}
  \caption{An example of a $(\lambda(t)=t^4,\rho(t)=t^5,N,w=3)$ spatially-coupled LDPC ensemble.
    Sockets in each variable- and check-node group are permuted ($\pi$ and $\pi'$ denote the permutations) and partitioned into $w$ groups, and connected as shown above.
    This results in some sockets of the check-node groups at the boundary unconnected.
  }
  \label{figure:tanner_graph_spatially_coupled_ldpc}
\end{figure*}
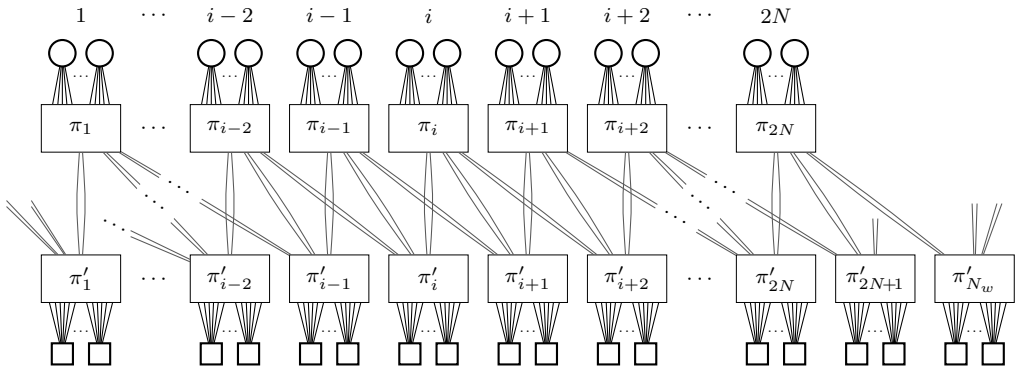

The potential theory for single systems is now extended to spatially-coupled systems.
Vectors of measures are denoted by underlines (e.g., $\msbx$) with $[\msbx]_{i} = \msx_{i}$.
Functionals operating on a single measure are distinguished from those operating on vectors by their input (i.e., $F(\msx)$ vs.~ $F(\msbx)$).
Also, for vectors $\msbx'$ and $\msbx$, we write $\msbx' \degreq \msbx$ if $\msx'_i \degreq \msx_i$ for all $i$, and $\msbx' \degr \msbx$ if $\msbx'_i \degreq \msbx_i$ for all $i$ and $\msx'_i \degr \msx_i$ for some $i$.

The ideas underlying spatial coupling now appear to be quite general.
The local coupling in the system allows the effect of the perfect information, provided at the boundary, to propagate throughout the system.
In the large-system limit, these coupled systems show a significant performance improvement.
The spatially-coupled system model is now described.

The $(\lambda,\rho,N,w)$ spatially-coupled LDPC ensemble is defined as follows.
As before, the node perspective degree distributions are denoted by $L$, $R$, and
\begin{align*}
  L(t) &=\sum_{n=0}^{\mathrm{deg}(L)}L_n t^n, & R(t)&= \sum_{n=0}^{\mathrm{deg}(R)} R_n t^n .
\end{align*}
A collection of $2N$ variable-node groups are placed at all positions in $\mc{N}_v=\{ 1, 2, \ldots, 2N \}$ and a collection of $2N+(w-1)$ check-node groups are placed at all positions in $\mc{N}_c=\{ 1, 2, \ldots, 2N+(w-1) \}$.
For notational convenience, the rightmost check-node group index is denoted by $\chend \triangleq 2N+ (w-1)$.
For the below construction of a spatially-coupled LDPC ensemble, we assume all $L_n$, $R_n$ are rational.

The integer $M$ is chosen large enough so that i) $ML_{i}$, $ML'(1)R_{j}/R'(1)$ are natural numbers for $1 \leq i \leq \mathrm{deg}(L)$, $1 \leq j \leq \mathrm{deg}(R)$, and ii) $ML'(1)$ is divisible by $w$.
At each variable-node group, $ML_{i}$ nodes of degree $i$  are placed for $1\leq i \leq \mathrm{deg}(L)$.
Similarly, at each check-node group, $ML'(1)R_{j}/R'(1)$ nodes of degree $j$ are placed for $1\leq j \leq \mathrm{deg}(R)$.
At each variable-node and check-node group, the $ML'(1)$ edge sockets are partitioned into $w$ equal-sized groups using a uniform random permutation.
Denote these partitions, respectively, by $\mathcal{P}^{v}_{i,k}$ and $\mathcal{P}^{c}_{j,k}$ at variable-node and check-node groups, where $1 \leq i \leq 2N$, $1 \leq j \leq \chend$ and $1 \leq k \leq w$.
The spatially-coupled system is constructed by connecting the sockets in $\mathcal{P}^{v}_{i,k}$ to sockets in $\mathcal{P}^{c}_{i+k-1,k}$ using uniform random permutations.
This construction leaves some sockets of the check-node groups at the boundaries unconnected and these sockets are assigned the binary value 0 (i.e., the socket and edge are removed).
These 0 values form the perfect information that gets decoding started.
A Tanner graph example of a spatially-coupled LDPC ensemble depicting these connections is provided in Fig.~\ref{figure:tanner_graph_spatially_coupled_ldpc}. 

The analysis below is valid for any spatially-coupled system whose density evolution is given by (\ref{equation:scde_update_0_ldpc}).
For the random ensemble described in \cite[Section II-B]{Kudekar-it11}, and for the $(\lambda,\rho,N,w)$ ensemble described above, the asymptotic density evolution is indeed described by (\ref{equation:scde_update_0_ldpc}).
Thus, our analysis holds for both these ensembles.
However, this is no longer true for the protograph construction described in \cite[Section II-A]{Kudekar-it11}.

Let $\msxvn_{i}^{(\ell)}$ be the variable-node output distribution at node $i$ after $\ell$ iterations of message passing.
Then, the input distribution to the $i$-th check-node group is the normalized sum of averaged variable-node output distributions,
\begin{align}
  \label{equation:coupling_ldpc}
  \msx_{i}^{(\ell)} = \frac{1}{w} \sum_{k=0}^{w-1} \msxvn_{i-k}^{(\ell)}.
\end{align}
The averaging in the reversed direction (i.e.~from check-node to the variable-node) follows naturally from this setup and is essentially the transpose of the forward averaging for the check-node output distributions.
This model uses uniform coupling over a fixed window, but in a more general setting window size and coefficient weights could vary from node to node.
By virtue of the fixed boundary condition, $\msxvn_{i}^{(\ell)} = \cnunit$ for $i \notin \mc{N}_v$ and all $\ell$, and from the relation in (\ref{equation:coupling_ldpc}), this implies $\msx_{i}^{(\ell)} = \cnunit$ for $i \notin \mc{N}_c$ and all $\ell$.

Generalizing~\cite[Eqn. 12]{Kudekar-it13} to irregular codes gives the evolution of the variable-node output distributions,
\begin{align}
  \label{equation:scde_update_0_ldpc}
  \msxvn_{i}^{(\ell+1)} = \msc \vnop \lambda^{\vnop} \left(\frac{1}{w} \sum_{j=0}^{w-1} \rho^{\cnop} \left( \frac{1}{w}  \sum_{k=0}^{w-1} \msxvn_{i+j-k}^{(\ell)} \right) \right).
\end{align}

Making a change of variables, the variable-node output distribution evolution in (\ref{equation:scde_update_0_ldpc}) can be rewritten in terms of check-node input distributions
\begin{align}
  \label{equation:scde_update_1_ldpc}
  \msx_{i}^{(\ell+1)} = \frac{1}{w} \sum_{k=0}^{w-1} \msc_{i-k} \vnop \lambda^{\vnop} \left(  \frac{1}{w} \sum_{j=0}^{w-1} \rho^{\cnop} \left( \ms{x}_{i-k+j}^{(\ell)} \right) \right) ,
\end{align}
for $i \in \mc{N}_c$, where $\msc_{i} = \msc$ when $i \in \mc{N}_v$ and $\msc_{i}=\cnunit$ otherwise. 
While (\ref{equation:scde_update_0_ldpc}) is a more natural representation for the underlying system, (\ref{equation:scde_update_1_ldpc}) is more mathematically tractable and easily yields a coupled potential functional.
As such, we adopt the system characterized by (\ref{equation:scde_update_1_ldpc}) and refer to it as the $(\lambda,\rho,N,w)$ \emph{spatially-coupled LDPC system}.

Borrowing notation from the single system, when the spatially-coupled system with channel $\msc$ is initialized with $\msba$ (i.e.~$\msx^{(0)}_{i}=\msa_i$), the check-node input distribution after $\ell$ iterations of message-passing is denoted by $\dec^{(\ell)}(\msba;\msc)$.
One iteration of this message-passing is also denoted by $\dec(\msba;\msc)$.
With this new notation, (\ref{equation:scde_update_1_ldpc}) can be written compactly as
\begin{align*}
  \msx_{i}^{(\ell+1)} &= \dec(\msbx^{(\ell)};\msc)_i .
\end{align*}
If the sequence of measure vectors $\{\dec^{(\ell)}(\msba;\msc)\}_{\ell=1}^\infty$ converges pointwise, then its limit is denoted by $\dec^{(\infty)}(\msba;\msc)$.
The following proposition establishes certain monotonicity properties of $\dec^{(\ell)}$. 

\begin{lemma}
  \label{lemma:scde_monotonicity_ldpc}
  The operator $\dec^{(\ell)}\colon \probs^{\chend} \times \probs \to \probs^{\chend}$ satisfies the following for all $1 \leq \ell < \infty$.
  \begin{enumerate}[i)]
  \item If $\msba_1 \degreq \msba_2$, then $\dec^{(\ell)}(\msba_1;\msc) \degreq \dec^{(\ell)}(\msba_2;\msc)$ for all $\msc \in \probs$.
  \item If $\msc_1 \degreq \msc_2$, then $\dec^{(\ell)}(\msba;\msc_1) \degreq \dec^{(\ell)}(\msba;\msc_2)$ for all $\msba \in \probs^{\chend}$.
  \item If $\dec(\msba;\msc) \upgreq \msba$, then $\dec^{(\ell+1)}(\msba;\msc) \upgreq \dec^{(\ell)}(\msba;\msc)$.
    Also, the limit $\dec^{(\infty)}(\msba;\msc)$ exists and satisfies $\dec^{(\infty)}(\msba;\msc) \upgreq \dec^{(\ell)}(\msba;\msc)$,
    \begin{align*}
      \dec(\dec^{(\infty)}(\msba;\msc);\msc)=\dec^{(\infty)}(\msba;\msc).
    \end{align*}
  \item If $\dec(\msba;\msc) \degreq \msba$, then $\dec^{(\ell+1)}(\msba;\msc) \degreq \dec^{(\ell)}(\msba;\msc)$.
    Also, the limit $\dec^{(\infty)}(\msba;\msc)$ exists and satisfies $\dec^{(\infty)}(\msba;\msc) \degreq \dec^{(\ell)}(\msba;\msc)$,
    \begin{align*}
      \dec(\dec^{(\infty)}(\msba;\msc);\msc)=\dec^{(\infty)}(\msba;\msc).
    \end{align*}
  \end{enumerate}
\end{lemma}
\begin{IEEEproof}
  The proof is almost identical to the proof of Lemma \ref{lemma:de_monotonicity_ldpc}.
  We skip the details for brevity.
\end{IEEEproof}

\begin{figure*}[!t]
  \begin{align}
    \label{equation:coupled_potential_ldpc}
    \potc(\msbx ; \msc) \triangleq L'(1) \sum_{i=1}^{\chend}\left[ \frac{1}{R'(1)}  \ent{ R^{\cnop}( \ms{x}_{i}) } + \ent{ \rho^{\cnop}( \ms{x}_{i}) } - \ent{ \ms{x}_{i} \cnop \rho^{\cnop}(\msx_{i}) } \right] - \sum_{i=1}^{2N}   \ent{ \msc \vnop L^{\vnop} \Big{(} \frac{1}{w}  \sum_{j=0}^{w-1} \rho^{\cnop} ( \ms{x}_{i+j}) \Big{)} } 
  \end{align}
  \noindent\makebox[\linewidth]{\rule{17.8cm}{0.4pt}}
\end{figure*}

\begin{figure*}[!t]
  \begin{align}
    \label{equation:second_derivative_coupled_system_ldpc}
    & \dderi{\msbx} \potc(\msbx ; \msc)[\msby,\msbz] =  \\
    & L'(1) \sum_{i=1}^{\chend}  \left[   \rho''(1) \ent{ \dec(\msbx;\msc)_{i} \cnop \frac{\rho''^{\cnop}(\msx_i)}{\rho''(1)} \cnop \msy_i \cnop \msz_i} - \rho''(1) \ent{\msx_{i} \cnop \frac{\rho''^{\cnop}(\msx_i)}{\rho''(1)} \cnop \msy_i \cnop \msz_i} -  \rho'(1) \ent{ \frac{ \rho'^{\cnop}(\msx_{i}) }{\rho'(1)} \cnop \msy_{i} \cnop \msz_{i} }  \right] \notag \\
    & - \frac{L'(1)\lambda'(1) \rho'(1)^{2}}{w} \sum_{i=1}^{\chend} \sum_{ m = \max\{i-(w-1),1\} }^{ \min \{i+(w-1),\chend \}   } \mathrm{H} \Bigg{(} \frac{1}{w} \sum_{k=0}^{w-1}\msc_{i-k} \vnop \tfrac{ \lambda'^{\vnop} \left( \frac{1}{w} \sum\limits_{j=0}^{w-1} \rho^{\cnop}( \ms{x}_{i-k+j})\right)}{\lambda'(1)}  \notag \vnop \Big{[} \tfrac{\rho'^{\cnop}(\msx_{i})}{\rho'(1)}  \cnop \msz_{i} \Big{]}  \vnop \Big{[} \tfrac{\rho'^{\cnop}( \msx_{m})}{\rho'(1)}  \cnop \msy_{m} \Big{]} \Bigg{)} \notag
  \end{align}
  \noindent\makebox[\linewidth]{\rule{17.8cm}{0.4pt}}
\end{figure*}

\begin{figure}[!tb]
  \centering
  \setlength\tikzheight{5cm}
  \setlength\tikzwidth{6cm} 
  \begin{tikzpicture}[scale=0.2][domain=-20:20]
  \draw[very thin,color=gray] (-9,-1.25) -- (-9,1.25) ; 
  \draw[very thin,color=gray] (9,-1.25) -- (9,1.25) ; 
  \draw[very thin,color=gray] (-16,0) -- (16,0) ; 

  \draw[color=gray,very thick,dashed] 
  (-9,1.02746)--(-8.6,2.40972)--(-8.2,4.03658)--(-7.8,5.80014)--(-7.4,7.62622)--(-7,9.4749)--(-6.6,10.3015)--(-6.2,10.7752)--(-5.8,11.0051)--(-5.4,11.0985)--(-5,11.1294)--(-4.6,11.1378)--(-4.2,11.1407)--(-3.8,11.1418)--(-3.4,11.1421)--(-3,11.1422)--(-2.6,11.1423)--(-2.2,11.1423)--(-1.8,11.1423)--(-1.4,11.1423)--(-1,11.1423)--(-0.6,11.1423)--(-0.2,11.1423)--(0.2,11.1423)--(0.6,11.1423)--(1,11.1423)--(1.4,11.1423)--(1.8,11.1423)--(2.2,11.1423)--(2.6,11.1423)--(3,11.1423)--(3.4,11.1423)--(3.8,11.1423)--(4.2,11.1423)--(4.6,11.1423)--(5,11.1423)--(5.4,11.1423)--(5.8,11.1423)--(6.2,11.1423)--(6.6,11.1423)--(7,11.1423)--(7.4,11.1423)--(7.8,11.1423)--(8.2,11.1423)--(8.6,11.1423)--(9,11.1423);
  \draw[color=gray,very thick,dashed] (9,0.25) -- (16,0.25);
  \draw[color=gray,very thick,dashed] (-16,0.25) -- (-9,0.25);
  
  \draw[color=black,very thick,yshift=-0.7cm] 
  (-9,1.02746)--(-8.6,2.40972)--(-8.2,4.03658)--(-7.8,5.80014)--(-7.4,7.62622)--(-7,9.4749)--(-6.6,10.3015)--(-6.2,10.7752)--(-5.8,11.0051)--(-5.4,11.0985)--(-5,11.1294)--(-4.6,11.1378)--(-4.2,11.1407)--(-3.8,11.1418)--(-3.4,11.1421)--(-3,11.1422)--(-2.6,11.1423)--(-2.2,11.1423)--(-1.8,11.1423)--(-1.4,11.1423)--(-1,11.1423)--(-0.6,11.1423)--(-0.2,11.1423)--(0.2,11.1423)--(0.6,11.1423)--(1,11.1423)--(1.4,11.1423)--(1.8,11.1423)--(2.2,11.1423)--(2.6,11.1423)--(3,11.1422)--(3.4,11.1421)--(3.8,11.1418)--(4.2,11.1407)--(4.6,11.1378)--(5,11.1294)--(5.4,11.0985)--(5.8,11.0051)--(6.2,10.7752)--(6.6,10.3015)--(7,9.4749)--(7.4,7.62622)--(7.8,5.80014)--(8.2,4.03658)--(8.6,2.40972)--(9,1.02746);
  \draw[color=black,very thick] (9,0) -- (16,0);
  \draw[color=black,very thick] (-16,0) -- (-9,0);

  \node at (-16,0) [below=20pt] {$\cdots$};
  \draw[color=black] (-13,-0.5) -- (-13,0.5) node[below=7pt] {$\begin{array}{c} 0 \\ \updownarrow \\ \Delta_\infty \end{array}$} ; 
  \draw[color=black] (-9,-0.5) -- (-9,0.5) node[below=7pt] {$\begin{array}{c} 1 \\ \updownarrow \\ \mathsf{x}_1 \end{array}$};
  \draw[color=black] (-5,-0.5) -- (-5,0.5) node[below=7pt] {$\begin{array}{c} 2 \\ \updownarrow \\ \mathsf{x}_2 \end{array}$};
  \node at (-2.5,0) [below=20pt]  {$\cdots$};
  \draw[color=black] (0,-0.5) -- (0,0.5) node[below=7pt] {$\begin{array}{c} i_0 \\ \updownarrow \\ \mathsf{x}_{i_0}  \end{array}$}; 
  \node at (4.5,0) [below=20pt] {$\cdots$};
  \draw[color=black] (9,-0.5) -- (9,0.5) node[below=7pt] {$\begin{array}{c} \chend \\ \updownarrow \\ \mathsf{x}_{\chend} \end{array}$};
  \draw[color=black] (13,-0.5) -- (13,0.5) node[below=7pt] {$\begin{array}{c} \chend\!\!+\!1 \\ \updownarrow \\ \Delta_{\infty} \end{array}$ };
  \node at (17,0) [below=20pt] {$\cdots$};
\end{tikzpicture}
  \caption{This figure depicts the entropies of $\msx_1,\cdots,\msx_{\chend}$ in a typical iteration. 
    The solid line corresponds to the spatially-coupled system and the dashed line to the modified system.
    The distributions of the modified system are always degraded with respect to the spatially-coupled system, hence a higher entropy.
    The distributions outside the set $\{1,\cdots,\chend\}$ are fixed to $\Delta_\infty$ for both the systems.
  }
  \label{figure:system_comparison_ldpc}
\end{figure}
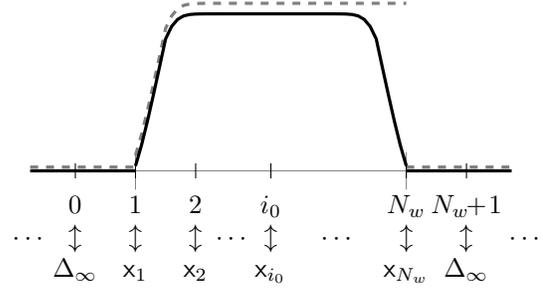

When the spatially-coupled system is initialized with
\begin{align*}
  \msx_{i}^{(0)} = \vnunit, \quad 1 \leq i \leq \chend ,
\end{align*}
the uniform coupling coefficients and symmetric boundary conditions induce left-right symmetry on $\msbx^{(\ell)}$.
In particular, the spatially-coupled system is fully described by only half the distributions because
\begin{align*}
  \msx_{i}^{(\ell)} = \msx_{2N+w-i}^{(\ell)},
\end{align*}
for all $\ell$.
As density evolution progresses, the perfect information from the boundary propagates inward.
This propagation induces a non-decreasing degradation ordering on positions $1, \ldots, \lceil \chend / 2 \rceil$ and a non-increasing degradation ordering on positions $ \lceil \chend / 2 \rceil + 1, \ldots, \chend$.
For example, see Fig.~\ref{figure:system_comparison_ldpc}.

This ordering introduces a degraded maximum at $i_{0} \triangleq N+ \lceil \frac{w-1}{2} \rceil$, and this maximum allows one to define a modified recursion that upper bounds the spatially-coupled system.
\begin{definition}
  The \emph{modified system} is a modification of (\ref{equation:scde_update_1_ldpc}) defined by fixing the values of positions outside $\mc{N}_c ' \triangleq \{1,2,\ldots,i_0\}$, where $i_{0}$ is defined as above.
  As before, the boundary is fixed to $\cnunit$, that is $\msx_{i}^{(\ell)} = \cnunit$ for $i \not \in \mc{N}_{c}$ and all $\ell$.
  More importantly, it fixes the values $\msx_{i}^{(\ell)} = \msx_{i_0}^{(\ell)}$ for $ i_0 < i \leq \chend $ and all $\ell$.
\end{definition}

The DE update of the modified system is identical to (\ref{equation:scde_update_1_ldpc}) for the first $i_{0}$ terms, $1, \ldots, i_{0}$, but a secondary update is required to impose the saturation constraint, $\msx_{i} = \msx_{i_{0}}$ for $i_{0} < i \le \chend$.
Repeated iterations for this system require that this saturation constraint is applied at every step.
The distributions of modified system are degraded with respect to that of spatially-coupled system, thus the modified system serves as a convenient upper bound for the spatially-coupled system.
Both the spatially-coupled system and the modified system are collectively referred to as \emph{coupled systems}.
 
In Fig.~\ref{figure:system_comparison_ldpc}, the entropies of the two systems are illustrated in a typical iteration.
We emphasize that the operator $\dec$ refers to the spatially-coupled system, \emph{not} the modified system.
However, the DE update for the modified system also satisfies the same monotonicity properties of $\dec$ in Lemma \ref{lemma:scde_monotonicity_ldpc}. 

If either spatially-coupled system or modified system is initialized with $\msbx^{(0)}=\underline{\vnunit} \triangleq \{\vnunit,\ldots,\vnunit\}$, then the sequence of measure vectors $\{\msbx^{(\ell)}\}$, by Lemma \ref{lemma:scde_monotonicity_ldpc}, satisfies $\msbx^{(\ell+1)} \upgreq \msbx^{(\ell)}$ and converges to a fixed point $\msbx$.
Thus, for the spatially-coupled system,
\begin{align*}
  \msbx = \dec(\msbx;\msc) .
\end{align*}
Such a fixed point for the modified system satisfies an additional property, stated in the following lemma.

\begin{lemma}
  \label{lemma:fp_monotonicity_modifiedsystem_ldpc}
  The fixed point $\msbx$ resulting from initializing the modified system with $\underline{\vnunit}$ satisfies 
  \begin{align*}
    \msx_{i} \degreq \msx_{i-1}, \quad 2 \leq i \leq \chend
  \end{align*}
\end{lemma}
\begin{IEEEproof}
  See Appendix \ref{appendix:proof_fp_monotonicity_modifiedsystem_ldpc}.
\end{IEEEproof}

Now, we define the coupled potential. 
The definitions below pertain to both spatially-coupled and modified system.

\begin{definition}
  \label{definition:coupled_potential_ldpc}
  The coupled potential functional $\potc\colon \mc{X}^{\chend} \times \mc{X} \rightarrow \mathbb{R}$ is given in (\ref{equation:coupled_potential_ldpc}).
\end{definition}

\begin{lemma}
  \label{lemma:first_derivative_coupled_system_ldpc}
  The directional derivative of the potential functional in (\ref{equation:coupled_potential_ldpc}) with respect to $\msbx \in \probs^{\chend}$, evaluated in the direction $\msby \in \dpros^{\chend}$ is given by
 \begin{align}
   \label{equation:first_derivative_coupled_system_ldpc}
   & \deri{\msbx} \potc(\msbx ; \msc)[\msby] \notag \\
   & \qquad = L'(1) \sum_{i=1}^{\chend} \ent{ \left( \dec(\msbx;\msc)_{i} - \msx_{i}\right) \cnop \rho'^{\cnop}(\msx_i) \cnop \msy_i } .
 \end{align}
\end{lemma}
\begin{IEEEproof}
  See Appendix \ref{appendix:proof_first_derivative_coupled_system_ldpc}.
\end{IEEEproof}

\begin{lemma}
  \label{lemma:second_derivative_coupled_system_ldpc}
  The second-order directional derivative of the potential functional in (\ref{equation:coupled_potential_ldpc}) with respect to $\msbx$, evaluated in the direction $[\msby,\msbz] \in \dpros^{\chend} \times \dpros^{\chend}$ is given in (\ref{equation:second_derivative_coupled_system_ldpc}).
\end{lemma}
\begin{IEEEproof}
  See Appendix \ref{appendix:proof_second_derivative_coupled_system_ldpc}.
\end{IEEEproof}

\section{Threshold Saturation for LDPC Ensembles}
\label{section:maintheorem_ldpc}

\subsection{Achievability of Threshold Saturation}
\label{subsection:thresholdsaturation_ldpc}

We now prove threshold saturation for spatially-coupled LDPC ensembles.
For a family of BMS channels, we will show that, if $\h < \h^*$, then the only fixed point of the modified system is $\underline{\cnunit}$.
Since the modified system is an upper bound on the spatially-coupled system, we then conclude that the only fixed point of the spatially-coupled system is $\underline{\cnunit}$.

Consider a modified system with potential functional $\potc$ as in Definition~\ref{definition:coupled_potential_ldpc}, and a non-trivial fixed point $\msbx$.
Also, consider a parameterization $\phi \colon [0,1] \rightarrow \mathbb{R}$, where
\begin{align*}
  \phi(t) = \potc(\msbx + t(\msbx' - \msbx);\msc(\h)) .
\end{align*}
The path endpoint $\msbx'$ is chosen to be a small perturbation of $\msbx$.
For all channels $\msc(\h)$ with $\h<\h^*$, at $\msbx$, it can be shown that the potential functional decreases, at least by a constant independent of the modified system, along the perturbation $\msbx'$.
Moreover, a fixed point is also a stationary point of the potential functional.
Also, at the fixed point, the second-order variations in the potential can be made arbitrarily small by choosing a large coupling parameter $w$.  
Thus, all variations in the potential functional up to second-order can be made arbitrarily small.

By calculating the change in potential at a non-trivial fixed point in two different ways: first by explicit calculation of change in the potential and second by the first- and second-order variations, one obtains a contradiction to the existence of a non-trivial fixed point from the second-order Taylor expansion of $\phi(t)$, for all $\msc(\h$) with $\h < \h^*$.

These ideas are formalized below.
A right shift is chosen for the perturbation and the shift operator $\shft(\cdot)$ is defined in Definition \ref{definition:shift_operator_ldpc}.
In Lemma \ref{lemma:potential_shift_bound_ldpc}, we bound the change in potential due to shift.
Lemmas \ref{lemma:modifiedsystem_potential_fixedpoint_ldpc} and \ref{lemma:second_derivative_bound_ldpc} characterize the first- and second-order variations, respectively, along the shift direction $[\shft(\msbx) - \msbx]$, for a non-trivial fixed point $\msbx$.
Finally, Theorem \ref{theorem:threshold_saturation_ldpc} proves threshold saturation.

\begin{definition}
  \label{definition:shift_operator_ldpc}
  The shift operator $\shft \colon \probs^{\chend}  \rightarrow \probs^{\chend}$ is defined pointwise by
  \begin{align*}
    [\shft(\msbx)]_{1} &\triangleq \cnunit, & [\shft(\msbx) ]_i &\triangleq \msx_{i-1}, \quad \text{$2 \leq i \leq \chend$}.
  \end{align*}
\end{definition}

\begin{lemma}
  \label{lemma:potential_shift_bound_ldpc}
  Let $\msbx \in \probs^{\chend}$ be such that $\msx_{i} = \msx_{i_0}$, for $i_0 \leq i \leq \chend$.
  Then the change in the potential functional for a modified system associated with the shift operator is bounded by
  \begin{align*}
    \potc(\shft(\msbx) ; \msc) - \potc(\msbx ; \msc) \leq - \pots(\msx_{i_0} ; \msc).
  \end{align*}
\end{lemma}
\begin{IEEEproof}
  See Appendix \ref{appendix:proof_potential_shift_bound_ldpc}.
\end{IEEEproof}

\begin{lemma}
  \label{lemma:modifiedsystem_potential_fixedpoint_ldpc}
  If $\msbx \degr \underline{\cnunit} \triangleq [\cnunit,\ldots,\cnunit]$ is a fixed point of the modified system resulting from $\underline{\vnunit}$ initialization, then
  \begin{align*}
    \deri{\msbx} \potc( \msbx ; \msc)[\shft(\msbx) - \msbx] = 0,
  \end{align*}
  and moreover $\msx_{i_0}$ is not in the basin of attraction to $\cnunit$ (i.e., $\msx_{i_0} \notin \mc{V}(\msc)$).
\end{lemma}
\begin{IEEEproof}
  See Appendix \ref{appendix:proof_modifiedsystem_potential_fixedpoint_ldpc}.
\end{IEEEproof}

The above two lemmas together with Definition \ref{definition:basinofattraction_energygap_ldpc}(ii) imply that for a non-trivial fixed point $\msbx$ resulting from initializing the modified system with $\underline{\vnunit}$,
\begin{align*}
    \potc(\shft(\msbx) ; \msc) - \potc(\msbx ; \msc) \leq - \pots(\msx_{i_0} ; \msc) \leq - \Delta E(\msc).
\end{align*}
Thus, when $\Delta E(\msc)>0$, the absolute change in potential due to shift is lower bounded by a constant independent of $\msbx$, $N$, $w$, and hence of the coupled system.

\begin{lemma}
  \label{lemma:second_derivative_bound_ldpc}
  Suppose $\msbx$ is a fixed point of the modified system resulting from $\underline{\vnunit}$ initialization.
  The second-order directional derivative of $\potc(\msbx_{1} ; \msc)$ with respect to $\msbx_{1}$, evaluated along $[\shft(\msbx) - \msbx, \shft(\msbx) - \msbx]$, can be absolutely bounded with 
  \begin{align*}
    \abs{ \dderi{\msbx_{1}} \potc(\msbx_{1} ; \msc)[\shft(\msbx) - \msbx,\shft(\msbx) - \msbx] } \leq \frac{K_{\lambda,\rho}}{w},
  \end{align*}
  where the constant
  \begin{align*}
    K_{\lambda,\rho} \triangleq  L'(1) \left( 2\rho''(1) + \rho'(1) + 2\lambda'(1) \rho'(1)^{2} \right)
  \end{align*}
  is independent of $N$ and $w$.
\end{lemma}
\begin{IEEEproof}
  See Appendix \ref{appendix:proof_second_derivative_bound_ldpc}.
\end{IEEEproof}

\begin{theorem}
  \label{theorem:threshold_saturation_ldpc}
  Fix a family of BMS channels $\msc(\h)$, and the LDPC$(\lambda,\rho)$ ensemble.
  For $\h < \h^{\ast}$, all $N$, and any $w > K_{\lambda,\rho}  / (2 \Delta E(\msc(\h)))$, the only fixed point of density evolution for the spatially-coupled LDPC $(\lambda,\rho,N,w)$ ensemble with channel $\msc(\h)$ is $\underline{\cnunit}$.
\end{theorem}
\begin{IEEEproof}
  First, since $\h < \h^*$, $\Delta E(\msc(\h)) >0$.
  Consider a modified system with a fixed $w>K_{\lambda,\rho}  / (2 \Delta E(\msc(\h)))$ and any $N$.
  Suppose $\msbx$ is a fixed point of modified system resulting from $\underline{\vnunit}$ initialization.
  If $\msbx=\underline{\cnunit}$, by the monotonicity of the DE update resulting from $\underline{\vnunit}$ initialization, there is no other fixed point for the modified system.
  Suppose instead that $\msbx \degr \underline{\cnunit}$.
  In this case, we will arrive at a contradiction in the following.

  Let $\msby = \shft(\msbx) - \msbx$ and define $\phi \colon [0,1] \rightarrow \mbb{R}$ by
  \begin{align*}
    \phi(t) = \potc(\msbx + t \msby ; \msc(\h)) .
  \end{align*}
  This is well defined because, for all $t \in [0,1]$, $\msbx + t \msby = (1-t) \msbx + t \shft(\msbx) $ is a vector of probability measures.
  As in Proposition \ref{proposition:polynomial_structure_example}, $\phi$ is a polynomial in $t$, and thus infinitely differentiable over the entire unit interval.
  Hence, the second-order Taylor series expansion about $t=0$, evaluated at $t=1$, provides
  \begin{align}
    \label{equation:maintheorem_phi_taylor_ldpc}
    \phi(1) = \phi(0) + \phi'(0)(1-0) + \tfrac{1}{2} \phi''(t_0)(1-0)^{2},
  \end{align}
  for some $t_0 \in [0,1]$.
  The first and second derivatives of $\phi$ are characterized by the first- and second-order directional derivatives of $\potc$:
  \begin{align*}
    \phi'(t) &= \lim_{\delta \rightarrow 0} \frac{ \potc( \msbx  + (t+\delta)\msby ; \msc(\h) ) - \potc(\msbx + t \msby ; \msc(\h)) }{\delta} \\
    &= \deri{\msbx_1} \potc( \msbx_1; \msc(\h))[\msby] \bvert{\msbx_1=\msbx + t \msby},
  \end{align*}
  and similarly,
  \begin{align*}
    \phi''(t) = \dderi{\msbx_1} \potc(\msbx_1 ; \msc(\h))[\msby,\msby] \bvert{\msbx_1=\msbx + t \msby}.
  \end{align*}
  Substituting and rearranging terms in (\ref{equation:maintheorem_phi_taylor_ldpc}) provides
  \allowdisplaybreaks{
    \begin{align*}
      & \tfrac{1}{2} \dderi{\msbx_1} \potc (\msbx_1 ; \msc(\h) ) [\msby,\msby]  \bvert{\msbx_1=\msbx + t_0 \msby}\\
      & \quad = \potc(\shft(\msbx) ; \msc(\h)) - \potc(\msbx ; \msc(\h))- \deri{\msbx} \potc( \msbx ; \msc(\h))[\shft(\msbx) - \msbx] \\
      & \quad = \potc(\shft(\msbx) ; \msc(\h)) - \potc(\msbx ; \msc(\h)) \quad \text{(Lemma \ref{lemma:modifiedsystem_potential_fixedpoint_ldpc})}  \\
      & \quad \leq - \pots(\msx_{i_{0}} ; \msc)   \quad \text{(Lemma \ref{lemma:potential_shift_bound_ldpc})} \\
      & \quad \leq - \Delta E(\msc(\h)). \quad \text{(Lemma \ref{lemma:modifiedsystem_potential_fixedpoint_ldpc} and Definition \ref{definition:basinofattraction_energygap_ldpc}(ii))}
  \end{align*}
  }
  Taking the absolute value and applying the second order directional derivative bound from Lemma \ref{lemma:second_derivative_bound_ldpc} gives
  \begin{align*}
    \Delta E(\msc(\h)) \leq \frac{K_{\lambda,\rho}}{2 w}  \quad \Longrightarrow \quad    w \leq \frac{K_{\lambda,\rho}}{2 \Delta E(\msc(\h))},
  \end{align*}
  a contradiction.
  Hence the only fixed point of the modified system is $\underline{\cnunit}$.
  The distributions of the modified system are degraded with respect to the spatially-coupled system, and therefore, the only fixed point of the spatially-coupled system is also $\underline{\cnunit}$.
\end{IEEEproof}
As an immediate consequence, for the $(\lambda,\rho,N,w)$ spatially-coupled ensemble with $0<K_{\lambda,\rho}  / (2 \Delta E(\msc(\h)))<w<\infty$ and any $N$, its BP threshold is at least $\h$.
Therefore, the BP threshold of the $(\lambda,\rho,N,w)$ spatially-coupled ensemble, by first taking the limit $N \to \infty$ and then $w \to \infty$, is at least $\h^*$.
Below, Theorem \ref{theorem:converse_threshold_saturation_ldpc} establishes that, under $\h^*<\hstab$, the BP threshold of the spatially-coupled ensemble in the limits given above is at most $\h^*$, which establishes the equality of the BP threshold to $\h^*$ in the above limits.

\subsection{Converse to Threshold Saturation}
\label{subsection:converse_ldpc}
We begin by establishing two monotonicity results. 

\begin{lemma}
  \label{lemma:scde_order_ldpc}
  Consider $\msbx_1 \in \probs^{N_w}$ and $\msbx_2=\dec \left( \msbx_1 ; \msc \right)$.
  \begin{enumerate}[i)]
    \item If $\msbx_2 \degreq \msbx_1$,
      \begin{align*}
        \dec \left( \msbx_1 + t ( \msbx_2 - \msbx_1) ; \msc \right) &\degreq \msbx_1 + t (\msbx_2 - \msbx_1) .
      \end{align*}
    \item If $\msbx_2 \upgreq \msbx_1$,
      \begin{align*}
        \dec \left( \msbx_1 + t ( \msbx_2 - \msbx_1) ; \msc \right) &\upgreq \msbx_1 + t (\msbx_2 - \msbx_1) .
      \end{align*}
  \end{enumerate}
\end{lemma}
\begin{IEEEproof}
  \begin{enumerate}[i)]
  \item If $\msbx_2 \degreq \msbx_1$, then for all $0 \le t \le 1$,
    \begin{align*}
      \msbx_2 \degreq \msbx_1 + t (\msbx_2 - \msbx_1) \degreq \msbx_1 .
    \end{align*}
    Since $\dec$ is order-preserving by Lemma \ref{lemma:scde_monotonicity_ldpc},
    \begin{align*}
      \dec \left( \msbx_1 + t (\msbx_2 - \msbx_1) ; \msc \right) & \degreq \dec \left( \msbx_1 ; \msc\right) \\
      &= \msbx_2 \degreq \msbx_1 + t(\msbx_2-\msbx_1) .
    \end{align*}
  \item Follows by symmetry.
  \end{enumerate}
\end{IEEEproof}

\begin{lemma}
  \label{lemma:potential_property_ldpc}
  Let $\msbx_1 \in \probs^{N_w}$, $\msbx_2=\dec \left( \msbx_1 ; \msc\right)$, and suppose $\msbx_{2} \degreq \msbx_{1}$ or $\msbx_{2} \upgreq \msbx_{1}$, then $\potc(\msbx_2 ; \msc) \leq \potc(\msbx_1 ; \msc)$.
\end{lemma}
\begin{IEEEproof}
Assume $\msbx_{2} \degreq \msbx_{1}$. 
Let $\phi\colon [0,1] \rightarrow \mathbb{R}$ be defined by
\begin{align*}
  \phi(t) = \potc(\msbx_1 + t (\msbx_2 - \msbx_1) ; \msc) .
\end{align*}
Observe that $\phi$ is a polynomial in $t$ as in Proposition \ref{proposition:polynomial_structure_example}, with $\phi(0)=\potc(\msbx_1;\msc)$ and $\phi(1)=\potc(\msbx_2;\msc)$.
Moreover,
\begin{align}
  \label{potential_property_ldpc_phi}
  \phi'(t) = \deri{\msbx} \potc(\msbx ; \msc) [\msbx_2-\msbx_1] \bvert{\msbx=\msbx_1+t(\msbx_2-\msbx_1)} .
\end{align}
By Lemma \ref{lemma:scde_order_ldpc}, 
\begin{align*}
  \dec \left( \msbx_1 + t (\msbx_2 - \msbx_1) ;\msc  \right) \degreq \msbx_1 + t (\msbx_2 - \msbx_1) ,
\end{align*} 
and observing (\ref{equation:first_derivative_coupled_system_ldpc}), the derivative in (\ref{potential_property_ldpc_phi}) is a sum of terms of the form
\begin{equation*}
  L'(1) \ent{ \left[ \msx'_{3} - \msx_{3} \right] \cnop \msx_{4} \cnop \left[ \msx'_{5} - \msx_{5} \right] },
\end{equation*}
where $\msx'_{3} \degreq \msx_{3}$ and $\msx'_{5} \degreq \msx_{5}$, which is negative by Proposition \ref{proposition:entropy_dpros_properties}(iii).
For the case $\msbx_{2} \upgreq \msbx_{1}$, we can write a similar expression with $\msx_3' \upgreq \msx_3$ and $\msx_5' \upgreq \msx_5$.
In either case, $\phi'(t) \leq 0$ for all $t \in [0,1]$.
Thus, $\potc(\msbx_2 ; \msc) = \phi(1) \leq \phi(0) = \potc(\msbx_1 ; \msc)$.
\end{IEEEproof}

\begin{theorem}
  \label{theorem:converse_threshold_saturation_ldpc}
Fix a family of BMS channels $\msc(\h)$ and the LDPC$(\lambda,\rho)$ ensemble with $\h^* < \hstab$.
Also, consider the spatially-coupled LDPC $(\lambda,\rho,N,w_0)$ ensemble with a fixed coupling window $w_0$, and a channel $\msc(\h)$ with $\h>\h^*$.
Then, there exists an $N_{0}$ such that, for any $N > N_{0}$, the fixed point of density evolution resulting from $\underline{\vnunit}$ initialization satisfies
\begin{align*}
    \dec^{(\infty)}(\underline{\vnunit};\msc(\h)) \degr \underline{\cnunit} .
  \end{align*}
\end{theorem}
\begin{IEEEproof}
  First, choose $\h > \h^*$.
  Since $\pots(\cdotp;\msc(\h))\colon \probs \rightarrow \mathbb{R}$ is continuous and $\probs$ is compact, $\pots(\cdotp;\msc(\h))$ attains its minimum.
  Let $\msa_{*}$ be a minimizer of $\pots(\cdotp;\msc(\h))$.
  By Lemma \ref{lemma:fixed_stationary_ldpc}, $\msa_{*}$ is a fixed point of the single system DE.
  By assumption $\hstab>\h^*$, and $\h>\h^*$. 
  Hence, by Lemma \ref{lemma:beyond_potential_threshold_negativity}, $\pots(\msa_{*} ; \msc(\h)) < 0$.
  Initialize the spatially-coupled LDPC $(\lambda,\rho,N,w_0)$ system with $\msba_{*}=[\msa_{*},\ldots,\msa_{*}]$.
  Since $\msa_{*}$ is a fixed point of the single system,
  \begin{align*}
    \dec(\msba_{*};\msc(\h))_{i} &= \frac{1}{w} \sum_{k=0}^{w-1} \msc(\h)_{i-k} \vnop \lambda^{\vnop} \Big{(}  \frac{1}{w} \sum_{j=0}^{w-1} \rho^{\cnop} \left( \msa_{*} \right) \Big{)}  \\
    &\upgreq \frac{1}{w} \sum_{k=0}^{w-1} \msc(\h) \vnop \lambda^{\vnop} \Big{(}  \frac{1}{w} \sum_{j=0}^{w-1} \rho^{\cnop} \left( \msa_{*} \right) \Big{)} \\
    &= \msc(\h) \vnop \lambda^{\vnop}(\rho^{\cnop}(\msa_{*})) = \msa_{*} .
  \end{align*}
  That is, $\dec(\msba_{*};\msc(\h)) \upgreq \msba_{*}$.
  Therefore, from the monotonicity of $\dec$ by Lemma \ref{lemma:scde_monotonicity_ldpc}, $\dec^{(\infty)}(\msba_{*};\msc(\h))$ exists and
  \begin{align*}
    \dec^{(\infty)}(\msba_{*};\msc(\h)) \upgreq \dec^{(\ell+1)}(\msba_{*};\msc(\h)) \upgreq \dec^{(\ell)}(\msba_{*};\msc(\h)) \upgreq \msba_{*} .
  \end{align*}
  By Lemma \ref{lemma:potential_property_ldpc} and the continuity of $\potc(\cdot;\msc(\h))$,
  \begin{align*}
    & \potc(\dec^{(\infty)}(\msba_{*};\msc(\h));\msc(\h)) \leq \potc(\dec^{(\ell+1)}(\msba_{*};\msc(\h)) ; \msc(\h)) \\
    & \qquad \leq \potc(\dec^{(\ell)}(\msba_{*};\msc(\h)) ; \msc(\h)) \leq \potc(\msba_{*} ; \msc(\h)) .
  \end{align*}
  Also, since all entries of $\msba_{*}$ are equal,
  \begin{align*}
    & \potc(\msba_{*} ; \msc(\h)) \\
    &\qquad = (2N+(w_0-1)) \pots(\msa_{*} ; \msc(\h)) \\
    &\qquad \qquad \qquad + (w_0-1) \ent{\msc(\h) \vnop L^{\vnop}(\rho^{\cnop}(\msa_*))} \\
    &\qquad \leq (2N+(w_0-1)) \pots(\msa_{*} ; \msc(\h)) + w_0-1 .
  \end{align*}
  Since $\pots(\msa_{*};\msc(\h))<0$, we can choose large enough $N_0$ such that for all $N>N_0$, $\potc(\msba_{*} ; \msc(\h)) < 0$.
  Therefore,
  \begin{align*}
    \potc(\dec^{(\infty)}(\msba_*;\msc(\h)) ; \msc(\h)) \leq \potc(\msba_{*};\msc(\h)) < 0 ,
  \end{align*}
  and, since $\potc(\underline{\cnunit};\msc(\h))=0$, this implies that $\dec^{(\infty)}(\msba_* ; \msc(\h)) \neq \underline{\cnunit}$.
  Since $\underline{\vnunit} \degreq \msba_{*}$,
  \begin{align*}
    \dec^{(\infty)}(\underline{\vnunit} ; \msc(\h)) \degreq \dec^{(\infty)}(\msba_{*} ; \msc(\h)).    
  \end{align*}
  Hence, $\dec^{(\infty)}(\underline{\vnunit} ; \msc(\h)) \degr \underline{\cnunit}$.
\end{IEEEproof}

\section{Low-Density Generator-Matrix Ensembles}

\label{section:ldgm}

\subsection{Single System}
\label{subsection:single_system_ldgm}

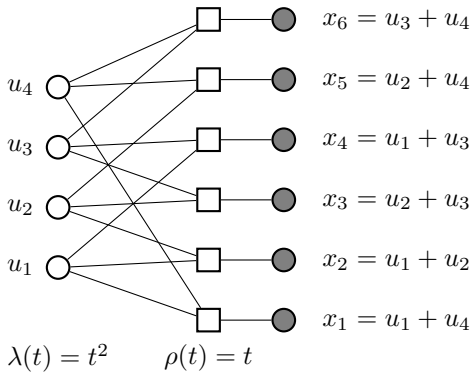
\begin{figure}[!tb]
  \centering
  \setlength\tikzheight{5cm}
  \setlength\tikzwidth{6cm} 
  \begin{tikzpicture}
  [
  node distance = 3mm, draw=black, thick, >=stealth',
  bitnode/.style={circle, inner sep = 0pt, minimum size = 3mm, draw=black},
  checknode/.style={rectangle, inner sep = 0pt, minimum size = 3mm, draw=black},
  bitnode2/.style={circle, inner sep = 0pt, minimum size = 3mm, draw=black, fill=black!50},
  ]
  
  \foreach \y in {1,2,...,6} {
    \node[bitnode2] (bb\y) at (3,0.8*\y-5) {};
  }

  \foreach \y in {1,2,...,6} {
    \node[checknode] (c\y) at (2,0.8*\y-5) {};
    \draw[thin] (c\y) -- (bb\y);
  }

  \foreach \y in {1,2,...,4} {
    \node[bitnode] (b\y) at (0,0.8*\y-4.3) {};
  }

  \draw[thin] (c1) -- (b4);
  \draw[thin] (c1) -- (b1);
  \draw[thin] (c2) -- (b1);
  \draw[thin] (c2) -- (b2);
  \draw[thin] (c3) -- (b3);
  \draw[thin] (c3) -- (b2);
  \draw[thin] (c4) -- (b1);
  \draw[thin] (c4) -- (b3);
  \draw[thin] (c5) -- (b2);
  \draw[thin] (c5) -- (b4);
  \draw[thin] (c6) -- (b3);
  \draw[thin] (c6) -- (b4);

  \foreach \y in {1,2,...,4} {
    \node at (-0.5,0.8*\y-4.3) {$u_{\y}$};
  }

  \node at (4.5,0.8*1-5) {$x_{1}=u_1+u_4$};
  \node at (4.5,0.8*2-5) {$x_{2}=u_1+u_2$};
  \node at (4.5,0.8*3-5) {$x_{3}=u_2+u_3$};
  \node at (4.5,0.8*4-5) {$x_{4}=u_1+u_3$};
  \node at (4.5,0.8*5-5) {$x_{5}=u_2+u_4$};
  \node at (4.5,0.8*6-5) {$x_{6}=u_3+u_4$};

  \node at (0,-4.7) {$\lambda(t)=t^2$};
  \node at (2,-4.7) {$\rho(t)=t$};

\end{tikzpicture}
  \caption{The Tanner graph representation of an LDGM code with left-degree 3 and right-degree 2.
    The leftmost nodes $u_i$'s are the information-nodes and the square nodes are generator-nodes.
    The rightmost nodes in gray represent the code-bits.
  }
  \label{figure:tanner_graph_ldgm}
\end{figure}

Low-density generator-matrix (LDGM) ensembles are a class of linear codes that have a sparse generator-matrix representation.
An example of a Tanner graph representation of an LDGM code is provided in Fig.~\ref{figure:tanner_graph_ldgm}.
The term LDGM$(\lambda,\rho)$ denotes the LDGM ensemble with information-node degree distribution $\lambda$ and generator-node degree distribution $\rho$ from the edge perspective.
An equivalent representation in terms of the node perspective degree distributions $L$, $R$ is given by
\begin{align*}
  \lambda(t) &= \frac{L'(t)}{L'(1)}, & \rho(t) &= \frac{R'(t)}{R'(1)} .
\end{align*}

LDGM codes are amenable to techniques similar to that of their counterpart, LDPC codes.
However, a key issue here is that these codes have non-negligible error floors.
One mathematical difficulty that arises from this is that the desired fixed point of DE is non-trivial and depends on the channel parameter.
This poses a great challenge when characterizing thresholds, convergence, etc.
Nevertheless, LDGM codes are an attractive option for rateless codes \cite{Luby-focs02}, \cite{Shokrollahi-06}, and in lossy source compression \cite{Wainwright-10}, \cite{Aref-arxiv13}.
See Section \cite[Section 7.5]{RU-2008} for an introduction to LDGM codes.

The analysis of LDGM codes, and their coupled variant, is very similar to that of the LDPC codes.
Thus, we keep the same notation for analogous quantities.

The evolution of message distributions is characterized by the DE described by
\begin{align}
  \label{equation:de_update_ldgm}
  \msxvn^{(\ell+1)} = \lambda^{\vnop}(\msc \cnop \rho^{\cnop}(\msxvn^{(\ell)})),
\end{align}
where $\msxvn^{(\ell)}$ denotes the message distribution at the output of information-nodes after $\ell$ iterations of message-passing, and $\msc$ represents the channel LLR distribution.
When the iterative system in (\ref{equation:de_update_ldgm}) is initialized with $\msa$, the information-node output after $\ell$ iterations is denoted by $\des^{(\ell)}(\msa;\msc)$.
The distribution after one iteration is therefore $\des^{(1)}(\msa;\msc)$, or shortly, $\des(\msa;\msc)$.
If the sequence of measures $\{\des^{(\ell)}(\msa;\msc)\}$ converges in $(\probs,\disth)$, then its limit is denoted by $\des^{(\infty)}(\msa;\msc)$.

The DE update operator $\des$ satisfies exactly the same monotonicity properties as in Lemma \ref{lemma:de_monotonicity_ldpc}.
To avoid repetition, we do not state them explicitly.

We note that $\cnunit$ is \emph{not} a fixed point of (\ref{equation:de_update_ldgm}), which is in stark contrast to LDPC codes.
If this system is initialized with $\cnunit$, then $\des(\cnunit;\msc) \degreq \cnunit$.
As such, the sequence $\{ \des^{(\ell)}(\cnunit;\msc)\}$ converges to the fixed point $\des^{(\infty)}(\cnunit;\msc)$.
If $\msx$ is any fixed point of (\ref{equation:de_update_ldgm}), since $\msx \degreq \cnunit$, by the monotonicity of $\des$, $$\msx=\des^{(\infty)}(\msx;\msc) \degreq \des^{(\infty)}(\cnunit;\msc).$$
Thus, $\des^{(\infty)}(\cnunit;\msc)$ is the \emph{minimal} fixed point.
 
\begin{definition}
  The \emph{minimal fixed point} for the LDGM$(\lambda,\rho)$ ensemble with channel $\msc$ is defined to be
  \begin{align*}
    \minf(\msc) \triangleq \des^{(\infty)}(\cnunit;\msc) .
  \end{align*}
  We also denote this by $\minf$ when the context is clear.
\end{definition}

The following definition of the potential functional is essentially the negative of the trial-entropy or the replica-symmetric free entropy calculated in \cite[Equation 6.2]{Montanari-it05}.
Also, in Appendix \ref{appendix:subsection_free_entropy_ldgm}, we briefly show the calculations to derive this potential from the Bethe formalism.
\begin{definition}
  \label{definition:uncoupled_potential_ldgm}
  The potential functional $\pots\colon \probs \times \probs \to \mbb{R}$ for the LDGM$(\lambda,\rho)$ ensemble with a channel $\msc$ is defined as
  \begin{align*}
    \pots(\msx;\msc) &=\tfrac{L'(1)}{R'(1)} \ent{\msc \cnop R^{\cnop}(\msx)}  - L'(1) \ent{\msx \cnop \msc \cnop \rho^{\cnop}(\msx) } \\
    & \quad + L'(1)\ent{\msc \cnop \rho^{\cnop}(\msx) } - \ent{L^{\vnop}(\msc \cnop \rho^{\cnop}(\msx))} \\
    & \quad - \tfrac{L'(1)}{R'(1)} \ent{\msc} .
  \end{align*}
\end{definition}

\begin{figure}[!tb]
  \centering
  \setlength\tikzheight{5cm}
  \setlength\tikzwidth{6cm} 
  \input{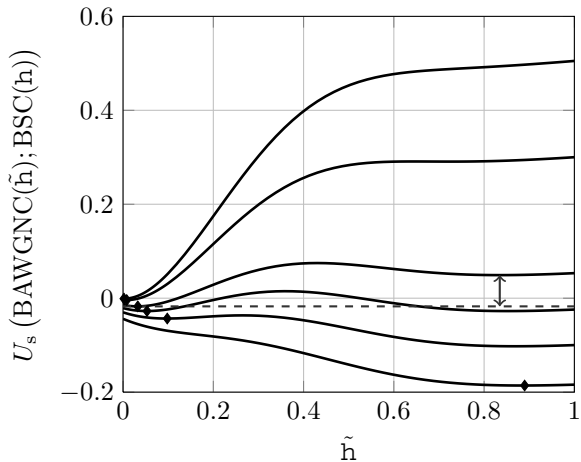}
  \caption{Potential functional for an LDGM$(\lambda,\rho)$ ensemble with $\lambda(t)=t^8$ and $\rho(t)=\tfrac{3}{50}+\tfrac{6}{50}t+\tfrac{9}{50}t^2+\tfrac{12}{50}t^3+\tfrac{20}{50}t^4$ over a binary symmetric channel with entropy $\h$.
    The values of $\h$ for these curves are, from the top to bottom, $0.37, 0.4529, 0.56, 0.5902, 0.62,0.66$.
    The other input to the potential functional is the binary AWGN channel (BAWGNC) with entropy $\tilde{\h}$.
    The choice of BAWGNC distribution for the first argument in $\pots(\cdotp;\cdot)$ is arbitrary.
    The marked points denote the minimal fixed points $\minf$.
  }
  \label{figure:potential_functional_ldgm}
\end{figure}

The directional derivative of the potential functional gives rise to the DE update in (\ref{equation:de_update_ldgm}).
Using Proposition \ref{proposition:duality_difference}, we have the following result similar to Lemma \ref{lemma:firstderivative_singlesystem_ldpc}.
\begin{lemma}
  The directional derivative of the potential functional with respect to $\msx \in \probs$, in the direction $\msy \in \dpros$, is given by
  \begin{align*}
    \deri{\msx} \pots(\msx;\msc) [\msy]  &= L'(1) \ent{ \left[ \des(\msx;\msc) - \msx \right] \cnop \left[ \msc \cnop \rho'^{\cnop}(\msx) \cnop \msy \right] } .
  \end{align*}
\end{lemma}
Similar to Lemma \ref{lemma:fixed_stationary_ldpc}, we can also show that the minimum of the potential functional for a fixed $\msc$ occurs at a fixed point of the DE.

\begin{definition}
  \label{definition:basinofattraction_energygap_ldgm}
  For the LDGM$(\lambda,\rho)$ ensemble with a channel $\msc \in \probs$, define
  \begin{enumerate}[i)]
  \item The basin of attraction to $\minf(\msc)$ as the set 
    \begin{align*}
      \mc{V}(\msc) = \{ \msx \in \probs \mid \des^{(\infty)}(\msx;\msc)=\minf(\msc)  \} .
    \end{align*}
  \item The \emph{energy gap} as
  \begin{align*}
    \Delta E (\msc) \triangleq \inf_{\msx \in \probs \setminus \mc{V}(\msc) } \pots(\msx;\msc) - \pots(\minf(\msc) ; \msc) ,
  \end{align*}
  with the convention that the infimum over the empty set is $\infty$.
  \end{enumerate}
\end{definition}

Fig.~\ref{figure:potential_functional_ldgm} illustrates the potential functional of an LDGM ensemble over a BSC channel with 
\begin{align*}
  \lambda(t) &= t^8, & \rho(t) &= \frac{3}{50}+\frac{6}{50}t+\frac{9}{50}t^2+\frac{12}{50}t^3+\frac{20}{50}t^4 .  
\end{align*}
A few observations are in order.
At small values of $\h$, the minimal fixed point $\minf(\msc(\h))$ determines the error floor of these ensembles.
As we increase $\h$ beyond $0.4529$, another fixed point appears in the right (from initializing DE with $\vnunit$), and this fixed point governs the DE performance. 
For $\h<0.5902$, the energy gap $\Delta E (\msc(\h))>0$ stays positive.
The range of $\h$ for which the energy gap stays positive is important, as this characterizes the performance of spatially-coupled codes.
For large values of $\h$, the fixed point resulting from $\vnunit$ initialization and the minimal fixed point coincide.
We emphasize that these observations are \emph{only qualitative} as this two-dimensional illustration does not characterize the behavior of $\pots(\cdotp;\msc)$ over all $\probs$.

By Definition \ref{definition:basinofattraction_energygap_ldgm}(ii), $\Delta E (\msc(\h))$ is a difference of two functions varying in $\h$. 
For general LDGM ensembles, whether the energy gap is monotone as a function of $\h$ is not known.
This poses a difficulty when defining the potential threshold.
We circumvent this by stating the threshold saturation theorem differently, and perhaps less elegantly, than LDPC ensembles.
More precisely, the result we have for LDGM ensembles is the following (Theorem \ref{theorem:threshold_saturation_ldgm}): If $\Delta E (\msc) > 0$, then, for a large enough coupling window $w$, any DE fixed point of the spatially-coupled system is elementwise better (in the degradation order) than the \emph{minimal fixed point} of the single system, $\minf(\msc)$.

It is conjectured \cite[Section X]{Montanari-it05} that the region where $\Delta E(\msc)>0$ characterizes the MAP decoding performance.
Accordingly, when $\Delta E(\msc)>0$, the potential functional is minimized at $\minf(\msc)$ and therefore the value of $L^{\vnop}(\msc \cnop \rho^{\cnop}(\minf(\msc)))$ under the error probability functional \cite[Definition 4.53]{RU-2008} characterizes the bit-error rate of the MAP decoder.
Moreover, when $\Delta E(\msc)<0$, the MAP decoder performance is strictly worse than the one characterized by $L^{\vnop}(\msc \cnop \rho^{\cnop}(\minf(\msc)))$.
Thus, if the conjecture in \cite[Section X]{Montanari-it05} is true, then the BP performance of the spatially-coupled ensemble and the MAP performance of the single system coincide.

\subsection{Coupled System}
\label{subsection:coupledsystem_ldgm}

The construction of spatially-coupled LDGM ensemble is similar to that of spatially-coupled LDPC ensembles and we refer the reader to Section \ref{subsection:coupled_system_ldpc} for an elaborate treatment.
A performance analysis of spatially-coupled LDGM ensembles first appeared in \cite{Aref-itw11}.
The information-node groups are placed at positions in $\mc{N}_v=\{1,2,\cdots,2N\}$, and the generator-node groups at $\mc{N}_c=\{1,2,\cdots,\chend\}$, where $\chend=2N+w-1$.
The DE update at generator-node inputs is given by 
\begin{align}
  \label{equation:scde_update_1_ldgm}
  \msx_{i}^{(\ell+1)} = \frac{1}{w} \sum_{k=0}^{w-1} \lambda^{\vnop} \left(\frac{1}{w} \sum_{j=0}^{w-1} \msc \cnop \rho^{\cnop}(\msx^{(\ell)}_{i-k+j}) ; \epsilon_{i-k} \right) ,
\end{align}
for $i \in \mc{N}_c$, where $\msx_{i}=\cnunit$ when $i \not \in \mc{N}_c$ and the shorthand $\lambda^{\vnop}(\msx;\epsilon_{i})$ denotes
\begin{align*}
  \lambda^{\vnop}(\msx;\epsilon_{i}) =
  \begin{cases}
    \lambda^{\vnop}(\msx) & \text{if $i \in \mc{N}_v$,} \\
    \cnunit & \text{otherwise.}
  \end{cases}
\end{align*}
We refer to the system characterized by (\ref{equation:scde_update_1_ldgm}) as the $(\lambda,\rho,N,w)$ spatially-coupled LDGM ensemble. 


\begin{figure*}[!t]
  \begin{align}
    \label{equation:coupled_potential_ldgm}
    \potc(\msbx ; \msc) & \triangleq L'(1) \sum_{i=1}^{\chend}\left[ \frac{1}{R'(1)}  \ent{ \msc \cnop R^{\cnop}(\ms{x}_{i}) } - \frac{1}{R'(1)} \ent{\msc}   - \ent{ \msx_i \cnop \msc \cnop \rho^{\cnop}( \ms{x}_{i}) } + \ent{ \msc \cnop \rho^{\cnop}(\msx_{i}) } \right] \\
    &  - \sum_{i=1}^{2N}   \ent{ L^{\vnop} \Big{(} \frac{1}{w}  \sum_{j=0}^{w-1} \msc \cnop \rho^{\cnop} ( \ms{x}_{i+j}) \Big{)} } - L'(1) \sum_{i=1}^{w-1} \left[ \frac{w-i}{w} \ent{\minf \vnop \left[ \msc \cnop \rho^{\cnop}(\msx_i) \right]} + \frac{i}{w} \ent{\minf \vnop \left[ \msc \cnop \rho^{\cnop} (\msx_{2N+i}) \right]} \right] \notag
  \end{align}
  \noindent\makebox[\linewidth]{\rule{17.8cm}{0.4pt}}
\end{figure*}

\begin{figure*}[!t]
  \begin{align}
    \label{equation:first_derivative_coupled_system_ldgm}
    \deri{\msbx} \potc(\msbx ; \msc)[\msby] =  L'(1) \sum_{i=1}^{\chend} \ent{ \left[ \frac{1}{w} \sum_{k=0}^{w-1} \lambda^{\vnop} \left( \frac{1}{w}  \sum_{j=0}^{w-1} \msc \cnop \rho^{\cnop} (\msx_{i-k+j}) ; \delta_{i-k}\right) - \msx_{i} \right] \cnop \left[ \msc \cnop \rho'^{\cnop}(\msx_i) \cnop \msy_i \right] }
  \end{align}
  \noindent\makebox[\linewidth]{\rule{17.8cm}{0.4pt}}
\end{figure*}

\begin{figure*}[!t]
  \begin{align}
    \label{equation:second_derivative_coupled_system_ldgm}
    & \dderi{\msbx} \potc(\msbx ; \msc)[\msby,\msbz] = \\  
    & L'(1) \rho''(1) \sum_{i=1}^{\chend} \ent{ \Bigg{[} \frac{1}{w} \sum_{k=0}^{w-1} \lambda^{\vnop} \Bigg{(} \frac{1}{w} \sum_{j=0}^{w-1} \msc \cnop \rho^{\cnop}( \msx_{i-k+j}) ; \delta_{i-k} \Bigg{)} \cnop \msc \cnop \frac{\rho''^{\cnop}(\msx_{i})}{\rho''(1)} \Bigg{]} \cnop \msy_{i} \cnop \msz_{i} }  \notag \\
    & - L'(1) \rho''(1) \sum_{i=1}^{\chend} \ent{ \left[ \msx_{i} \cnop \msc \cnop \frac{\rho''^{\cnop}(\msx_{i})}{\rho''(1)} \right] \cnop \msy_{i} \cnop \msz_{i} } - L'(1) \rho'(1) \sum_{i=1}^{\chend} \ent{ \msc \cnop \frac{ \rho'^{\cnop}(\msx_{i}) }{\rho'(1)} \cnop \msy_{i} \cnop \msz_{i} }  \notag \\
    & - \frac{L'(1)\lambda'(1) \rho'(1)^{2}}{w} \sum_{i=1}^{\chend} \sum_{ m = \max\{i-(w-1),1\} }^{ \min \{i+(w-1),\chend \}   } \mathrm{H} \Bigg{(} \frac{1}{w} \sum_{k=0}^{w-1}\tfrac{ \lambda'^{\vnop} \left( \frac{1}{w} \sum\limits_{j=0}^{w-1} \msc \cnop \rho^{\cnop}( \ms{x}_{i-k+j}) ; \delta_{i-k} \right)}{\lambda'(1)} \!\vnop\! \Big{[} \msc \!\cnop\! \tfrac{\rho'^{\cnop}(\msx_{i})}{\rho'(1)}  \!\cnop\! \msy_{i} \Big{]}  \!\vnop\! \Big{[} \msc \!\cnop\! \tfrac{\rho'^{\cnop}( \msx_{m})}{\rho'(1)}  \!\cnop\! \msz_{m} \Big{]} \Bigg{)} \notag
  \end{align}
  \noindent\makebox[\linewidth]{\rule{17.8cm}{0.4pt}}
\end{figure*}


A few of the terms that appear in the summation on the RHS of (\ref{equation:scde_update_1_ldgm}) will be $\cnunit$ and these represent the boundary condition that gets decoding started.
When the spatially-coupled LDGM system is initialized with $\msbx=\underline{\vnunit}$, the information at the boundary propagates inward and this induces a nondecreasing degradation ordering on positions $1,\ldots,\lceil \chend/2 \rceil$ and a nonincreasing degradation ordering on positions $\lceil \chend/2 \rceil +1,\ldots,\chend$.
This ordering results in a degraded maximum at position $i_0=N+\lceil \frac{w-1}{2} \rceil$.

As seen in Section \ref{subsection:single_system_ldgm}, the minimal fixed point $\minf$ plays a crucial role in the performance of the LDGM ensembles under iterative decoding.
Spatially-coupled LDGM ensembles are no exception.
The minimal fixed point $\minf$ of the single system is also crucial for the spatially-coupled system.
Changing the boundary in (\ref{equation:scde_update_1_ldgm}) from $\cnunit$ to $\minf$ therefore facilitates the proof of threshold saturation for these ensembles. 

\begin{definition}
  The \emph{modified system} is defined by the following update,
  \begin{align*}
    \msx_{i}^{(\ell+1)} = \frac{1}{w} \sum_{k=0}^{w-1} \lambda^{\vnop} \left(\frac{1}{w} \sum_{j=0}^{w-1} \msc \cnop \rho^{\cnop}(\msx^{(\ell)}_{i-k+j}) ; \delta_{i-k} \right),
  \end{align*}
  for $i \in \{1,\ldots,i_0\}$, and $\msx^{(\ell+1)}_i = \msx^{(\ell+1)}_{i_0}$ for $i_0 < i \leq \chend$, $\msx_{i}=\minf$ when $i \not \in \mc{N}_c$.
  The shorthand $\lambda^{\vnop}(\msx;\delta_{i})$ represents
  \begin{align*}
    \lambda^{\vnop}(\msx;\delta_{i}) =
    \begin{cases}
      \lambda^{\vnop}(\msx) & \text{if $i \in \mc{N}_v$,} \\
      \minf & \text{otherwise.}
    \end{cases}
  \end{align*}
\end{definition}
In comparison to (\ref{equation:scde_update_1_ldgm}), the modified system here differs both in the boundary condition and the saturation constraint $\msx_i=\msx_{i_0}$ for $i_0 < i \leq \chend$.
When the modified system and spatially-coupled system have the same initialization, as DE progresses, the distributions of the modified system will be degraded with respect to that of spatially-coupled system in (\ref{equation:scde_update_1_ldgm}).
Again, the modified system serves as an upper bound to the spatially-coupled system.
The DE updates for both spatially-coupled and modified system satisfy the monotonicity properties listed in Lemma \ref{lemma:scde_monotonicity_ldpc}.
For brevity, we do not state them explicitly.

If the modified system is initialized with $\msbx^{(0)}=\underline{\vnunit}$, then $\msbx^{(\ell+1)} \upgreq \msbx^{(\ell)}$ and $\msbx^{(\ell)} \degreq \minfb$ for all $\ell$.
To see this, suppose $\msbx^{(\ell)} \degreq \minfb$ for some $\ell$ (e.g., this is automatically true when $\ell=0$).
Observing the modified system DE update for $1 \leq i \leq i_0$,
  \begin{align*}
    \msx_{i}^{(\ell+1)} &= \frac{1}{w} \sum_{k=0}^{w-1} \lambda^{\vnop} \Big{(}\frac{1}{w} \sum_{j=0}^{w-1} \msc \cnop \rho^{\cnop}(\msx^{(\ell)}_{i-k+j}) ; \delta_{i-k} \Big{)} \\
    & \overset{(a)}{\degreq} \frac{1}{w} \sum_{k=0}^{w-1} \lambda^{\vnop} \Big{(}\frac{1}{w} \sum_{j=0}^{w-1} \msc \cnop \rho^{\cnop}(\minf ) ; \delta_{i-k} \Big{)} \\
    & = \frac{1}{w} \sum_{k=0}^{w-1} \lambda^{\vnop} \Big{(} \msc \cnop \rho^{\cnop}(\minf) ; \delta_{i-k} \Big{)} \\
    & \overset{(b)}{=} \lambda^{\vnop} \left( \msc \cnop \rho^{\cnop}(\minf)\right) \overset{(c)}{=} \minf ,
\end{align*}
where $(a)$ follows since $\msbx^{(\ell)} \degreq \minfb$, while $(b)$ and $(c)$ follow since $\minf$ is a fixed point of the single system DE.
Thus, the sequence of measure vectors $\{\msbx^{(\ell)}\}$ satisfies $\msbx^{(\ell)} \degreq \msbx^{(\ell+1)}$, $\msbx^{(\ell)} \degreq \minfb$, and consequently $\{ \msbx^{(\ell)}\}$ converges to a fixed point $\msbx$ with $\msbx \degreq \minfb$.
We also have the following result analogous to Lemma \ref{lemma:fp_monotonicity_modifiedsystem_ldpc}.
\begin{lemma}
  \label{lemma:fp_monotonicity_modifiedsystem_ldgm}
  The fixed point $\msbx$ of the modified system resulting from $\underline{\vnunit}$ initialization satisfies
  \begin{align*}
    \msx_{i} \degreq \msx_{i-1} \degreq \minf, \quad 2 \leq i \leq \chend .
  \end{align*}
\end{lemma}

Below, we define the coupled potential for LDGM ensembles.
Unlike LDPC codes, the coupled potential here and the properties that follow pertain exclusively to the modified system due to the difference in boundary conditions.
The key difference in our proof strategy for LDGM codes is to tweak the coupled potential to reflect the modified boundary and show that this modified potential still has the desired properties.

\begin{definition}
  The coupled potential functional $\potc\colon \mc{X}^{\chend} \times \mc{X} \rightarrow \mathbb{R}$ for a modified system is defined in (\ref{equation:coupled_potential_ldgm}).
\end{definition}
The last two terms of (\ref{equation:coupled_potential_ldgm}) are not present in (\ref{equation:coupled_potential_ldpc}).
These additional terms are necessary to reflect the modified boundary.
Proofs of Lemmas \ref{lemma:first_derivative_coupled_system_ldgm}, \ref{lemma:second_derivative_coupled_system_ldgm} are nearly identical to their analogues, Lemmas \ref{lemma:first_derivative_coupled_system_ldpc}, \ref{lemma:second_derivative_coupled_system_ldpc}, respectively.

\begin{lemma}
  \label{lemma:first_derivative_coupled_system_ldgm}
  The directional derivative of the potential functional in (\ref{equation:coupled_potential_ldgm}) with respect to $\msbx \in \probs^{\chend}$, evaluated in the direction $\msby \in \dpros^{\chend}$ is given in (\ref{equation:first_derivative_coupled_system_ldgm}).
\end{lemma}

\begin{lemma}
  \label{lemma:second_derivative_coupled_system_ldgm}
  The second-order directional derivative of the potential functional in (\ref{equation:coupled_potential_ldgm}) with respect to $\msbx$, evaluated in the direction $[\msby,\msbz] \in \dpros^{\chend} \times \dpros^{\chend}$ is given by (\ref{equation:second_derivative_coupled_system_ldgm}), where $\lambda'^{\vnop}(\msx;\delta_i)$ denotes 
\begin{align*}
  \lambda'^{\vnop}(\msx;\delta_{i}) =
  \begin{cases}
    \lambda'^{\vnop}(\msx) & \text{if $i \in \mc{N}_v$,} \\
    0 & \text{otherwise.}
  \end{cases}
\end{align*}
\end{lemma}

\section{Threshold Saturation for LDGM Ensembles}

\label{section:maintheorem_ldgm}

The proof strategy for threshold saturation of spatially-coupled LDGM ensembles is similar to that of spatially-coupled LDPC ensembles.
It is clear that $\minf$ plays a role similar to that of $\cnunit$ for LDPC ensembles.
The shift operator in Definition \ref{definition:shift_operator_ldgm} is adjusted accordingly.
Explicit characterization of the change in coupled potential due to shift is stated in Lemma \ref{lemma:potential_shift_bound_ldgm}.
The proof for this lemma is considerably different from that of its counterpart in LDPC section, and it is detailed in Appendix \ref{appendix:proof_potential_shift_bound_ldgm}.

Lemmas \ref{lemma:modifiedsystem_potential_fixedpoint_ldgm} and \ref{lemma:second_derivative_bound_ldgm} characterize the first- and second-order variations in the coupled potential at a non-trivial fixed point.
Theorem \ref{theorem:threshold_saturation_ldgm} states the threshold saturation result.
Proofs of Lemma \ref{lemma:modifiedsystem_potential_fixedpoint_ldgm}, Lemma \ref{lemma:second_derivative_bound_ldgm} and Theorem \ref{theorem:threshold_saturation_ldgm} are nearly identical to that of their counterparts in LDPC section, requiring only straightforward changes from $\cnunit$ to $\minf$.
We skip these proofs for brevity.

\begin{definition}
  \label{definition:shift_operator_ldgm}
  The shift operator $\shft : \probs^{\chend}  \rightarrow \probs^{\chend}$ is defined pointwise by
  \begin{align*}
    [\shft(\msbx)]_{1} &\triangleq \minf, & [\shft(\msbx) ]_i &\triangleq \msx_{i-1}, \quad \text{$2 \leq i \leq \chend$}.
  \end{align*}
\end{definition}

\begin{lemma}
  \label{lemma:potential_shift_bound_ldgm}
  Let $\msbx \in \probs^{\chend}$ be such that $\msbx \degreq \minfb \triangleq [\minf,\cdots,\minf]$ and $\msx_{i} = \msx_{i_0}$, for $i_0 \leq i \leq \chend$.
  Also suppose $i_0 \leq 2N$.
  Then the change in the potential functional for a modified system associated with the shift operator is bounded by

  \begin{align}
    \potc(\shft(\msbx) ; \msc) - \potc(\msbx ; \msc) \leq \pots(\minf; \msc) - \pots(\msx_{i_0} ;\msc)
  \end{align}
\end{lemma}
\begin{IEEEproof}
  See Appendix \ref{appendix:proof_potential_shift_bound_ldgm}.
\end{IEEEproof}

\begin{lemma}
  \label{lemma:modifiedsystem_potential_fixedpoint_ldgm}
  If $\msbx \degr \underline{\minf} $ is a fixed point of the modified system resulting from $\underline{\vnunit}$ initialization, then
  \begin{align*}
    \deri{\msbx} \potc(\msbx ; \msc)[\shft(\msbx) - \msbx] = 0,
  \end{align*}
  and moreover, $\msx_{i_0}$ is not in the basin of attraction to $\minf$ (i.e., $\msx_{i_0} \notin \mc{V}(\msc)$).
\end{lemma}

Lemma \ref{lemma:potential_shift_bound_ldgm}, Lemma \ref{lemma:modifiedsystem_potential_fixedpoint_ldgm}, and Definition \ref{definition:basinofattraction_energygap_ldgm}(ii) therefore imply that, for a non-trivial fixed point $\msbx$ resulting from initializing the  modified system with $\underline{\vnunit}$,
\begin{align*}
    \potc(\shft(\msbx) ; \msc) - \potc(\msbx ; \msc) \leq \pots(\minf;\msc) - \pots(\msx_{i_0} ; \msc) \leq - \Delta E(\msc).
\end{align*}
We note that while the shift bound in Lemma \ref{lemma:potential_shift_bound_ldgm} requires $i_0 \leq 2N$, which is satisfied by choosing $N > \lceil \tfrac{w-1}{2} \rceil$, this restriction has no bearing on Theorem \ref{theorem:threshold_saturation_ldgm}.
This is because for a fixed $w$, distributions of spatially-coupled systems with larger $N$ are degraded with respect to that of systems with smaller $N$.

\begin{lemma}
  \label{lemma:second_derivative_bound_ldgm}
  Suppose $\msbx$ is a fixed point of the modified system resulting from $\underline{\vnunit}$ initialization.
  Then
  \begin{align*}
    \abs{ \dderi{\msbx_{1}} \potc(\msbx_{1} ; \msc)[\shft(\msbx) - \msbx,\shft(\msbx) - \msbx] } \leq \frac{K_{\lambda,\rho}}{w},
  \end{align*}
  where the constant
  \begin{align*}
    K_{\lambda,\rho} \triangleq  L'(1) \left( 2\rho''(1) + \rho'(1) + 2\lambda'(1) \rho'(1)^{2} \right)
  \end{align*}
  is independent of $N$ and $w$.
\end{lemma}

\begin{theorem}
  \label{theorem:threshold_saturation_ldgm}
  Fix the LDGM$(\lambda,\rho)$ ensemble and a BMS channel $\msc$ with $\Delta E(\msc)>0$.
  For the $(\lambda,\rho,N,w)$ spatially-coupled LDGM ensemble with $w  > K_{\lambda,\rho}  / (2 \Delta E(\msc))$, any fixed point $\msbx$ of density evolution satisfies 
  \begin{align*}
    \msx_i \upgreq \minf(\msc), \quad 1 \leq i \leq \chend .
  \end{align*}
\end{theorem}

\section{Conclusions}
In this paper, a proof of threshold saturation, based on potential functions, is provided for spatially-coupled codes over BMS channels.
In particular, we show that for spatially-coupled irregular LDPC codes over a BMS channel, the belief-propagation decoding threshold saturates to the conjectured MAP threshold.
For LDGM codes, although the notion of thresholds is not systematically defined, a similar result holds. 
A converse to the threshold saturation result is also provided for LDPC codes.
This result reiterates the generality of the threshold saturation phenomenon, which is now evident from many observations and proofs that span a wide variety of systems.

The approach taken in this paper can be seen as analyzing the average Bethe free entropy in the large-system limit.
We also believe that this approach can be extended to more general graphical models by computing their average Bethe free entropy.

\section*{Acknowledgments}
The authors thank R{\"u}diger Urbanke, Arvind Yedla, and Yung-Yih Jian for a number of very useful discussions during the early stages of this research.

\appendices

\section{A Metric Topology on $\probs$}
\label{appendix:topology}
This section establishes a metric topology on $\probs$ that is homeomorphic to the weak topology on the set of probability measures on $[0,1]$.
The given metric is closely related to the entropy functional. 
The reader is assumed to be familiar with the notation in Section \ref{section:preliminaries}.

For $\msx \in \probs$, recall from Proposition \ref{proposition:entropy_symmetric_measures},
\begin{align*}
  \ent{\msx} = 1 - \sum_{k=1}^{\infty} \gamma_k M_{k}(\msx), \quad \text{where $\gamma_k=\frac{(\log 2)^{-1}}{2k(2k-1)}$} .
\end{align*}
The entropy distance $\disth \colon \probs \times \probs \to \mathbb{R}$ is defined as 
\begin{align*}
  \disth(\msx_1,\msx_2) \triangleq \sum_{k=1}^{\infty} \gamma_k \abs{M_k(\msx_1)-M_k(\msx_2)} .
\end{align*}

Endow the space of extended real numbers $\extR=[-\infty,\infty]$ with the metric given by
\begin{align*}
  d_{\extR}(\alpha_1,\alpha_2)=\abs{ \tanh(\alpha_1) - \tanh(\alpha_2)} .
\end{align*}
Under this metric, $\extR$ is \emph{compact}.
We begin by establishing a bijection between the set of symmetric probability measures on $\extR$, $\probs$, and the set of probability measures on $[0,1]$, denoted by $\mbb{P}([0,1])$.
This bijection is useful when characterizing the properties of the entropy distance $\disth$.
\begin{remark*}
The role of the entropy distance $\disth$ is similar to that of the Wasserstein metric in \cite[Section II-H]{Kudekar-it13}.
In fact, one could easily define a weighted Wasserstein metric where, like $\disth$, the distance between $\msx_1$ and $\msx_2$ is equal to $\ent{\msx_1-\msx_2}$ if $\msx_1 \degreq \msx_2$.
The relationship between such a weighted Wasserstein metric and $\disth$ warrants further attention.
\end{remark*}

The function defined by $\psi\colon [-\infty,\infty] \rightarrow [0,1]$, $\psi(\alpha)=\tanh^2(\tfrac{\alpha}{2})$ is continuous.
Consider the pushforward measure from $\probs$ to $\mathbb{P}([0,1])$ induced by $\psi$,
\begin{align*}
  \Psi\colon \probs &\rightarrow \mbb{P}([0,1]) \\
  \msx  &\mapsto \hat{\msx},
\end{align*}
where $\hat{\msx}(A)=\msx(\psi^{-1}(A))$ for all Borel sets $A \in \mc{B}([0,1])$.
Below, for any $\msx \in \probs$, we denote $\hat{\msx}$ for $\Psi(\msx)$.
For any measurable $f\colon[0,1] \rightarrow \mbb{R}$,
\begin{align*}
  \int f \diff{\hat{\msx}} = \int (f \circ \psi) \diff{\msx} .
\end{align*}
This immediately implies that
\begin{align*}
  \int \alpha^k \hat{\msx}(\diff{\alpha}) = \int  \tanh^{2k}\left(\tfrac{\alpha}{2}\right) \msx(\diff{\alpha}) .
\end{align*}
Thus, $k$-th moments of $\hat{\msx}$ are given by $M_k(\msx)$.

\begin{lemma*}
  The function $\Psi\colon\probs \rightarrow \mbb{P}([0,1])$ defined above is a bijection.
\end{lemma*}
\begin{IEEEproof}
  For injectivity of $\Psi$, consider $\msx_1,\msx_2 \in \probs$ such that $\hat{\msx}_1=\hat{\msx}_2$.
  Clearly, $\msx_1(\{0\})=\msx_2(\{0\})$.
  Suppose $E$ is a Borel set in $\mc{B}((0,\infty])$ and $A_{E}=\psi(E)$.
  We have $\msx_1(\psi^{-1}(A_{E}))=\msx_2(\psi^{-1}(A_{E}))$, which implies 
  \begin{align*}
    \int_{-E} \msx_1(\diff{\alpha})  + \int_{E} \msx_1(\diff{\alpha}) &= \int_{-E} \msx_2(\diff{\alpha}) + \int_{E} \msx_2(\diff{\alpha}) , \\
    \int_{E} (1+e^{-\alpha}) \msx_1(\diff{\alpha}) &= \int_{E} (1+e^{-\alpha}) \msx_2(\diff{\alpha}) ,
  \end{align*}
  due to symmetry.
  Since $1+e^{-\alpha}$ is non-zero, $\msx_1(E)=\msx_2(E)$ for all $E \in \mc{B}((0,\infty])$.
  Again by symmetry,
  \begin{align*}
    \msx_1(-E) = \int_{E} e^{-\alpha} \msx_1(\diff{\alpha}) = \int_{E} e^{-\alpha} \msx_2(\diff{\alpha}) = \msx_2(-E) .
  \end{align*}
  This implies that $\msx_1(E)=\msx_2(E)$ for all $E \in \mc{B}(\extR)$, and consequently, $\msx_1=\msx_2$.
  Hence, $\Psi$ is injective.

  For surjectivity, suppose $\mu \in \mbb{P}([0,1])$.
  Define measures $\msx_1,\msx_2$ on $[0,\infty]$ such that for $E \in \mc{B}([0,\infty])$, 
  \begin{align*}
    \msx_1(E)&=\mu(\psi(E)), & \msx_2(E) = \int_{E} \frac{1}{1+e^{-\alpha}} \msx_1(\diff{\alpha}) .
  \end{align*}
  Extend $\msx_2$ to $[-\infty,\infty]$ by defining $\msx$ as 
  \begin{align*}
    \msx(E)&=\msx_2(E), \quad \text{for $E \in \mc{B}((0,\infty])$}, \\
    \msx(\{0\})&= 2\msx_2(\{0\}), \\
    \msx(E)&=\int_{-E} e^{-\alpha} \msx_2(\diff{\alpha}), \quad \text{for $E \in \mc{B}([-\infty,0))$} .
  \end{align*}
  Then, $\msx$ is a symmetric probability measure on $[-\infty,\infty]$, and $\hat{\msx}=\mu$.
  Hence $\Psi$ is surjective.
\end{IEEEproof}

\begin{proposition*}
  The set of symmetric probability measures with the entropy distance $(\probs,\disth)$ is a metric space.
\end{proposition*}
\begin{IEEEproof}
  It is easy to see that $\disth(\cdot,\cdot)$ is non-negative, symmetric, and satisfies the triangle inequality.
  For $\disth$ to be a metric, it suffices to show that $\disth(\msx_1,\msx_2)=0$ implies $\msx_1=\msx_2$. 
  Let $\disth(\msx_1,\msx_2)=0$.
  Note that $\disth(\msx_1,\msx_2)=0$ iff $M_{k}(\msx_1)=M_{k}(\msx_2)$ for all $k \in \mbb{N}$.
  Thus
  \begin{align*}
    \int \alpha^k \hat{\msx}_1(\diff{\alpha}) = \int \alpha^k \hat{\msx}_2(\diff{\alpha}) , \quad \text{for all $k \in \mbb{N}$}.
  \end{align*}
  By the Hausdorff moment problem \cite[Theorem VII.3.1]{Feller-1971-2}, $\hat{\msx}_1=\hat{\msx}_2$.
  By injectivity of $\Psi$, $\msx_1=\msx_2$.
  Thus $\disth$ is a metric on $\probs$.
\end{IEEEproof}

\begin{proposition*}
  The metric topology $(\probs,\disth)$ is homeomorphic to the weak topology on $\mathbb{P}([0,1])$.
\end{proposition*}
\begin{IEEEproof}
  It suffices to show that $\Psi$ and $\Psi^{-1}$ are continuous.
  Suppose $\mu_n \rightarrow \mu$ weakly in $\mbb{P}([0,1])$.
  Since $x^k\colon[0,1] \rightarrow [0,1]$ is a bounded continuous function for $k \in \mbb{N}$,
  \begin{align*}
    \int \alpha^k \mu_n(\diff{\alpha}) \rightarrow \int \alpha^k \mu(\diff{\alpha}) .
  \end{align*}
  But this implies $M_k(\Psi^{-1}(\mu_n)) \rightarrow M_k(\Psi^{-1}(\mu))$.
  Hence $\Psi^{-1}$ is continuous.
  
  For the continuity of $\Psi$, let $\msx_n \xrightarrow{\disth} \msx$ in $\probs$.
  That is
  \begin{align*}
    \int \alpha^k \hat{\msx}_n(\diff{\alpha}) \rightarrow \int \alpha^k \hat{\msx}(\diff{\alpha}) ,
  \end{align*}
  and consequently,
  \begin{align*}
    \int p(\alpha) \hat{\msx}_n(\diff{\alpha}) \rightarrow \int p(\alpha) \hat{\msx}(\diff{\alpha}) ,
  \end{align*}
  for any polynomial $p\colon[0,1] \rightarrow \mbb{R}$.
  By an application of the Stone-Weirstrass theorem \cite[Theorem 4.45]{Folland-1999}, polynomials are dense in the set of continuous functions on $[0,1]$ under the supremum norm, $C[0,1]$.
  This implies 
  \begin{align*}
    \int f(\alpha) \hat{\msx}_n(\diff{\alpha}) \rightarrow \int f(\alpha) \hat{\msx}(\diff{\alpha}) ,
  \end{align*}
  for any $f \in C([0,1])$. 
  Thus $\hat{\msx}_n \rightarrow \hat{\msx}$ weakly, and this establishes the continuity of $\Psi$.
\end{IEEEproof}

\begin{corollary*}
  The metric topology $(\probs,\disth)$ is compact and separable.
  Since compact metric spaces are complete, it is also a Polish space.
\end{corollary*}

\begin{proposition*}
  The functionals $\mathrm{H}\colon\probs \rightarrow \mbb{R}$ and $M_k\colon\probs \rightarrow \mbb{R}$ are continuous. 
\end{proposition*}
\begin{IEEEproof}
  The continuity of $\mathrm{H}$ follows since 
  \begin{align*}
    \abs{\ent{\msx_1}-\ent{\msx_2}} \leq \disth(\msx_1,\msx_2) , 
  \end{align*}
  while the continuity of $M_k(\cdot)$ follows from
  \begin{align*}
    \abs{M_k(\msx_1)-M_k(\msx_2)} \leq \frac{1}{\gamma_k}\disth(\msx_1,\msx_2) .
  \end{align*}
\end{IEEEproof}

\begin{proposition*}
  If we endow $\probs \times \probs$ with the product topology, then the operators $\cnop\colon\probs \times \probs \rightarrow \probs$ and $\vnop\colon\probs \times \probs \rightarrow \probs$ are continuous.
\end{proposition*}
\begin{IEEEproof}
  Suppose $\msx_{n,1} \xrightarrow{\disth} \msx_1$ and $\msx_{n,2} \xrightarrow{\disth} \msx_{2}$.
  Below, we will show that $\msx_{n,1} \cnop \msx_{n,2} \xrightarrow{\disth} \msx_1 \cnop \msx_2$ and $\msx_{n,1} \vnop \msx_{n,2} \xrightarrow{\disth} \msx_1 \vnop \msx_2$.
  First, consider the operator $\cnop$.
  \allowdisplaybreaks{
    \begin{align*}
      & \disth(\msx_{n,1}\cnop\msx_{n,2},\msx_1\cnop\msx_2)  \\
      & \qquad = \sum_{k=1}^\infty \gamma_k \abs{M_k(\msx_{n,1})M_k(\msx_{n,2}) - M_k(\msx_1)M_k(\msx_2)} \\
      & \qquad \leq \sum_{k=1}^{\infty} \gamma_k \abs{M_k(\msx_{n,1})-M_k(\msx_1)} M_k(\msx_{n,2}) \\
      & \hspace{2cm} + \sum_{k=1}^{\infty} \gamma_k \abs{M_k(\msx_{n,2})-M_k(\msx_2)} M_k(\msx_{1}) \\
      & \qquad \leq \disth(\msx_{n,1},\msx_1) + \disth(\msx_{n,2},\msx_2) \rightarrow 0.
    \end{align*}
  }
  Thus $\cnop$ is continuous.
  For the operator $\vnop$, note that $\hat{\msx}_{n,1} \rightarrow \hat{\msx}_1$ weakly and $\hat{\msx}_{n,2} \rightarrow \hat{\msx}_2 $ weakly.
  Let $\mu_n=\Psi( \msx_{n,1} \vnop \msx_{n,2})$.
  We have
  \begin{align*}
    M_k(\msx_{n,1}\vnop\msx_{n,2})  &= \int \tanh^{2k}\left(\tfrac{\alpha}{2}\right) (\msx_{n,1}\vnop\msx_{n,2})(\diff{\alpha})  \\
    &= \int \alpha^k \mu_n(\diff{\alpha}) \\
    &= \iint f_{\vnop,k}(\alpha_1,\alpha_2) \hat{\msx}_{n,1}(\diff{\alpha_1}) \hat{\msx}_{n,2}(\diff{\alpha_2}) ,
  \end{align*}
  where the kernel $f_{\vnop,k}\colon[0,1] \times [0,1] \rightarrow \mbb{R}$ is the continuous function given by
  \begin{align*}
    & f_{\vnop,k}(\alpha_1,\alpha_2)\\
    & \quad = \tfrac{1+\sqrt{\alpha_1\alpha_2}}{2} \tanh^{2k}\left(\tanh^{-1}(\sqrt{\alpha_1})+\tanh^{-1}(\sqrt{\alpha_2})\right) \\
    & \qquad + \tfrac{1-\sqrt{\alpha_1\alpha_2}}{2} \tanh^{2k}\left(\tanh^{-1}(\sqrt{\alpha_1})-\tanh^{-1}(\sqrt{\alpha_2})\right) .
  \end{align*}
  Since $f_{\vnop,k}$ is continuous and $\{ \hat{\msx}_{n,1}\}$, $\{ \hat{\msx}_{n,2} \}$ converge weakly,
  \begin{align*}
    & \iint f_{\vnop,k}(\alpha_1,\alpha_2) \hat{\msx}_{n,1}(\diff{\alpha_1}) \hat{\msx}_{n,2}(\diff{\alpha_2}) \\
    & \qquad \rightarrow \iint f_{\vnop,k}(\alpha_1,\alpha_2) \hat{\msx}_{1}(\diff{\alpha_1}) \hat{\msx}_{2}(\diff{\alpha_2}) = \int \alpha^k \mu(\diff{\alpha}) \\
    & \qquad = \int \tanh^{2k}\left(\tfrac{\alpha}{2}\right) (\msx_1 \vnop \msx_2)(\diff{\alpha}) = M_k(\msx_1 \vnop \msx_2) ,
  \end{align*}
  where $\mu = \Psi(\msx_1 \vnop \msx_2)$.
  Thus $M_k(\msx_{n,1} \vnop \msx_{n,2}) \rightarrow M_k(\msx_{1} \vnop \msx_{2})$, and consequently,
  \begin{align*}
    \msx_{n,1} \vnop \msx_{n,2} \xrightarrow{\disth} \msx_1 \vnop \msx_2 .
  \end{align*}
  This establishes the continuity of $\vnop$.
\end{IEEEproof}

\begin{proposition*}
  If a sequence of measures $\{\msx_n\}_{n=1}^\infty$ satisfies $\msx_{n+1} \degreq \msx_{n}$ (respectively, $\msx_{n+1} \upgreq \msx_{n}$), then $\msx_n \xrightarrow{\disth} \msx$, for some $\msx \in \probs$ which satisfies $\msx \degreq \msx_n$ (respectively, $\msx \upgreq \msx_{n}$) for all $n$.
\end{proposition*}
\begin{IEEEproof}
  We suppose $\msx_{n+1} \degreq \msx_{n}$ for $n \in \mbb{N}$; the case where $\msx_{n+1} \upgreq \msx_{n}$ follows similarly.
  Since the entropy functional preserves the order by degradation, $\ent{\msx_{n+1}} \geq \ent{\msx_n}$.
  Since $0 \leq \ent{\msx} \leq 1$ for $\msx \in \probs$, $\{\ent{\msx_n}\}$ is a Cauchy sequence.
  For any $m>n$, since $\msx_m \degreq \msx_n$, 
  \begin{align*}
    \disth(\msx_m,\msx_n)=\ent{\msx_m}-\ent{\msx_n} \to 0 \quad \text{as $m,n \to \infty$} .
  \end{align*}
  Thus, the sequence $\{\msx_n\}$ is Cauchy and as $(\probs,\disth)$ is complete, $\msx_n \xrightarrow{\disth} \msx$ for some $\msx \in \probs$.

  To show $\msx \degreq \msx_n$, in view of Definition \ref{definition:degradation}, let $f$ be a concave non-increasing function on $[0,1]$.
  Then, necessarily, $f$ is continuous on $[0,1)$.
  First suppose $f$ is continuous on $[0,1]$.
  We discuss the case where
  \begin{align*}
    f(1) < \lim_{\alpha \to 1} f(\alpha)
  \end{align*}
  separately.
  Since $\msx_{n+1} \degreq \msx_{n}$, for any $m>n$, $\msx_{m} \degreq \msx_{n}$.
  This implies
  \begin{align*}
    \int f \left( \abs{\tanh \left(\tfrac{\alpha}{2}\right)} \right) \msx_{m}(\diff{\alpha}) &\geq  \int f \left( \abs{\tanh \left(\tfrac{\alpha}{2}\right) }\right) \msx_{n}(\diff{\alpha}) , \\
    \int (f \circ \sqrt{\cdot}) \diff{\hat{\msx}_{m}} &\geq  \int (f \circ \sqrt{\cdot}) \diff{\hat{\msx}_{n}} , \\
    \lim_{m \to \infty} \int (f \circ \sqrt{\cdot}) \diff{\hat{\msx}_{m}} &\geq  \int (f \circ \sqrt{\cdot}) \diff{\hat{\msx}_{n}} ,
  \end{align*}
  and, since $\hat{\msx}_{m} \to \hat{\msx}$ weakly and $f \circ \sqrt{\cdot}$ is continuous on $[0,1]$,
  \begin{align*}
    \lim_{m \to \infty} \int (f \circ \sqrt{\cdot}) \diff{\hat{\msx}_{m}} = \int (f \circ \sqrt{\cdot}) \diff{\hat{\msx}} .
  \end{align*}
  Thus,
  \begin{align*}
    \int (f \circ \sqrt{\cdot}) \diff{\hat{\msx}} &\geq \int (f \circ \sqrt{\cdot}) \diff{\hat{\msx}_n} , \\
    \int f \left( \abs{\tanh \left(\tfrac{\alpha}{2} \right)} \right) \msx(\diff{\alpha}) &\geq  \int f \left(\abs{ \tanh \left(\tfrac{\alpha}{2}\right) } \right) \msx_{n}(\diff{\alpha}) .
  \end{align*}
  
  Now suppose $f$ is a concave, non-increasing function on $[0,1]$, but discontinuous at $1$.
  Since $f$ is bounded, to show 
  \begin{align*}
    \int (f \circ \sqrt{\cdot}) \diff{\hat{\msx}} \geq \int (f \circ \sqrt{\cdot}) \diff{\hat{\msx}_n} ,
  \end{align*}
  we can assume $f$ is non-negative by adding a suitable constant.
  Also, there exists a sequence of functions $\{ f_m \}_{m=1}^{\infty}$ that are non-negative, non-increasing, continuous, concave and
  \begin{align*}
    f_m & \leq f_{m+1}, & f_m \to f \quad \text{pointwise} .
  \end{align*}
  By the monotone convergence theorem \cite[Theorem 2.14]{Folland-1999},
  \begin{align*}
    \int (f \circ \sqrt{\cdot}) \diff{\hat{\msx}} &= \lim_{m \to \infty} \int (f_m \circ \sqrt{\cdot}) \diff{\hat{\msx}}, \\
    \int (f \circ \sqrt{\cdot}) \diff{\hat{\msx}_n} &= \lim_{m \to \infty} \int (f_m \circ \sqrt{\cdot}) \diff{\hat{\msx}_n} .  
  \end{align*}
  Since $f_m$ is continuous, from the arguments above, 
  \begin{align*}
    \int (f_m \circ \sqrt{\cdot}) \diff{\hat{\msx}} \geq \int (f_m \circ \sqrt{\cdot}) \diff{\hat{\msx}_n} .
  \end{align*}
  Consequently, 
  \begin{align*}
    \int (f \circ \sqrt{\cdot}) \diff{\hat{\msx}} \geq \int (f \circ \sqrt{\cdot}) \diff{\hat{\msx}_n} .
  \end{align*}
  Hence $\msx \degreq \msx_n$ for any $n$.
\end{IEEEproof}

We state the following result without proof as it is similar to the previous proposition.
\begin{proposition*}
  If  $\{\msx'_n\}_{n=1}^\infty$, $\{\msx_n\}_{n=1}^\infty$ satisfy $\msx'_{n} \degreq \msx_{n}$ and $\msx'_n \xrightarrow{\disth} \msx'$, $\msx_n \xrightarrow{\disth} \msx$, then $\msx' \degreq \msx$.
\end{proposition*}

\section{Proofs from Section \ref{section:preliminaries}}
\label{appendix:section_preliminaries}

\subsection{Proof of Proposition \ref{proposition:tanh_symmetry}}
\label{appendix:proof_tanh_symmetry}

By symmetry and since $f(0)=0$ for an odd function,
\begin{align*}
  \int f(\alpha) \msx(\diff{\alpha})
  &= f(0) \msx(\{0\}) \!+\! \int\limits_{(0,\infty]}\! \left[ f(\alpha) \!+\! f(-\alpha) e^{- \alpha} \right] \msx(\diff{\alpha}) \\
  &= \int_{(0,\infty]} f(\alpha) (1 - e^{-\alpha}) \, \msx(\diff{\alpha}) \\
  &= \int_{(0,\infty]} f(\alpha) \tanh \left( \tfrac{\alpha}{2} \right) (1 + e^{- \alpha}) \, \msx(\diff{\alpha}) \\
  &= \int f(\alpha) \tanh \left( \tfrac{\alpha}{2} \right) \, \msx(\diff{\alpha}).
\end{align*}

\subsection{Proof of Proposition \ref{proposition:Mk_properties}}
\label{appendix:proof_Mk_properties}
\begin{enumerate}[i)]
\item Follows from $0 \le \tanh^{2k}(\alpha) \leq 1$.
\item Note that $f(\alpha)=-\alpha^{2k}$ is a concave decreasing function over $[0,1]$.
  Since $\msx_1 \degreq \msx_2$, Definition \ref{definition:degradation} implies that
  \begin{align*}
    -M_k(\msx_1) = I_f(\msx_1) \geq I_f(\msx_2) = -M_k(\msx_2) .
  \end{align*}
  Thus, $M_k(\msx_1) \leq M_k(\msx_2)$.
\item
  By the equivalent characterization of the operator $\cnop$,
  \begin{align*}
    &  M_k(\msx_1 \cnop \msx_2)  \\
    &\quad = \int \tanh^{2k}(\tfrac{\alpha}{2}) (\msx_1 \cnop \msx_2) (\diff{\alpha})  \\
    & \quad \overset{(a)}{=} \!\iint\! \tanh^{2k}\!\left(\!\tfrac{\tau^{-1}(\tau(\alpha_1)\tau(\alpha_2))}{2}\!\right)\! \msx_1(\diff{\alpha_1})\msx_2(\diff{\alpha_2})  \\
    & \quad = \iint \!\tanh^{2k}\!\left(\tfrac{\alpha_1}{2}\right)\tanh^{2k}\!\left(\tfrac{\alpha_2}{2}\right)\!\msx_1(\diff{\alpha_1})\msx_2(\diff{\alpha_2}) \\
    & \quad = M_k(\msx_1) M_k(\msx_2) ,
  \end{align*}
  where $\tau(\alpha)=\tanh(\tfrac{\alpha}{2})$ in the RHS of $(a)$.
\item
  If $\msx=\cnunit$ (respectively, $\msx=\vnunit$), then it is easy to see that $M_k(\msx)=1$ (respectively, $M_k(\msx)=0$) for all $k$.
  The other direction follows from
  \begin{align*}
    0 &< \tanh^{2k}(\alpha) \quad \text{if $\alpha \neq 0$} , \\
    1 &> \tanh^{2k}(\alpha) \quad \text{if $\alpha \neq \pm \infty$} ,
  \end{align*}
  and since the symmetry of the measure implies
  \begin{align*}
    \msx(\{ -\infty \}) = e^{-\infty} \msx(\{ \infty \}) = 0 .
  \end{align*}
\end{enumerate}

\subsection{Proof of Proposition \ref{proposition:entropy_dpros_properties}}
\label{appendix:proof_entropy_dpros_properties}

\begin{enumerate}[i)]
\item Using Proposition \ref{proposition:entropy_symmetric_measures} and $(\msy_1\cnop \msy_2)(\extR)=0$ when $\msy_1,\msy_2 \in \dpros$, we have the result.
  
\item With the observation
  \begin{align*}
    \ent{\msy_1 \vnop \msy_2} = - \ent{\msy_1 \cnop \msy_2} 
  \end{align*}
  from Proposition \ref{proposition:duality_difference}, the inequalities are trivial.
  It remains to show that $\msy=0$ when $\ent{\msy \cnop \msy}=0$.
  For this, let $\msy=\msx_1-\msx_2$ with $\msx_1,\msx_2 \in \probs$, and observe that
  \begin{align*}
    \ent{\msy \cnop \msy}=0 \Longleftrightarrow M_k(\msx_1)=M_k(\msx_2) \quad \text{for all $k$} .  
  \end{align*}
  The fact that $M_k(\msx_1)=M_k(\msx_2)$ for all $k$ iff $\msx_1=\msx_2$ follows as a consequence of the metric properties of the entropy functional; see Definition \ref{definition:entropy_distance} and Proposition \ref{proposition:entropy_distance_properties}.

\item Using the first part of this proposition and the inequalities $M_k(\msx'_1) \leq M_k(\msx_1)$ and $M_k(\msx'_2) \leq M_k(\msx_2)$, we have the result.
\item Assume $\msx_1 \degr \msx_2$ and consider $\msx_3 \neq \cnunit$.
  To show $\ent{\msx_1 \vnop \msx_3} > \ent{\msx_2 \vnop \msx_3}$, observe that
  \begin{align*}
    & \ent{\msx_1 \vnop \msx_3} - \ent{\msx_2 \vnop \msx_3} \\
    &  \qquad = \ent{[\msx_1 -\msx_2] \vnop [\msx_3 - \cnunit]} \\
    &  \qquad = - \ent{[\msx_1 -\msx_2] \cnop [\msx_3 - \cnunit]} \quad \text{(Proposition \ref{proposition:duality_difference})}\\
    &  \qquad = \sum_{k=0}^{\infty} \gamma_k [M_k(\msx_2) \!-\! M_k(\msx_1)][1\!-\!M_k(\msx_3)] \\
    &  \qquad > 0 .
  \end{align*}
  The last inequality follows since $M_k(\msx_3)<1$ for all $k \in \mbb{N}$ (from Proposition \ref{proposition:Mk_properties}(iv)) and $M_k(\msx_2) > M_k(\msx_1)$ for some $k \in \mbb{N}$ (see the proof of part ii of this proposition).

  Now, consider $\msx_3 \neq \vnunit$.
  Again, we observe that
  \begin{align*}
    & \ent{\msx_1 \cnop \msx_3} - \ent{\msx_2 \cnop \msx_3} \\
    &  \qquad = \ent{[\msx_1 -\msx_2] \cnop \msx_3} \\
    &  \qquad = \sum_{k=0}^{\infty} \gamma_k [M_k(\msx_2) \!-\! M_k(\msx_1)]M_k(\msx_3) \\
    &  \qquad > 0 ,
  \end{align*}
  where the last inequality follows since $M_k(\msx_3)>0$ for all $k$ and $M_k(\msx_2) > M_k(\msx_1)$ for some $k$.

\end{enumerate}

\subsection{Proof of Proposition \ref{proposition:entropy_vnop_decay_rate}}
\label{appendix:proof_entropy_vnop_decay_rate}

From \cite[Problems 4.60-61]{RU-2008},
\begin{align*}
  2 \mf{E}(\msx) \leq \ent{\msx} \leq \mf{B}(\msx) ,
\end{align*}
where $\mf{E}(\cdot)$ is the error functional
\begin{align*}
  \mf{E}(\msx) \triangleq \frac{1}{2}\int  e^{-(\alpha+\abs{\alpha})/2} \msx(\diff{\alpha}) .
\end{align*}
From \cite[Lemma 4.66]{RU-2008}, for $n \geq 2$,
\begin{align*}
  \frac{\alpha \mf{B}(\msx)^{3/2}}{\sqrt{n}} \mf{B}(\msx)^n \leq 2 \mf{E}(\msx^{\vnop n}) \leq \mf{B}(\msx)^n ,
\end{align*}
for a constant $\alpha>0$.
The above relations, together with $\mf{B}(\msx^{\vnop n})=\mf{B}(\msx)^n$, imply that
\begin{align*}
  \lim_{n \rightarrow \infty} \frac{1}{n} \log \ent{\msx^{\vnop n}} &= \log \mf{B}(\msx) .
\end{align*}

\section{Proofs From Section \ref{section:ldpc}}
\label{appendix:section_ldpc}

\subsection{Proof of Lemma \ref{lemma:fixed_stationary_ldpc}}
\label{appendix:proof_fixed_stationary_ldpc}

The first statement follows from Lemma \ref{lemma:firstderivative_singlesystem_ldpc}.

For the second part, suppose $\msx$ is not a fixed point of single system DE.
We discuss the cases $\msx \neq \vnunit$ and  $\msx=\vnunit$ separately.
First, consider $\msx \neq \vnunit$.
The derivative in Lemma \ref{lemma:firstderivative_singlesystem_ldpc} in the direction $\des(\msx;\msc) - \msx$ is
\begin{align*}
  \deri{\msx} \pots( \msx ; \msc )  [\des(\msx;\msc) \!-\! \msx] = L'(1) \mathrm{H}\Big{(}  \left( \des(\msx;\msc) \!-\! \msx \right)^{\cnop 2} \!\cnop\! \rho'^{\cnop}(\msx) \Big{)}.
\end{align*}
From Proposition \ref{proposition:entropy_dpros_properties}(ii), the above equation is strictly negative if $\msx \neq \des(\msx;\msc)$ and $\msx \neq \vnunit$.
Thus, if $\msx \neq \des(\msx;\msc)$ and $\msx \neq \vnunit$, 
\begin{align*}
  \deri{\msx} \pots(\msx;\msc) [\des(\msx;\msc) - \msx] < 0.
\end{align*}
By definition,
\begin{align*}
  \lim_{\delta \to 0}\frac{\pots(\msx+\delta [\des(\msx;\msc) - \msx] ; \msc) - \pots(\msx;\msc)}{\delta} < 0 .
\end{align*}
Thus, there exists a $t \in (0,1]$ such that 
\begin{align*}
  \pots \left( \msx + t \left[ \des(\msx;\msc) - \msx \right] ; \msc \right) < \pots ( \msx ; \msc ) .
\end{align*}
Therefore, $\pots(\msx;\msc)$ cannot be a minimum if $\msx$ is not a fixed point and $\msx \neq \vnunit$.

Now, we consider the case $\msx=\vnunit$.
Since $\msx$ is not a fixed point, $\des(\vnunit;\msc) \upgr \vnunit$.
For notational convenience, let 
\begin{align*}
  \msx_t= \des(\vnunit;\msc)+t[\vnunit-\des(\vnunit;\msc)] \quad \text{for $t \in [0,1]$} .  
\end{align*}
This implies for $t \in (0,1)$, $\msx_0 \upgr \msx_t \upgr \vnunit$, and by the monotonicity of the operator $\des$,
\begin{align*}
  \des(\msx_t ; \msc)  \upgreq \des(\vnunit;\msc) = \msx_0 \upgr \msx_t .
\end{align*}
Define $\phi\colon [0,1] \to \mbb{R}$, $\phi(t) = \pots(\msx_t ; \msc)$.
As in Proposition \ref{proposition:polynomial_structure_example}, for $t \in (0,1)$,
\begin{align*}
  \phi'(t) &= \deri{\msx_t} \pots(\msx_t ; \msc) [\vnunit - \msx_0] \\
  &=  - L'(1) \ent{ [\msx_t - \des(\msx_t;\msc)] \cnop [\vnunit-\msx_0] \cnop \rho'^{\cnop}(\msx_t) } \\
  &= L'(1) \ent{ [\msx_t - \des(\msx_t;\msc)] \cnop \msx_0 \cnop \rho'^{\cnop}(\msx_t) } \\
  &> 0,
\end{align*}
by Proposition \ref{proposition:degradation_preservation}(ii), since $\msx_t \degr \des(\msx_t;\msc)$, $\msx_0 \neq \vnunit$ and $\rho'^{\cnop}(\msx_t) \neq \vnunit$.
Thus, $\pots(\vnunit;\msc) = \phi(1) > \phi(0) = \pots(\msx_0;\msc)$.
As such, $\pots(\vnunit;\msc)$ cannot be a minimum of $\pots(\cdotp;\msc)$.

Hence, the minimum of $\pots(\cdotp;\msc)$ can only occur at a density evolution fixed point.

\subsection{Proof of Lemma \ref{lemma:monotonicity_ldpc}}
\label{appendix:proof_monotonicity_ldpc}
\begin{enumerate}[i)]
\item By Proposition \ref{proposition:entropy_dpros_properties}(iv),
  \begin{align*}
    \ent{\msc_1 \vnop \msx} > \ent{\msc_2 \vnop \msx} \quad \text{if $\msx \neq \cnunit$}.
  \end{align*}
  Thus, $\pots(\msx;\msc_1) < \pots(\msx;\msc_2)$ if $\msx \neq \cnunit$.
\item Using monotonicity of the DE operator,
  \begin{align*}
    \des^{(\ell)}(\msa;\msc_1)  \degreq \des^{(\ell)}(\msa;\msc_2) .
  \end{align*}
  Thus, if $\msa \in \mc{V}(\msc_1)$, then $\des^{(\infty)}(\msa;\msc_1)=\cnunit$.
  Then, it is easy to show that
  \begin{align*}
    \des^{(\ell)}(\msa;\msc_2) \xrightarrow{\disth} \cnunit .
  \end{align*}
  Thus $\des^{(\infty)}(\msa;\msc_2)=\cnunit$, and $\msa \in \mc{V}(\msc_2)$.
\item Follows from parts i and ii.
\end{enumerate}

\subsection{Proof of Lemma \ref{lemma:stability_threshold_properties}}
\label{appendix:proof_stability_threshold_properties}

\begin{enumerate}[i)]
\item
  If $\hstab=1$, then the result is trivial; therefore we assume $\hstab < 1$.
  Consider any $\h>\hstab$.
  From \cite[Section 4.9.2]{RU-2008}, $\mc{V}(\msc(\h))=\{\cnunit\}$, and by the continuity of $\pots(\cdotp;\msc(\h))$ at $\cnunit$, $\Delta E (\msc(\h)) \leq 0$.
  This implies $\h \geq \h^*$ by Definition \ref{definition:thresholds_ldpc}(iii).
  Thus $\h^* \leq \hstab$.
\item
  If $\h < \hstab$, there exists an $\epsilon>0$ such that for all $\msx$ with $\ent{\msx}<\epsilon$, $\msx \in \mc{V}(\msc(\h))$ \cite[Section 4.9.2]{RU-2008}.
  Thus, if $\disth(\msx,\cnunit)<\epsilon$, then $\disth(\cnunit,\msx)=\ent{\msx}<\epsilon$, and hence $\msx \in \mc{V}(\msc(\h))$.
  Thus, there is an $\epsilon$-ball around $\cnunit$ which is in $\mc{V}(\msc(\h))$.
\end{enumerate}

\subsection{Proof of Lemma \ref{lemma:beyond_potential_threshold_negativity}}
\label{appendix:proof_beyond_potential_threshold_negativity}
If $\h^*=1$, then the statement of the lemma is vacuous; suppose $\h^*<1$.
Let $\h > \h^*$.
By assumption, $\h^* < \hstab$, and thus there exists $\h' < \h$ such that $\h^{\ast} < \h' < \hstab$.
Since $\h' < \hstab$, by Lemma \ref{lemma:stability_threshold_properties},
\begin{align*}
  \Delta_{\infty} \in (\mc{V}(\msc(\h')))^{o}  \qquad \Longrightarrow \qquad \Delta_{\infty} \not\in \overline{\probs \backslash \mc{V}(\msc(\h'))}.
\end{align*}
Moreover, $\overline{\probs \backslash \mc{V}(\msc(\h'))}$ is compact and $\pots(\cdotp;\msc(\h'))$ is continuous.
Therefore, the infimum
\begin{align*}
  \inf_{\msx \in \overline{\probs \setminus \mc{V}(\msc(\h'))}} \pots(\msx ; \msc(\h'))
\end{align*}
is achieved at some $\msa \neq \cnunit$.
By Lemma \ref{lemma:monotonicity_ldpc}(i), $\pots(\msa;\msc(\h))$ is strictly decreasing in $\h$.
Therefore,
\begin{align*}
  \min_{\msx \in \probs} \pots(\msx ; \msc(\h)) &\leq \pots(\msa ; \msc(\h)) \\
  &< \pots(\msa;\msc(\h')) \quad \text{(Since $\h' < \h$)} \\
  &=\inf_{\msx \in \overline{\probs \setminus \mc{V}(\msc(\h'))}} \pots(\msx ; \msc(\h')) \\
  &\leq\inf_{ \msx \in \probs \setminus \mc{V}(\msc(\h')) } \pots(\msx ; \msc(\h')) \\
  &=\Delta E (\msc(\h')) \leq 0 \quad \text{(Since $\h' > \h^*$)}.
\end{align*}
Hence,
\begin{align*}
  \min_{\msx \in \probs} \pots(\msx ; \msc(\h)) < 0,
\end{align*}
and there exists an $\msx \in \probs$ such that $\pots(\msx;\msc(\h))<0$.

\subsection{Proof of Lemma \ref{lemma:fp_monotonicity_modifiedsystem_ldpc}}
\label{appendix:proof_fp_monotonicity_modifiedsystem_ldpc}

Since the modified system is initialized with $\msbx^{(0)}=\underline{\vnunit}$, $\msx_{i}^{(0)} \degreq \msx_{i-1}^{(0)}$.
Suppose at some iteration $\ell$, $\msx_{i}^{(\ell)} \degreq \msx_{i-1}^{(\ell)}$.
If $i>i_0$, then due to the saturation constraint in the modified system, $\msx_{i}^{(\ell+1)}=\msx_{i_0}^{(\ell+1)}$, $\msx_{i}^{(\ell+1)} \degreq \msx_{i-1}^{(\ell+1)}$.
For $1\leq i \leq i_0$, by observing (\ref{equation:scde_update_1_ldpc}),
\begin{align*}
  \msx^{(\ell+1)}_{i} - \msx^{(\ell+1)}_{i-1} &= \frac{1}{w} \msc_{i} \vnop \lambda^{\vnop} \Big{(} \frac{1}{w} \sum_{j=0}^{w-1} \rho^{\cnop}(\msx^{(\ell)}_{i+j}) \Big{)} \\
  & \qquad - \frac{1}{w} \msc_{i-w} \vnop \lambda^{\vnop} \Big{(} \frac{1}{w} \sum_{j=0}^{w-1} \rho^{\cnop}(\msx^{(\ell)}_{i-w+j}) \Big{)} .
\end{align*}
Note that $\msc_i=\msc$ if $i \in \mc{N}_v$ and $\msc_i=\cnunit$ otherwise.
At this point, we need to consider two cases: 1) $2N \geq i_0$ 2) $2N < i_0$.

When $2N \geq i_0$, for any $1 \leq i \leq i_0$, $i \in \mc{N}_v$, which implies $\msc_i = \msc$ and $\msc_i \degreq \msc_{i-w}$.
Since $\msx_{i}^{(\ell)} \degreq \msx_{i-1}^{(\ell)}$, we see that $\msx_{i}^{(\ell+1)} \degreq \msx_{i-1}^{(\ell+1)}$.

When $2N < i_0$, for $2N < i \leq i_0$, we note that $\msc_i=\cnunit$.
However, $2N < i_0=N+\lfloor \tfrac{w}{2} \rfloor$ implies $N < \lfloor \tfrac{w}{2} \rfloor$.
Thus, if $2N < i \leq i_0$, then we have
\begin{align*}
  2N-w < i-w &\leq i_0-w \\
  &= N + \lfloor \tfrac{w}{2} \rfloor - w \\
  & \leq N - \lfloor \tfrac{w}{2} \rfloor \qquad \text{(Using $2 \lfloor \tfrac{w}{2} \rfloor \leq w$)} \\
  & < 0.
\end{align*}
As such, $\msc_{i-w}=\cnunit$.
Here again, $\msc_{i} \degreq \msc_{i-w}$ and $\msx_{i}^{(\ell+1)} \degreq \msx_{i-1}^{(\ell+1)}$.

By letting $\ell \to \infty$, we have $\msx_{i} \degreq \msx_{i-1}$ by Proposition \ref{proposition:entropy_distance_properties}, where $\msbx$ is the limit of $\{\msbx^{(\ell)}\}$.

\subsection{Proof of Lemma \ref{lemma:first_derivative_coupled_system_ldpc}}
\label{appendix:proof_first_derivative_coupled_system_ldpc}
The linearity of the entropy functional and the properties of the operators $\vnop$ and $\cnop$ (e.g., see Proposition \ref{proposition:directional_derivative_single_operator}) allow one to write
\begin{align*}
  \deri{\msbx} \potc( \msbx ; \msc ) [\msby] = \sum_{i=1}^{\chend} \deri{\msx_{i}} \potc (\msbx ; \msc) [\msy_i] .
\end{align*}
As in the proof of Lemma \ref{lemma:firstderivative_singlesystem_ldpc}, using the duality rule for entropy for differences of symmetric measures, the derivatives of the first three terms of $\potc$ in (\ref{equation:coupled_potential_ldpc}) are
\begin{align*}
  \deri{\msx_i} \ent{ R^{\cnop}(\msx_i)} [\msy_i] &= R'(1) \ent{ \rho^{\cnop}(\msx_i) \cnop \msy_i }, \\
  \deri{\msx_i} \ent{ \rho^{\cnop}(\msx_i) }[\msy_i] &= \ent{ \rho'^{\cnop}(\msx_i) \cnop \msy_i }, \\    
  \deri{\msx_i}\ent{ \msx_i \cnop \rho^{\cnop}(\msx_i) } [\msy_i] &= \ent{ \rho^{\cnop}(\msx_i) \cnop \msy_i } \!+\! \ent{  \rho'^{\cnop}(\msx_i) \cnop \msy_i } \\
  & \qquad - \ent{ \msx_i \vnop \left[ \rho'^{\cnop}(\msx_i) \cnop \msy_i \right]} .
\end{align*}
For the final term in (\ref{equation:coupled_potential_ldpc}), observe that if $w \leq i \leq 2N$, since there are exactly $w$ components containing $\msx_i$, its derivative with respect to $\msx_i$ is
\begin{align*}
  \tfrac{L'(1)}{w} \sum_{k=0}^{w-1} \mathrm{H} \Big{(}\msc \!\vnop\! \lambda^{\vnop} \Big{(} \frac{1}{w} \sum_{j=0}^{w-1} \rho^{\cnop}(\msx_{i-k+j}) \Big{)} \!\vnop\! (\rho'^{\cnop}(\msx_i) \!\cnop\! \msy_i) \Big{)} .
\end{align*}
If $1 \leq i < w$, derivative of the final term in (\ref{equation:coupled_potential_ldpc}) with respect to $\msx_i$ is
\begin{align*}
  \tfrac{L'(1)}{w} \sum_{k=0}^{i-1} \mathrm{H} \Big{(}\msc \!\vnop\! \lambda^{\vnop} \Big{(} \frac{1}{w} \sum_{j=0}^{w-1} \rho^{\cnop}(\msx_{i-k+j}) \Big{)} \!\vnop\! (\rho'^{\cnop}(\msx_i) \!\cnop\! \msy_i) \Big{)} .
\end{align*}
This can be written as
\begin{align*}
  \tfrac{L'(1)}{w} \!\sum_{k=0}^{w-1} \!\mathrm{H} \Big{(}\msc_{i-k} \!\vnop\! \lambda^{\vnop} \Big{(} \frac{1}{w} \!\sum_{j=0}^{w-1} \rho^{\cnop}(\msx_{i-k+j}) \Big{)} \!\vnop\! (\rho'^{\cnop}(\msx_i) \!\cnop\! \msy_i) \Big{)} ,
\end{align*}
where $\msc_{i}=\msc$ when $1 \leq i \leq 2N$ and $\msc_{i}=\cnunit$ otherwise.
This is because $\ent{\cnunit \vnop \msx}=0$ for any $\msx$, and hence the additional terms that are added evaluate to zero.
A similar expression holds when $2N < i \leq \chend$.
Combining these observations, the derivative of the final term in (\ref{equation:coupled_potential_ldpc}) with respect to $\msx_i$ for $1 \leq i \leq \chend$ is
\begin{align*}
  \tfrac{L'(1)}{w} \sum_{k=0}^{w-1} \!\mathrm{H} \Big{(}\msc_{i-k} \!\vnop\! \lambda^{\vnop} \Big{(} \frac{1}{w} \!\sum_{j=0}^{w-1} \rho^{\cnop}(\msx_{i-k+j}) \Big{)} \!\vnop\! (\rho'^{\cnop}(\msx_i) \!\cnop\! \msy_i) \Big{)} ,
\end{align*}
which is 
\begin{align*}
  L'(1) \ent{ \dec (\msbx ; \msc)_i \vnop (\rho'^{\cnop}(\msx_i) \cnop \msy_i)} .
\end{align*}
Consolidating these four terms and using Proposition \ref{proposition:duality_difference} results in (\ref{equation:first_derivative_coupled_system_ldpc}).

\subsection{Proof of Lemma \ref{lemma:second_derivative_coupled_system_ldpc}}
\label{appendix:proof_second_derivative_coupled_system_ldpc}
We have
\begin{align*}
  \dderi{\msbx} \potc( \msbx ; \msc ) [\msby, \msbz] &= \sum_{m=1}^{\chend} \sum_{i=1}^{\chend} \deri{\msx_{m}} \left( \deri{\msx_{i}} \potc (\msbx ; \msc) [\msy_i] \right) [\msz_{m}] .
\end{align*}
Using the calculations for $\deri{\msx_{i}} \potc (\msbx ; \msc) [\msy_i]$ in Appendix \ref{appendix:proof_first_derivative_coupled_system_ldpc}, it is tedious but straightforward to obtain the desired result. 

\section{Proofs From Section \ref{section:maintheorem_ldpc}}
\label{appendix:section_maintheorem_ldpc}

\subsection{Proof of Lemma \ref{lemma:potential_shift_bound_ldpc}}
\label{appendix:proof_potential_shift_bound_ldpc}
Due to the boundary condition $\msx_{i}=\msx_{i_0}$, for $i_0 \leq i \leq \chend$, the only terms that contribute to $\potc(\shft(\msbx) ; \msc) - \potc(\msbx ; \msc)$ are given by 
\begin{align*}
  & \potc(\shft(\msbx) ; \msc) - \potc(\msbx ; \msc)= \\
  & - \tfrac{L'(1)}{R'(1)} \ent{ R^{\cnop}(\msx_{\chend}) } - L'(1) \ent{ \rho^{\cnop}(\msx_{\chend}) } \\
  & +\!L'(1) \ent{ \msx_{\chend} \!\cnop\! \rho^{\cnop}(\msx_{\chend}) } \!+\! \mathrm{H}\Big{(}\!\msc \!\vnop\! L^{\vnop} \Big{(} \frac{1}{w}\!\sum_{j=0}^{w-1}\!\rho^{\cnop}(\msx_{2N+j}) \Big{)} \Big{)} \\
  & - \mathrm{H}\Big{(} \msc \vnop L^{\vnop} \Big{(} \frac{1}{w}  \sum_{j=0}^{w-1} \rho^{\cnop} ( \ms{x}_{j}) \Big{)} \Big{)} , \quad \text{where $\msx_{0}=\cnunit$.}
\end{align*}
Since $\msx_{2N+j} \upgreq \msx_{\chend}=\msx_{i_0}$ for $0\leq j \leq w-1$ and the contribution from the last term is negative,
\begin{align*}
  & \potc(\shft(\msbx) ; \msc) - \potc(\msbx ; \msc) \\
  & \leq - \tfrac{L'(1)}{R'(1)} \ent{ R^{\cnop}(\msx_{\chend}) } - L'(1) \ent{ \rho^{\cnop}(\msx_{\chend}) } \\
  & \quad +\!L'(1) \ent{ \msx_{\chend} \!\cnop\! \rho^{\cnop}(\msx_{\chend}) } \!+\! \mathrm{H}\Big{(}\!\msc \!\vnop\! L^{\vnop} \Big{(} \!\rho^{\cnop}(\msx_{\chend}) \Big{)} \Big{)} \\
  & = -\pots(\msx_{\chend};\msc) = -\pots(\msx_{i_0};\msc) .
\end{align*}

\subsection{Proof of Lemma \ref{lemma:modifiedsystem_potential_fixedpoint_ldpc}}
\label{appendix:proof_modifiedsystem_potential_fixedpoint_ldpc}

Since $\msbx$ is a fixed point of the modified system,
\begin{align*}
  \msx_{i} = \dec(\msbx;\msc)_{i} ,
\end{align*}
for $1 \leq i \leq i_0 $.
Since $\msx_{i}=\msx_{i-1}$ for $i_{0} < i \le \chend$, we have $[\shft(\msbx) - \msbx]_i = 0$.
The first result follows from applying these relations to the directional derivative given in Lemma \ref{lemma:first_derivative_coupled_system_ldpc}.

Below, we show that $\msx_{i_0} \not \in \mc{V}(\msc)$.
By assumption, we know that $\msbx \degr \underline{\cnunit}$, and by Lemma \ref{lemma:fp_monotonicity_modifiedsystem_ldpc}, $\msx_{i} \degreq \msx_{i-1}$.
Thus, $\msx_{i_0} \degr \cnunit$.
Also, 
\begin{align*}
  \msx_{i_0} = \dec(\msbx;\msc)_{i_0} & = \frac{1}{w} \sum_{k=0}^{w-1} \msc_{i_0-k} \vnop \lambda^{\vnop} \Big{(} \frac{1}{w} \sum_{j=0}^{w-1} \rho^{\cnop} (\msx_{i_0-k+j})  \Big{)} \\
  & \upgreq \frac{1}{w} \sum_{k=0}^{w-1} \msc_{i_0-k} \vnop \lambda^{\vnop} \Big{(} \frac{1}{w} \sum_{j=0}^{w-1} \rho^{\cnop} (\msx_{i_0})  \Big{)} \\
  & \upgreq \msc \vnop \lambda^{\vnop} \left( \rho^{\cnop} \left( \msx_{i_0}  \right)  \right) \\
  & = \des(\msx_{i_0};\msc).
\end{align*}
Hence, by Lemma \ref{lemma:de_monotonicity_ldpc}, $\des^{(\infty)}(\msx_{i_0};\msc) \degreq \des(\msx_{i_0};\msc) \degreq \msx_{i_0} \degr \cnunit$.
Thus $\msx_{i_0} \notin \mc{V}(\msc)$.

\subsection{Proof of Lemma \ref{lemma:second_derivative_bound_ldpc}}
\label{appendix:proof_second_derivative_bound_ldpc}
Let $\msby = \shft(\msbx) - \msbx$, with componentwise decomposition
\begin{align*}
  \msy_{i} = [\shft(\msbx) - \msbx]_{i}  = \msx_{i-1} - \msx_{i},
\end{align*}
where $\msx_{i} = \cnunit$ for $i<1$.
Since $\msbx$ is a fixed point of the modified system, if $i>i_0$, due to the saturation constraint, $\msx_i=\msx_{i-1}$.
If $1 \leq i \leq i_0$, then using the update in (\ref{equation:scde_update_1_ldpc}) gives
\begin{align*}
  \msx_{i-1}-\msx_{i} &= \frac{1}{w} \msc_{i-w} \vnop \lambda^{\vnop} \Big{(} \frac{1}{w} \sum_{j=0}^{w-1} \rho^{\cnop}(\msx_{i-w+j}) \Big{)} \\
  & \qquad - \frac{1}{w} \msc_{i} \vnop \lambda^{\vnop} \Big{(} \frac{1}{w} \sum_{j=0}^{w-1} \rho^{\cnop}(\msx_{i+j}) \Big{)} .
\end{align*}
Thus, $\msy_i=\msx_{i-1}-\msx_i$ is of the form $\tfrac{1}{w}\msa_i-\tfrac{1}{w}\msb_i$, $\msa_i,\msb_i \in \probs$ for all $i$ (if $i>i_0$, $\msa_i=\msb_i$).
From Lemma \ref{lemma:second_derivative_coupled_system_ldpc} and (\ref{equation:second_derivative_coupled_system_ldpc}), the first three terms of the second-order directional derivative are of the form, for some $\msd \in \probs$, 
\begin{align*}
  & \ent{ \msd \cnop \msy_{i} \cnop \msy_{i} } = \frac{1}{w} \ent{ \msd_{3} \cnop (\msb_i - \msa_i)  \cnop (\msx_{i} - \msx_{i-1})} ,
\end{align*}
by linearity of the entropy functional.
From Lemma \ref{lemma:fp_monotonicity_modifiedsystem_ldpc}, $\msx_{i} \degreq \msx_{i-1}$, and by Proposition \ref{proposition:entropy_bound}, this term is absolutely bounded by
\begin{align*}
  & \abs{ \ent{ \msd \cnop \msy_{i} \cnop \msy_{i} } } \leq \frac{ 1 }{w}  \ent{ \msx_{i} - \msx_{i-1}}.
\end{align*}
The final term is of the form, for some $\msd_1,\msd_2,\msd_3,\msd_4,\msd_5 \in \probs$,
\begin{align*}
  & \abs{ \ent{ \msd_{1} \vnop \left[ \msd_{2} \cnop  \msy_{m} \right] \vnop \left[ \msd_{3} \cnop \msy_{i} \right] } }  \\
  & = \abs{ \ent{ \left[ \msd_{1} \vnop \left( \msd_{2} \cnop  \msy_{m} \right) \right] \cnop \left[ \msd_{3} \cnop \msy_{i} \right] } } \quad \text{(Proposition \ref{proposition:duality_difference})}\\
  & = \abs{ \ent{ \msd_{3} \cnop \left[ \msd_{1} \vnop \left( \msd_{2} \cnop  \msy_{m} \right) \right]   \cnop \msy_{i} } } \\
  & = \frac{1}{w}\abs{ \ent{ \msd_{3} \cnop \left[ \msd_5 \!-\! \msd_4 \right] \cnop [\msx_{i} \!-\! \msx_{i-1}]}} \quad (\msy_m\!=\!\tfrac{1}{w}\msa_m\!-\!\tfrac{1}{w}\msb_m)\\
  & \leq \frac{1}{w} \ent{\msx_{i} - \msx_{i-1}}. \quad \text{(Proposition \ref{proposition:entropy_bound})}
\end{align*}
By telescoping, one observes
\begin{align*}
  \sum_{i=1}^{\chend} \ent{\msx_{i} - \msx_{i-1}} = \ent{ \msx_{\chend} - \cnunit } \leq 1 .
\end{align*}
Combining these observations, the triangle inequality provides
\begin{align*}
  & \abs{\dderi{\msbx_1} \potc (\msbx_1 ; \msc) [\msby, \msby]} \\
  & \quad \leq L'(1)\left( 2 \rho''(1)\frac{1}{w} + \rho'(1)\frac{1}{w} + 2w \frac{\lambda'(1)\rho'(1)^2}{w} \frac{1}{w}  \right) \\
  & \quad = \frac{L'(1)\left( 2 \rho''(1) + \rho'(1) + 2\lambda'(1)\rho'(1)^2 \right)}{w} .
\end{align*}

\section{Proofs from Section \ref{section:maintheorem_ldgm}}
\label{appendix:section_maintheorem_ldgm}

\subsection{Proof of Lemma \ref{lemma:potential_shift_bound_ldgm}}
\label{appendix:proof_potential_shift_bound_ldgm}

Due to the boundary condition $\msx_{i}=\msx_{i_0}$ for $i_0 < i \leq \chend$ and by assumption $i_0 \leq 2N$, the terms that contribute to $\potc(\shft(\msbx) ; \msc) - \potc(\msbx ; \msc)$ are given by
\begin{align*}
  & \potc(\shft(\msbx) ; \msc) - \potc(\msbx ; \msc)  \\
  & \quad = \pots(\minf;\msc) - \pots(\msx_{i_0};\msc) \\
  & \quad + L'(1)\mathrm{H} \Big{(}\minf \vnop \Big{[} \frac{1}{w} \sum_{j=0}^{w-1} \msc \cnop \rho^{\cnop}(\msx_{j}) \Big{]} \Big{)} \\
  & \quad - L'(1) \ent{\minf \vnop \left[\msc \cnop \rho^{\cnop}(\minf) \right]} \\
  & \quad - \mathrm{H} \Big{(} L^{\vnop} \Big{(} \frac{1}{w}  \sum_{j=0}^{w-1} \msc \cnop \rho^{\cnop} ( \ms{x}_{j}) \Big{)} \Big{)} + \ent{L^{\vnop}(\msc \cnop \rho^{\cnop}(\minf))},
\end{align*}
where $\msx_{0}=\minf$.
It suffices to show that the contribution from the last four terms is negative.
Define $F\colon\probs^{w} \rightarrow \mbb{R}$ by
\begin{align*}
  F(\msbx) & =  L'(1) \mathrm{H}\Big{(} \minf \!\vnop\! \Big{[} \frac{1}{w} \sum_{j=0}^{w-1} \msc \cnop \rho^{\cnop}(\msx_{j}) \Big{]} \Big{)} \\
  & - L'(1)\ent{\minf \!\vnop\! \left[\msc \cnop \rho^{\cnop}(\minf) \right]} \\
  & - \mathrm{H} \Big{(} L^{\vnop} \Big{(} \frac{1}{w}  \sum_{j=0}^{w-1} \msc \!\cnop\! \rho^{\cnop} ( \ms{x}_{j}) \Big{)} \Big{)}  \!+\! \ent{L^{\vnop}(\msc \!\cnop\! \rho^{\cnop}(\minf))} .
\end{align*}
It is easy to see that $F(\minfb)=0$, where $\minfb=[\minf,\cdots,\minf]$.
For fixed $\msbx \degr \minfb$, define $\phi\colon[0,1] \rightarrow \mbb{R}$ as
\begin{align*}
  \phi(t) = F(\minfb + t(\msbx - \minfb)) .
\end{align*}
Then, $\phi(0)=F(\minfb)$, $\phi(1)=F(\msbx)$ and for $t\in [0,1]$,
\allowdisplaybreaks{
  \begin{align*}
    & \phi'(t) \\
    &=\deri{\msbx_1} F(\msbx_1)[\msbx - \minfb] \bvert{\msbx_1=\minfb + t(\msbx - \minfb)}\\
    &= \frac{L'(1)}{w} \sum_{i=0}^{w-1} \mathrm{H}\Bigg{(} \Big{[} \minf \!-\! \lambda^{\vnop}\Big{(} \frac{1}{w} \sum_{j=0}^{w-1}\msc \!\cnop\! \rho^{\cnop}(t \msx_j \!+\! (1-t)\minf) \Big{)} \Big{]} \\
    & \qquad \qquad \qquad \quad \vnop \Big{[} \msc \!\cnop\! \rho'^{\cnop}(t \msx_i+ (1-t)\minf) \cnop (\msx_{i} - \minf )  \Big{]}  \Bigg{)} \\
    &= \frac{L'(1)}{w} \sum_{i=0}^{w-1} \mathrm{H}\Bigg{(} \Big{[} \lambda^{\vnop}\Big{(} \frac{1}{w} \sum_{j=0}^{w-1}\msc \!\cnop\! \rho^{\cnop}(t \msx_j \!+\! (1-t)\minf) \Big{)} \!-\! \minf \Big{]} \\
    & \qquad \qquad \qquad \quad \cnop \Big{[} \msc \!\cnop\! \rho'^{\cnop}(t \msx_i + (1-t)\minf) \cnop (\msx_{i} - \minf )  \Big{]}  \Bigg{)} .
  \end{align*}
}
Also, since $\msbx \degr \minfb$, $\msx_i \degreq \minf$ and $t \msx_j + (1-t) \minf \degreq \minf$.
Thus,
\begin{align*}
  &\lambda^{\vnop}\Big{(} \frac{1}{w} \sum_{j=0}^{w-1}\msc \!\cnop\! \rho^{\cnop}(t \msx_j \!+\! (1-t)\minf) \Big{)} \\
  &\qquad \degreq \lambda^{\vnop}\Big{(} \frac{1}{w} \sum_{j=0}^{w-1}\msc \!\cnop\! \rho^{\cnop}(\minf) \Big{)} = \lambda^{\vnop} (\msc \cnop \rho^{\cnop}(\minf)) = \minf ,
\end{align*}
since $\minf$ is a fixed point.
By Proposition \ref{proposition:entropy_dpros_properties}(iii), $\phi'(t) \leq 0$.
Thus, $\phi(1)\leq \phi(0)$, which implies $F(\msbx) \leq F(\minfb)=0$ for any $\msbx \degr \minfb$.
Consequently,
\begin{align*}
  \potc(\shft(\msbx) ; \msc) - \potc(\msbx ; \msc) \leq \pots(\minf; \msc) - \pots(\msx_{i_0} ;\msc) .
\end{align*}

\section{Negativity of Potential Functional Beyond Potential Threshold}
\label{appendix:negativity_potential_functional}

In this section, we discuss negativity of the potential functional (Lemma \ref{lemma:beyond_potential_threshold_negativity}) beyond the potential threshold when $\h^*=\hstab$.

Suppose $\h^*=\hstab$.
Consider any $\h > \hstab$ and observe that 
\begin{align*}
  \lambda'(0)\rho'(1)\mf{B}(\msc(\h)) > 1.
\end{align*}
For some $\msx \in \probs$, define $\phi\colon[0,1] \to \mbb{R}$,
\begin{align*}
  \phi(t) = \pots(\cnunit + t(\msx - \cnunit) ; \msc(\h)) .
\end{align*}
According to Proposition \ref{proposition:polynomial_structure_example}, note that $\phi$ is a polynomial in $t$, and $\phi(0)=0$.
By Lemma \ref{lemma:firstderivative_singlesystem_ldpc}, since $\cnunit$ is a fixed point of single system DE, $\phi'(0)=0$.
Moreover,
\begin{align*}
  &\phi''(0) \\
  &= L'(1) \mathrm{H} \Big{(} \left[ \msy - \msc(\h) \vnop \lambda'^{\vnop} \left( \rho^{\cnop}(\cnunit) \right) \vnop [\rho'^{\cnop}(\cnunit) \cnop \msy] \right] \\
  & \qquad \qquad \vnop \left[ \rho'^{\cnop}(\cnunit) \cnop \msy \right] \Big{)} , \quad \text{where $\msy=\msx-\cnunit$} .\\
  &= L'(1)\rho'(1) \ent{\left[ \msy - \lambda'(0)\rho'(1) \msc(\h) \vnop \msy \right] \vnop \msy} \\
  &= L'(1)\rho'(1) \ent{ \msx \vnop \msx - \lambda'(0)\rho'(1) \msc(\h) \vnop \msx \vnop \msx } .
\end{align*}
For a family of BEC or binary input AWGN channels, we can choose $\msx \in \probs$ such that
\begin{align*}
  \msx^{\vnop 2}=\msc(\h)^{\vnop n} \quad \text{for any $n \in \mbb{N}$}.
\end{align*}
For such a choice of $\msx$,
\begin{align*}
  \phi''(0) &= L'(1)\rho'(1) \ent{ \msc(\h)^{\vnop n} - \lambda'(0)\rho'(1) \msc(\h)^{\vnop n+1} } \\
  &= \frac{L'(1)\rho'(1)}{(\lambda'(0)\rho'(1))^n} (f(n) - f(n+1)) ,
\end{align*}
where
\begin{align*}
  f(n) = \left( \lambda'(0) \rho'(1) \right)^n \ent{\msc(\h)^{\vnop n}} .
\end{align*}
Since $\lambda'(0)\rho'(1)\mf{B}(\msc(\h)) > 1$, by Proposition \ref{proposition:entropy_vnop_decay_rate},
\begin{align*}
  \lim_{n \to \infty} \frac{1}{n} \log f(n) = \lambda'(0)\rho'(1)\mf{B}(\msc(\h)) > 1 .
\end{align*}
As such,
\begin{align*}
  \lim_{n \to \infty} f(n) = \infty,
\end{align*}
and thus there exists $m \in \mbb{N}$ such that $f(m)<f(m+1)$. 
Thus, for a suitable choice of $\msx$ such that $\msx^{\vnop 2} = \msc(\h)^{\vnop m}$, we have $\phi''(0)<0$.
Since $\phi$ is a polynomial with $\phi(0)=\phi'(0)=0$, there exists a $t \in (0,1]$ such that $\phi(t)=\pots(\cnunit + t(\msx - \cnunit) ; \msc(\h))<0$.
Thus, we have produced a suitable $\msx$ for which $\pots(\msx ; \msc(\h)) < 0$.
This completes the discussion for BEC and binary input AWGN channels.

For general BMS channels, we can show the same result under the condition
\begin{align*}
  \lim_{n \to \infty} \frac{\ent{\msx^{\vnop n+1}}}{\ent{\msx^{\vnop n}}} = \mf{B}(\msx) .
\end{align*}
For this to hold, by Proposition \ref{proposition:entropy_vnop_decay_rate}, it suffices to show that the limit $\lim_{n \to \infty} \ent{\msx^{\vnop n+1}} / \ent{\msx^{\vnop n}}$ exists.
One way to guarantee the existence of such a limit is to show that the sequence of numbers $\{\ent{\msx^{\vnop n}} \}$ is log-convex,
\begin{align*}
  \ent{\msx^{\vnop n+1}} \ent{\msx^{\vnop n-1}} \geq \ent{\msx^{\vnop n}}^2 ,
\end{align*}
which itself follows by showing that the sequence $\{\ent{\msx^{\vnop n}}\}$ is completely monotonic \cite[Proposition 4.7, Appendix A]{Steutel-03}.
That is, the $k$-th differences of the sequence $\{\ent{\msx^{\vnop n}}\}$,
\begin{align*}
  \ent{\msx^{\vnop n} \vnop (\msx - \vnunit)^{\vnop k}} = (-1)^{k} \ent{\msx^{\vnop n} \vnop (\vnunit - \msx)^{\vnop k}},
\end{align*}
have the sign $(-1)^k$.
That first and second differences of this sequence have the sign $-1$ and $+1$, respectively, follows from Proposition \ref{proposition:entropy_dpros_properties}.
However, it remains to show
\begin{align*}
  \ent{\msx^{\vnop n} \vnop (\vnunit - \msx)^{\vnop k}} > 0 , \quad \text{for $k>2$.}
\end{align*}

\section{Connecting the Potential Functional and the Replica-Symmetric Free Entropy}
\label{appendix:potential_free_entropy}

The purpose of this section is to provide pedagogical insight into the potential functional.
As such, the following discussion is independent from the results of this article and the uninterested reader may skip this section of the appendix.

The potential functional in Definition \ref{definition:uncoupled_potential_ldpc} can be viewed as a Lyapunov function.
For the problem at hand, the negative of the replica-symmetric (RS) free entropy associated with the code ensemble is both a ``natural'' and an ``optimal'' Lyapunov function.
It is optimal in the sense that it allows one to prove threshold saturation up to the MAP threshold (as $w \to \infty$), and it is natural because of its connection to RS formulas of statistical physics.
Below, we first describe the RS free entropy for a general statistical mechanical system and then show how the corresponding expression for an LDPC ensemble reduces to the negative of the potential functional in Definition \ref{definition:uncoupled_potential_ldpc}.
We then briefly describe how the calculations change for LDGM ensembles.
The choice of the negative sign for the potential is a convention for consistency with \cite{Takeuchi-ieice12,Yedla-istc12,Yedla-itw12}.\footnote{This convention is also consistent with physics concepts: because parity-checks of LDPC codes are hard constraints, the RS free entropy is the negative of the RS free energy, thus the potential functional is the RS free energy. Moreover, in physics, entropies are maximized and energies, potentials are minimized.}

\subsection{RS Free Entropy of General Graphical Models}
\label{appendix:subsection_free_entropy_general}

Consider a graphical model on a bipartite graph $G=(V,C,E)$ with variable-node set $V$, a factor-node set $C$, and a set $E$ of edges connecting variable- and factor-nodes. 
Let $\mc{A}$ be a discrete alphabet (for example $\mc{A}= \{0, 1\}$).
Then, $\mc{A}^{|V|}$ is the set of all possible assignments to the variable-nodes.
For $i \in V$, we denote the neighborhood of $i$ $\partial i$ as the set of all factor-nodes $a$ such that $(i,a) \in E$; for $a \in C$, a similar definition is given for $\partial a$.
For $\underline{x} \in \mc{A}^{|V|}$ and a subset $U \subset V$, we write $(x_i)_{i \in U}$ for the collection of elements in $\{x_i | i \in U \}$.

Each variable-node $i \in V$ has an associated weight function $g_i\colon \mc{A} \to [0,\infty)$, and each factor-node $a \in C$ has an associated function $f_a\colon \mc{A}^{|\partial a|} \to [0,\infty)$, which is a mapping from assignments of variable-nodes in $\partial a$, i.e. a function acting on unordered sets.
One is generally interested in the marginals of the probability measure
\begin{align*}
  P(\underline x) = \frac{1}{Z}\prod_{a\in C} f_a((x_i)_{i\in \partial a})\prod_{i\in V} g_i(x_i) ,
\end{align*}
where the normalizing factor
\begin{align*}
  Z = \sum_{\underline x \in \mc{A}^{|V|}}\prod_{a\in C} f_a((x_i)_{i\in \partial a})\prod_{i\in V}g_i(x_i)
\end{align*}
is called the partition function. 
The free entropy is defined as
\begin{align*}
  \frac{1}{|V|} \log Z .
\end{align*}
The quantity $\log Z$ is closely related to the conditional entropy of the input in a communication channel given the output, and thus it naturally appears in a MAP decoding problem.
See \cite[Section 15.4]{Mezard-2009} for more details.

It is well known that when $G$ is a \emph{tree}, a recursive evaluation of the sums allows one to solve for the marginals and the partition function exactly using the message passing formulas:
\begin{align*}
  \mu_{i\to a}(x_i)&=\frac{g_i(x_i)\prod_{b\in \partial i\setminus a} \hat{\mu}_{b\to i}(x_i)}{\sum_{x_i\in\mathcal{A}}g_i(x_i)\prod_{b\in \partial i\setminus a} \hat{\mu}_{b\to i}(x_i)} \\
  \hat{\mu}_{a\to i}(x_i) & =  \frac{\sum_{(x_j)_{j\in \partial a\setminus i}}f_a((x_j)_{j\in \partial a}) \prod_{j\in \partial a\setminus i} \mu_{j\to a}(x_j)}{\sum_{(x_j)_{j\in \partial a}}f_a((x_j)_{j\in \partial a}) \prod_{j\in \partial a\setminus i} \mu_{j\to a}(x_j)} .
\end{align*}
On a tree, these formulas are solved by initializing the messages emanating from leaf nodes and then recursively computing all the other messages.
When a leaf node is the factor-node $a$, the outgoing message is $\hat{\mu}_{a\to i}(x_i)\! \propto \!f_a(x_i)$.
Note that the factor-node degree is one here.
When it is a variable-node $i$, the outgoing message is $\mu_{i\to a}(x_i) \! \propto \! g_i(x_i)$. 
The marginal distribution $\mu_i$ at variable-node $i \in V$ is then given by
\begin{align*}
  \mu_i(x_i) = \frac{g_i(x_i) \prod_{a \in \partial i} \hat{\mu}_{a \to i} (x_i)}{\sum_{x_i \in \mc{A}} g_i(x_i) \prod_{a \in \partial i} \hat{\mu}_{a \to i} (x_i)} .
\end{align*}

The free entropy on a tree is given by the Bethe formula
\begin{align}
  \label{equation:bethe_formula}
  \frac{1}{|V|}\log Z= \frac{1}{|V|}\Big{(} \sum_{i\in V} \varphi_i + \sum_{a\in C} \phi_a - \sum_{(i,a) \in E} \psi_{i,a} \Big{)} ,
\end{align}
where 
\begin{align*}
  \varphi_i&\triangleq \log \Big{(} \sum_{x_i \in \mc{A}} g_i(x_i)\prod_{b\in \partial i} \hat{\mu}_{b\to i}(x_i) \Big{)} , \\
  \phi_a&\triangleq \log \Big{(} \sum_{(x_i)_{i\in \partial a}} f_a((x_i)_{i\in \partial a}) \prod_{j\in \partial a} \hat{\mu}_{j\to a}(x_j)\Big{)} ,\\
  \psi_{i,a}&\triangleq \log \Big{(} \sum_{x_i\in \mc{A}} \mu_{i\to a}(x_i) \hat{\mu}_{a\to i}(x_i) \Big{)} .
\end{align*}
When $G$ is not a tree, it is usually difficult to calculate the free entropy exactly. 
In this case, \eqref{equation:bethe_formula} can be seen as the pseudo-dual of the Bethe free entropy~\cite{Walsh-com10}.
It also provides a first, a priori uncontrolled, approximation for the free entropy.

We now concentrate on \emph{random graphical models} where $G$ is an instance of a random bipartite graph. 
We assume that the functions $f_a$ and $g_i$ are realizations of possibly random functions $f$ and $g$.
For example, the weight function $g_i(x_i;Y_i)$ could be an implicit function of random observation $Y_i$.
An application to LDPC ensembles below will make this framework clear. 
Also, we denote by $\expt[\cdot]$, the expectation with respect to all random objects.

The RS free entropy functional is an average of the Bethe formula (\ref{equation:bethe_formula}) applied to the graph ensemble.
Fix a trial probability measure $\ms{m}$ over the simplex
\begin{align*}
  \left\{(\alpha_1,\cdots,\alpha_{|\mc{A}|}) \in [0,1]^{|\mc{A}|} \mid \sum_i \alpha_i = 1 \right\} .
\end{align*}
Let $\mu=(\mu(x))_{x \in \mc{A}}$ be a random variable distributed according to $\ms{m}$, where the random variables $\mu(x)$, for $x \in \mc{A}$, are its components.
Draw an integer $r_e$ from the \emph{edge-perspective} factor-node degree distribution.
Let $\mu_i$, for $i=1,\cdots,r_e-1$, be iid random variables distributed according to $\ms{m}$.
In the following, we define a new random variable $\hat{\mu}$, over the simplex given above, by its components:
\begin{align*}
  \hat{\mu}(x) \!\triangleq\! \frac{\sum\limits_{(x_1,\dots, x_{r_e-1})\in \mc{A}^{r_e-1}}\! f_a(x, x_1,\cdots, x_{r_e-1}) \prod\limits_{i=1}^{r_e-1} \mu_{i}(x_i)}{\sum\limits_{(x_0, x_1,\cdots, x_{r_e-1})\in \mc{A}^{r_e}}\! f_a(x_0, x_1,\cdots, x_{r_e-1})\! \prod\limits_{i=1}^{r_e-1} \mu_{i}(x_i)} .
\end{align*}
Draw integers $r$, $\ell$ from the \emph{node-perspective} factor- and variable-node degree distributions, respectively.
Let $\mu_i$ for $i=1,\cdots,r$ and $\hat{\mu}_i$ for $i=1,\cdots,\ell$ be independent copies of $\mu$ and $\hat{\mu}$, respectively.

Define the RS free entropy functional, a function of the trial distribution $\ms{m}$, as 
\begin{align}
  \label{equation:RS_functional_general}
  & \Phi_{\mr{RS}}(\ms{m}) \triangleq \expt \Big{[} \log \Big{(} \sum_{x \in \mc{A}} g(x) \prod_{j=1}^\ell \hat{\mu}_{j}(x)\Big{)}\Big{]} \nonumber \\
  & + \frac{L'(1)}{R'(1)}\expt \Big{[} \log \Big{(} \sum_{(x_1,\dots, x_r)\in \mc{A}^r} f(x_1, \cdots, x_r) \prod_{i=1}^r \mu_{i}(x_i) \Big{)} \Big{]} \nonumber \\
  & - L'(1) \expt \Big{[} \log \Big{(} \sum_{x\in \mc{A}} \mu(x) \hat{\mu}(x)\Big{)} \Big{]} .
\end{align}
Each successive term is an average of the variable, factor and edge sums in the Bethe formula \eqref{equation:bethe_formula}.
We note that $\expt[\ell]=L'(1)$ and $\expt[r]=R'(1)$.
The coefficient $L'(1)/R'(1)$ accounts for the average number of factor-nodes per variable-node in the second term, and $L'(1)$ accounts for the average number of edges per variable-node in the third term. 

The RS approximation for the free entropy of a random graphical model is given by the minimum of this functional over an appropriate class of trial measures $\ms{m}$.
This approximation, or it's more sophisticated versions, may or may not be exact. 
Exactness of the RS formulas, if true, is usually difficult to prove and is the subject of various conjectures.  

Finally, we point out that such formulas for sparse graph models were first derived in the framework of the replica method \cite{Wong-jpa87}.
Apart from the conceptual problems related to the replica method, the derivations are also quite algebraically involved for the case of sparse graphs.
The approach presented here via the Bethe formalism is better suited to sparse graphs and is of a more probabilistic nature.

\subsection{Application to LDPC ensembles}
\label{appendix:subsection_free_entropy_ldpc}

We now specialize the RS free entropy functional to the $\text{LDPC}(\lambda, \rho)$ ensemble. 
Here, the alphabet is binary, $\mathcal{A}\in \{0, 1\}$.
The quantity $P(\underline{x})$ is the posterior probability of the input vector given the output vector.
The parity check constraint functions are $f_a((x_i)_{i\in \partial a}) = \mb{1}(\oplus_{i\in \partial a} x_i=0)$, and the weight function at a variable-node is the prior from channel observations, $g_i(x_i) =\Pr(Y_i|x_i)/\Pr(Y_i|0)= e^{-l_i x_i}$, where $l_i$ is the LLR of the memoryless channel output assuming that $0$ was transmitted.\footnote{The random variable $l_i$ is distributed according to the BMS channel $\msc$.}

\begin{remark*}
  It is instructive to note that it is possible to choose different functions $g_i$ without changing $P(\underline{x})$, e.g. $g_i(x_i)=e^{l_i(1-2 x_i)/2}$ is chosen in \cite{Kudekar-it09,Macris-it07}.
  Depending on the choice of $g_i$, the Bethe free entropy may be different.
  However, the estimate of the conditional entropy can be adjusted accordingly and remains independent of the choice of the functions $g_i$.
\end{remark*}

Since the alphabet is binary, we can parameterize the vectors $(\mu(0), \mu(1))$ and $(\hat\mu(0), \hat\mu(1))$ by real valued random variables $\nu$ and $\hat{\nu}$ as follows:
\begin{align*}
  \nu &= \log \frac{\mu(0)}{\mu(1)}, & \hat{\nu}&= \log \frac{\hat{\mu}(0)}{\hat{\mu}(1)} .
\end{align*}
Equivalently, 
\begin{align*}
  \mu(x)&=\frac{1+(-1)^x \tanh \tfrac{\nu}{2}}{2}, & \hat{\mu}(x)&=\frac{1+(-1)^x \tanh \tfrac{\hat{\nu}}{2}}{2} .
\end{align*}

The random variable $\nu$ is distributed according to a trial measure $\ms{n}$.
By taking $r_e-1$ independent copies $\nu_1,\cdots, \nu_{r_e-1}$ of $\nu$, it is easy to show that $\hat{\nu}$ has the same distribution as
\begin{align}
  \label{equation:nuhat_distibution}
  \hat{\nu} \sim 2\tanh^{-1}\left( \prod_{i=1}^{r_e-1}\tanh \tfrac{\nu_i}{2} \right) .
\end{align}
Also, take $r$ independent copies $\nu_1, \cdots, \nu_r$ of $\nu$, and $\ell$ independent copies $\hat{\nu}_1, \cdots, \hat{\nu}_\ell$ of $\hat{\nu}$.
Straightforward algebra shows that the RS free entropy functional in (\ref{equation:RS_functional_general}) is given by
\begin{align}
  \label{equation:RS_functional_LDPC}
  &\Phi_{\mathrm{RS,LDPC}}(\ms{n})  = \nonumber \\ 
  & \expt \Big{[} \log \Big{(} \prod_{j=1}^\ell\frac{1}{2} \Big{[}1+\tanh \tfrac{\hat{\nu}_j}{2} \Big{]} + e^{-l} \prod_{j=1}^\ell \frac{1}{2} \Big{[}1-\tanh \tfrac{\hat{\nu}_j}{2} \Big{]} \Big{)} \Big{]} \nonumber \\
  & + \frac{L'(1)}{R'(1)}\expt \Big{[} \log \Big{(} \frac{1}{2}\Big{[} 1 + \prod_{i=1}^r \tanh \tfrac{\nu_i}{2} \Big{]} \Big{)} \Big{]} \nonumber \\ 
  & - L'(1)\expt \Big{[} \log \Big{(} \frac{1}{2} \Big{[} 1 + \tanh \tfrac{\nu}{2} \tanh \tfrac{\hat{\nu}}{2} \Big{]} \Big{)} \Big{]} ,
\end{align}
where the random variable $l$ is distributed according to the BMS channel $\msc$.
We note that the above expectation $\expt[\cdot]$ includes the average over the LDPC$(\lambda,\rho)$ ensemble via the integers $\ell$ and $r$ drawn according to the variable- and check-node degree distributions, respectively.

We will now relate (\ref{equation:RS_functional_LDPC}) to the potential functional in Definition \ref{definition:uncoupled_potential_ldpc}.
First note that the definitions of the operators $\vnop$ and $\cnop$ in Section \ref{section:preliminaries} imply for any $k\geq 1$
and symmetric measures $\msx_i$, $i=1, \dots, k$,
\begin{align}
  \label{equation:variable_operator_multi}
  \ent{\vnop_{i=1}^k\msx_i} = \int \log_2(1+ e^{-\sum_{i=1}^k\alpha_i}) \prod_{i=1}^k\msx_i(\diff \alpha_i) ,
\end{align}
\begin{align}
  \label{equation:check_operator_multi}
  \ent{\cnop_{i=1}^k\msx_i} = - \int \log_2\Big{(} \frac{1}{2} \Big{[} 1+\prod_{i=1}^k\tanh\tfrac{\alpha_i}{2} \Big{]}\Big{)} \prod_{i=1}^k\msx_i(\diff \alpha_i) .
\end{align}
First consider the second term in (\ref{equation:RS_functional_LDPC}).
Using (\ref{equation:check_operator_multi}), since $\nu_i$ is distributed according to $\ms{n}$,
\begin{align}
  \label{equation:second_term_RS_LDPC}
  & \frac{L'(1)}{R'(1)}\expt \Big{[} \log \Big{(} \frac{1}{2}\Big{[} 1 + \prod_{i=1}^r \tanh \tfrac{\nu_i}{2} \Big{]} \Big{)} \Big{]} \nonumber \\ 
  & \qquad = - (\log 2) \frac{L'(1)}{R'(1)}\ent{R^\cnop(\ms{n})} .
\end{align}
For the third term in (\ref{equation:RS_functional_LDPC}), since $\hat{\nu}$ is distributed according to (\ref{equation:nuhat_distibution}), using (\ref{equation:check_operator_multi}),
\begin{align}
  \label{equation:third_term_RS_LDPC}
  & L'(1)\expt \Big{[} \log \Big{(} \frac{1}{2} \Big{[} 1 + \tanh \tfrac{\nu}{2} \tanh \tfrac{\hat{\nu}}{2} \Big{]} \Big{)} \Big{]}  \nonumber \\ 
  & = L'(1)\expt \Big{[} \log \Big{(} \frac{1}{2} \Big{[} 1 + \tanh \tfrac{\nu}{2} \prod_{i=1}^{r_e-1}\tanh \tfrac{\nu_i}{2} \Big{]} \Big{)} \Big{]} \nonumber \\
  & = -(\log 2) L'(1)\ent{\ms{n} \cnop \rho^\cnop(\ms{n})} .
\end{align}
For the first term in (\ref{equation:RS_functional_LDPC}), we have 
\begin{align}
  \label{equation:first_term_RS_LDPC}
  & \expt \Big{[} \log \Big{(} \prod_{j=1}^\ell \frac{1}{2} \Big{[} 1+\tanh \tfrac{\hat{\nu}_j}{2} \Big{]} + e^{-l} \prod_{j=1}^\ell \frac{1}{2} \Big{[}1-\tanh \tfrac{\hat{\nu}_j}{2} \Big{]} \Big{)} \Big{]} \nonumber \\
  & = \expt \Big{[} \sum_{j=1}^\ell \log \Big{(} \tfrac{1}{2} \Big{[} 1\!+\!\tanh \tfrac{\hat{\nu}_j}{2} \Big{]} \Big{)} \Big{]} \!+\! \expt \Big{[} \log( 1 \!+\! e^{-l - \sum_{j=1}^\ell \hat{\nu}_j})\Big{]} \nonumber \\ 
  & = L'(1) \expt \Big{[} \log \Big{(} \tfrac{1}{2} \Big{[} 1\!+\!\tanh \tfrac{\hat{\nu}}{2} \Big{]} \Big{)} \Big{]} \!+\! \expt \big{[} \log (1\!+\!e^{-l - \sum_{j=1}^\ell \hat{\nu}_i})\Big{]} \nonumber \\
  & = - (\log 2) L'(1) \ent{\rho^\cnop(\ms{n})} + (\log 2) \ent{ \msc \vnop L^\vnop(\rho^\cnop(\ms{n}))} ,
\end{align}
where we used (\ref{equation:nuhat_distibution}), (\ref{equation:check_operator_multi}) and (\ref{equation:variable_operator_multi}) to get the last equality.

Collecting (\ref{equation:first_term_RS_LDPC}), (\ref{equation:second_term_RS_LDPC}), (\ref{equation:third_term_RS_LDPC}), we find that
\begin{align*}
  \Phi_{\mathrm{RS,LDPC}}(\ms{n}) = - (\log 2) \pots(\ms{n};\msc) ,
\end{align*}
which shows that the potential functional is the negative of the RS free entropy functional. 

For completeness, we point out that the conditional entropy $\ent{X^n | Y^n}$ of the input $X^n$ conditional on the output $Y^n$ is equal to the free entropy averaged over the noise realizations $\expt [\ent{X^n | Y^n}] = \expt [\log_2 Z]$.
For a detailed discussion of this relation, see \cite{Montanari-it05,Kudekar-it09,Macris-it07}.
Again, we note that due to different normalizations of the free entropy, additional nuisance terms may appear in these references.
As stated in Lemma \ref{lemma:mapandpotential_connection}, it is shown in these references that 
\begin{align*}
  \mathbb{E}[\ent{X^n|Y^n}] \geq - \inf_{\msx\in \probs} \pots(\msx; \msc(\h)) .
\end{align*}
It is conjectured that this is in fact an equality, and recently the equality has been proven for a class of regular codes and smooth channel families \cite{Giurgiu-arxiv13}.
This is a case where the replica formula allows an exact calculation of the average free entropy.

\subsection{Application to LDGM ensembles}
\label{appendix:subsection_free_entropy_ldgm}

We now briefly describe the calculations involved in obtaining the potential functional for LDGM ensembles in Definition \ref{definition:uncoupled_potential_ldgm}.
Observing the Tanner graph representation of an LDGM code in Fig.~\ref{figure:tanner_graph_ldgm}, each generator-node $a$ is connected to a code-bit $x_a$, and to each code-bit $x_a$ there is an associated observation $l_a$, which is the LLR of the channel output.
The parity-check constraint function at the generator-node $a$ is given by
\begin{align*}
  f_a((u_i)_{i\in \partial a}) = e^{-l_a x_a} \mb{1}(\oplus_{i\in \partial a} u_i \oplus x_a =0) .
\end{align*}
In the set $\partial a$ above, we do not include the neighbor $x_a$.
The weight function at an information-node is given by $g_i(u_i)=1$.

With the above functions, the RS free entropy in (\ref{equation:RS_functional_general}) for LDGM ensembles is given by
\begin{align}
  \label{equation:RS_functional_LDGM}
  &\Phi_{\mathrm{RS,LDGM}}(\ms{n})  = \nonumber \\ 
  & \expt \Big{[} \log \Big{(} \prod_{j=1}^\ell\frac{1}{2} \Big{[}1+\tanh \tfrac{\hat{\nu}_j}{2} \Big{]} + \prod_{j=1}^\ell \frac{1}{2} \Big{[}1-\tanh \tfrac{\hat{\nu}_j}{2} \Big{]} \Big{)} \Big{]} \nonumber \\
  & + \frac{L'(1)}{R'(1)}\expt \Bigg{[} \log \Bigg{(} \frac{ 1 + \!\prod\limits_{i=1}^r \!\tanh \tfrac{\nu_i}{2} + e^{-l} \Big{[} 1-\!\prod\limits_{i=1}^r\! \tanh \tfrac{\nu_i}{2} \Big{]} }{2} \Bigg{)} \Bigg{]} \nonumber \\ 
  & - L'(1)\expt \Big{[} \log \Big{(} \frac{1}{2} \Big{[} 1 + \tanh \tfrac{\nu}{2} \tanh \tfrac{\hat{\nu}}{2} \Big{]} \Big{)} \Big{]} ,
\end{align}
where the random variable $l$ is distributed according to $\msc$, and $\hat{\nu}$ has the same distribution as 
\begin{align*}
    \hat{\nu} \sim 2\tanh^{-1}\left( \tanh \tfrac{l}{2} \prod_{i=1}^{r_e-1}\tanh \tfrac{\nu_i}{2} \right) .
\end{align*}
Proceeding as in the LDPC case, the three terms in (\ref{equation:RS_functional_LDGM}) are, respectively, 
\begin{align*}
  & - (\log 2)L'(1)\ent{\msc \cnop \rho^{\cnop}(\ms{n})}+(\log 2)\ent{L^{\vnop}(\msc \cnop \rho^{\cnop}(\ms{n}))} , \\
  & (\log 2) \frac{L'(1)}{R'(1)} \ent{\msc}-(\log 2) \frac{L'(1)}{R'(1)} \ent{\msc \cnop R^{\cnop}(\ms{n})} , \\
  & (\log 2) L'(1) \ent{\ms{n} \cnop \msc \cnop \rho^{\cnop}(\ms{n})} ,
\end{align*}
which gives the relation
\begin{align*}
  \Phi_{\mathrm{RS,LDGM}}(\ms{n}) = - (\log 2) \pots(\ms{n};\msc) .
\end{align*}

\bibliographystyle{IEEEtran}
\bibliography{WCLabrv,WCLbib,WCLnewbib}

\end{document}